\documentclass[english,reprint,amsmath,amssymb,aps,superscriptaddress]{revtex4-2}
\usepackage[T1]{fontenc}
\usepackage[utf8]{inputenc}
\usepackage{float}
\usepackage{mathtools}
\usepackage{amsmath}
\usepackage{graphicx}

\makeatletter
\usepackage{hyperref}

\makeatother

\usepackage{babel}
\begin{document}
\title{Beyond-mean-field fluctuations for the solution of constraint satisfaction
problems}
\author{Niklas Foos}
\thanks{Equal contribution}
\affiliation{Institute for Advanced Simulation (IAS6), Computational and Systems
Neuroscience, Jülich Research Centre, Jülich, Germany}
\affiliation{RWTH Aachen University, Aachen, Germany}
\author{Bastian Epping}
\thanks{Equal contribution}
\affiliation{Institute for Advanced Simulation (IAS6), Computational and Systems
Neuroscience, Jülich Research Centre, Jülich, Germany}
\affiliation{RWTH Aachen University, Aachen, Germany}
\author{Jannik Grundler}
\affiliation{Institute for Advanced Simulation (IAS6), Computational and Systems
Neuroscience, Jülich Research Centre, Jülich, Germany}
\affiliation{RWTH Aachen University, Aachen, Germany}
\author{Alexandru Ciobanu}
\affiliation{Institute for Advanced Simulation (IAS6), Computational and Systems
Neuroscience, Jülich Research Centre, Jülich, Germany}
\affiliation{RWTH Aachen University, Aachen, Germany}
\affiliation{Neuromorphic Compute Nodes (PGI-14), Peter Grünberg Institut (PGI),
Jülich Research Centre, Jülich, Germany}
\author{Ajainderpal Singh}
\affiliation{Institute for Quantum Computing Analytics (PGI-12), Jülich Research
Centre, Jülich, Germany}
\author{Tim Bode}
\affiliation{Institute for Quantum Computing Analytics (PGI-12), Jülich Research
Centre, Jülich, Germany}
\author{Moritz Helias}
\affiliation{Institute for Advanced Simulation (IAS6), Computational and Systems
Neuroscience, Jülich Research Centre, Jülich, Germany}
\affiliation{Department of Physics, RWTH Aachen University, Aachen, Germany}
\author{David Dahmen}
\thanks{Correspondence: d.dahmen@fz-juelich.de}
\affiliation{Institute for Advanced Simulation (IAS6), Computational and Systems
Neuroscience, Jülich Research Centre, Jülich, Germany}
\begin{abstract}
Constraint Satisfaction Problems (CSPs) lie at the heart of complexity
theory and find application in a plethora of prominent tasks ranging
from cryptography to genetics. Classical approaches use Hopfield networks
to find approximate solutions while recently, modern machine-learning
techniques like graph neural networks have become popular for this
task. In this study, we employ the known mapping of MAX-2-SAT, a class
of CSPs, to a spin-glass system from statistical physics, and use
Glauber dynamics to approximately find its ground state, which corresponds
to the optimal solution of the underlying problem. We show that Glauber
dynamics outperforms the traditional Hopfield-network approach and
can compete with state-of-the-art solvers. A systematic theoretical
analysis uncovers the role of stochastic fluctuations in finding CSP
solutions: even in the absence of thermal fluctuations at $T=0$ a
significant portion of spins, which correspond to the CSP variables,
attains an effective spin-dependent non-zero temperature. These spins
form a subspace in which the stochastic Glauber dynamics continuously
performs flips to eventually find better solutions. This is possible
since the energy is degenerate, such that spin flips in this free-spin
space do not require energy. Our theoretical analysis leads to deterministic
solvers that effectively account for such fluctuations, thereby reaching
state-of-the-art performance.
\end{abstract}
\maketitle

\section{Introduction}

Constraint satisfaction problems (CSPs) play an important role in
complexity theory and arise in various applications such as type inference
in formal languages, coloring problems, scheduling, geometric modeling,
cryptography, genetics, but also in logical puzzles, such as Sudoku
\citep{pierce2000local,ge1999geometric,hell2008colouring,aceto2004complexity,RAMAMOORTHY20232539,brailsford1999constraint}.
Solving these problems requires finding variable assignments such
that a minimal number of constraints are left unsatisfied. The MAX-$K$-SAT
optimization problem is one class of CSPs in which $N$ binary variables
are combined in $M$ constraints (clauses) of $K$ variables (literals)
each, and the task is to find a variable assignment that satisfies
a maximum amount of constraints.

From the side of complexity theory, MAX-$K$-SAT is an interesting
problem, because it is NP-hard. This means it is at least as hard
as NP-complete problems (such as $K\geq3$-SAT \citep{Karp72_85}):
For the latter problems, there is no deterministic algorithm known
that solves them in polynomial time. MAX-$K$-SAT is an NP-hard optimization
problem even in the minimal case of $K=2$, where the corresponding
decision problem $2$-SAT (asking whether all clauses of a CSP can
be satisfied) is solvable in polynomial time (P) \citep{Mezard_Montanari09}.
For NP-complete problems only the verification of a solution can
be done in polynomial time, unless it were generally shown that the
P=NP \citep{goldreich2010p,garey1979computers}. Many tools have thus
been developed to solve NP-complete and NP-hard problems approximately.

For example, a classical line of works employs continuous Hopfield
networks \citep{Hopfield84} and Boltzmann machines to find approximate
optimal solutions to CSPs such as the traveling-salesman problem \citep{looi_neural_1992,hopfield85_141}.
To this effect, one maps the CSP to an Ising system, a model of interacting
magnetic spins pointing either up or down, and defines an energy function
which is minimized by the configuration of spins that corresponds
to the optimal solution of the underlying CSP. Hopfield networks or
Boltzmann machines are then used to find the (possibly degenerate)
energy minimum (also called ground state) approximately. In \citet{Peterson87_995},
the authors calculate the mean-field approximation of the Boltzmann
machine, finding it both to be computationally cheaper and better
performing, suggesting that a problem formulation in continuous variables
simplifies the problem. In a similar manner, the same authors used
the mean-field approximation of discrete Hopfield networks \citep{Hopfield82}
to approximately solve the minimum cut bisection problem \citep{peterson_neural_1988}.
In \citet{bilbro_optimization_1988}, the mean-field approximation
of the Ising formulation of a CSP was shown to be equivalent to the
Hopfield network, and thus their performance is equivalent. Due to
the good performance and its simple architecture, the continuous Hopfield
network retains its relevance to this day \citep{wen_review_2009,sathasivam_election_2020,kasihmuddin_hybrid_2017}.

In more recent studies, machine learning techniques such as graph
neural networks have been used to develop better approximate CSP solvers
(\citet{selsam2018learning,Toenshoff21_580607} and citations therein).
While these models exist in many different versions, they operate
on continuous spaces which are subsequently discretized to infer specific
states of the given CSP. While these algorithms reach good approximate
solutions, their progress and development is driven empirically and
their architecture does not make use of the characteristics of the
CSP's energy landscape.

While the aforementioned studies focus on developing efficient algorithms
to find approximate CSP solutions, it is also worthwhile to investigate
properties of the CSPs on their own. Insights into their structure
may then be used to develop better solvers. In this vein, another
line of works made use of methods from statistical physics developed
for disordered systems, such as spin glasses, to investigate properties
of the MAX-$K$-SAT problem as well as the decision problem, $K$-SAT
\citep{Mezard02_812,monasson1996entropy,kirkpatrick1983optimization,Krzakala07_10318}.
Partly this is done, as described for the above studies, by mapping
the (MAX)-$K$-SAT problem to an Ising system, and identifying its
energy function to represent the number of violated clauses for each
variable/spin configuration. For (MAX)-$K$-SAT, this was first done
in \citet{Monasson97_1357}. Studying MAX-$K$-SAT in the Ising setting
using equilibrium statistical mechanics tools has proven fruitful,
resulting in e.g. state-of-the-art approximate solvers \citep{PhysRevLett.94.197205,Mezard1985,PhysRevE.66.056126}
as well as estimates of the number of ``good\textquotedbl{} solutions
and quantifying the difficulty in finding them in dependence of the
problem's parameters \citep{Monasson97_1357,Mezard02_812}. For random
MAX-$K$-SAT problems, \citet{Monasson97_1357} investigated the phase
transition when going from the underconstrained to the overconstrained
regime of MAX-$K$-SAT, which occurs at a critical ratio between the
number of constraints and the number of variables in the limit of
infinitely large problems. A challenge shared among fixed-temperature
solvers in the Ising formulation is choosing the temperature $T$
to one that allows both to escape local minima and also to reach the
ground state \citep{kirkpatrick1983optimization}. The probability
of a state occurring is determined by the Boltzmann factor $e^{-\frac{E}{T}}$
where $E$ is the energy of the state (setting the Boltzmann constant
to one, $k_{B}=1$). Thus, the system is expected to be in its ground
state (or one of them, if the ground state is degenerate) if one were
to set the temperature to zero. However, the approximations used to
develop the solvers break at low temperatures, forbidding to go to
the zero-temperature case.

In this work, we use Glauber dynamics \citep{Glauber63_294} to simulate
the Ising model. We find these discrete dynamics to outperform the
traditional Hopfield-network approach on established benchmarks given
by random MAX-2-SAT instances. Moreover, we find that the Glauber
dynamics approach works best in the zero-temperature case, overcoming
the limitation to finite temperatures of previous solvers. By employing
a cumulant expansion of the Glauber dynamics \citep{Dahmen16_031024}
to investigate their properties, we find that Glauber dynamics in
their mean-field approximation are equivalent to Hopfield networks.
Including fluctuations in the form of the variances of individual
spins allows us to identify free subspaces of spins in the phase space
that are degenerate in energy; therefore no energy is needed to flip
spins and explore these spaces. This happens even in the absence of
thermal fluctuations due to the probabilistic nature of spin updates
in the Glauber dynamics. From this variance approximation we obtain
a new deterministic solver which reaches state-of-the-art performance
on random MAX-2-SAT instances. In principle, our approach allows us
to calculate the statistics of the Glauber dynamics to arbitrary order.
Here, we calculate the statistics up to the second cumulant and find
a better match between simulated Glauber dynamics and its approximate
version.

\section{MAX-2-SAT as a statistical physics problem}

Formally, MAX-$K$-SAT problems consist of $N$ binary variables $x_{i}\in\{0,1\}$
where $i=1,\dots,N$. Based on $x_{i}$ we can define literals $z_{i}$
which are either $x_{i}$ or their negation $\overline{x}_{i}=1-x_{i}$.
Clauses $C_{l}$ with $l=1,\dots,M$, also called constraints, are
logical \textbf{OR}-connections of these literals, $C_{l}=\vee_{k=1}^{K}z_{i_{k,l}}$
where $z_{i_{k,l}}$ is the $k$-th literal in clause $l$ with index
$i_{k,l}\in\{1,\dots,N\}$. In this work we are concerned with MAX-2-SAT
problems, meaning that in each clause there are at most two literals
(we will exclusively use clauses of length $K=2$). As shown in \citet{Monasson97_1357}
for the broader class of any (MAX-)$K$-SAT problem, this can be mapped
to an Ising model by introducing spin variables $S_{i}=2x_{i}-1$,
$i=1,\dots,N$, corresponding to the Boolean variables $x_{i}$ with
value $S_{i}=1$ in the case of $x_{i}=1$ and $S_{i}=-1$ in the
case of $x_{i}=0$, respectively. The energy function of the resulting
spin system can be written as

\begin{equation}
E(\boldsymbol{S})=\frac{M}{4}-\sum_{i=1}^{N}H_{i}S_{i}-\frac{1}{2}\sum_{i,j=1}^{N}J_{ij}S_{i}S_{j}\,,\label{eq:Energy_function}
\end{equation}
where the parameters $H$, $J$ and $M$ can be calculated from the
clauses $C_{l}$ in a straightforward manner \citep{Monasson97_1357}
(see appendix \ref{app:CSP-to-Ising}) by the condition that the energy
$E(\boldsymbol{S})$ is the number of violated clauses in the MAX-2-SAT
problem for the configuration that corresponds to the state $\boldsymbol{S}$
. From here on, we will use the broader term CSPs to refer to MAX-2-SAT
instances. Using the above mapping, solving a CSP, i.e. finding the
maximum number of satisfied constraints, is equivalent to finding
the energy ground state in its corresponding spin system. This can
be done approximately by using tools from statistical physics. In
doing so one introduces an artificial temperature $T$ and assigns
probabilities $p(\boldsymbol{S})\propto e^{-\frac{E(\boldsymbol{S})}{T}}$
to each state $\boldsymbol{S}$. We measure temperature in units of
energy, such that the Boltzmann constant is $k_{B}=1$. Then, one
can approximate averages $\boldsymbol{m}=\langle\boldsymbol{S}\rangle$
(magnetizations) over this probability distribution. For solving CSPs,
the most interesting temperatures are the ones close to zero, since
in that case the system is most likely to be found in its ground state
which in conclusion would dominate the average $\boldsymbol{m}$.
In practice, however, approximate methods to find $\boldsymbol{m}$
fail in the low temperature case due to the rough energy landscape
with many local minima \citep{Mezard87}. If one were to find the
true magnetizations $\boldsymbol{m}$ at zero temperature, these would
give a state $\boldsymbol{S}^{*}=\boldsymbol{m}$ that would optimally
solve the CSP. Since it is NP-hard to solve the CSP, it thus is also
NP-hard to find the magnetizations at zero temperature or to compute
the partition function exactly (since the magnetizations can easily
be computed from the partition function).

\subsection{Hopfield networks and Glauber dynamics}

In this study, we focus on investigating and comparing mainly two
different methods used to find the minimum energy state for the spin
system written above. For one, we take the traditional Hopfield network
approach \citep{Hopfield84}. This is a common approach for solving
CSPs \citep{wen2009review,talavan2002parameter,wang1995using,kasihmuddin_hybrid_2017,feng2000using}.
In the original work \citep{hopfield85_141}, this has been used successfully
for the traveling-salesman problem. For another, we will use Glauber
dynamics \citep{Glauber63_294} of the system (\ref{eq:Energy_function})
to sample states corresponding to minimal energy. Both methods are
introduced in this section, starting with the Hopfield network.

In the continuous Hopfield model, variables model binary neurons ($S_{i}=+1$
and $S_{i}=-1$ for firing and non-firing neurons respectively) instead
of spins. Then $m_{i}=\tanh(\beta\mu_{i})$ takes the role of an average
firing of neuron $i$ which depends on its input voltage $\mu_{i}$.
We will find later that, when considering spin systems, $\mu_{i}(\boldsymbol{m})$
acts as an effective local field for spin $i$. The coupling matrix
$J_{ij}$ takes the role of synapses linking the neurons, feeding
outputs $S_{i}$ of neurons as inputs to other neurons while the variables
$H_{i}$ are to be interpreted as an external input to neuron $i$.
The scaling factor $\beta$ originates in biological details of the
neurons and will later be identified with an inverse temperature.
Using these naming conventions as well as neuron-independent time
scales $\tau$ and $\tanh$ as non-linearity, the Hopfield dynamics
can be written as (see Appendix \ref{app:Hopfield-renaming} for details)
\begin{eqnarray}
\tau\frac{d\mu_{i}(t)}{dt} & = & -\mu_{i}(t)+\sum_{i=1}^{N}J_{ij}\tanh\left(\beta\mu_{j}(t)\right)+H_{i}\,.\label{eq:Hopfield_dynamics}
\end{eqnarray}
These dynamics are monotonically decreasing the energy function (\ref{eq:Energy_function})
if one is to replace the spins $S_{i}\in\{-1,+1\}$ with the average
firings $m_{i}\in[-1,+1]$ of the corresponding neurons. The solution
strategy is thus to start from a random initialization of firings
$m_{i}$ and let the Hopfield network evolve until convergence, guaranteeing
that one is in a local minimum of the energy function. However, one
cannot determine from the time evolution whether this is the global
minimum, or how close the energy one converged to is to the ground
state energy. After convergence, the spin values are chosen as $S_{i}=\mathrm{sign}\,m_{i}$.
This is motivated by the fact that the energy of the CSP and the energy
of the spin system (\ref{eq:Energy_function}) are the same for very
high/low firing averages $m_{i}=\pm1$.

The Glauber dynamics describe the physical evolution of the spin system
introduced above \citep{Glauber63_294}. They can be summarized as
follows: In each time interval $\delta t$ each spin $S_{i}$ has
the probability $\frac{1}{\tau}\delta t$ to be chosen for an update.
Here $\tau$ denotes the time scale over which the system evolves.
Throughout this manuscript all times are unitless i.e. considered
relative to some unit time. If a spin $i$ is chosen for an update,
we denote the probability of it being in the up state $S_{i}=1$ or
down state $S_{i}=-1$ after the update with $F^{+}(\boldsymbol{S}\backslash S_{i})$
and $F^{-}(\boldsymbol{S}\backslash S_{i})=1-F^{+}(\boldsymbol{S}\backslash S_{i})$
respectively where $\boldsymbol{S}\backslash S_{i}$ denotes the set
of all spins without the spin $S_{i}$. The expressions for $F^{+}(\boldsymbol{S}\backslash S_{i})$
and $F^{-}(\boldsymbol{S}\backslash S_{i})$ are given in Appendix
\ref{app:gain-function}. As we are ultimately interested in the limit
of infinitesimal small time intervals and the probability of two spins
being updated in the same interval is $\propto\delta t^{2}$ only
one spin at a time is updated at most. The net flux of probability
in a state $\boldsymbol{S}$ is given by the probability of another
state changing into the state $\boldsymbol{S}$ minus the probability
of itself changing into another state. The flux can thus be decomposed
into a sum of the probability fluxes of single spin flips
\begin{eqnarray}
\tau\frac{d}{dt}p(\boldsymbol{S},t) & =\sum_{i=1}^{N} & S_{i}\Big(p(\boldsymbol{S}_{i-})F_{i}^{+}(\boldsymbol{S}\backslash S_{i})\label{eq:Glauber_master}\\
 &  & -p(\boldsymbol{S}_{i+})F_{i}^{-}(\boldsymbol{S}\backslash S_{i})\Big).\nonumber 
\end{eqnarray}
Here, $p(\boldsymbol{S},t)$ is the time-dependent probability distribution
over the spins $\boldsymbol{S}$, and $\boldsymbol{S}_{i\pm}$ denotes
the spin configuration $\boldsymbol{S}$ with $S_{i}=\pm1$. Since
the Glauber dynamics describe a physical process, we can guarantee
that this distribution is well behaved at all times $t$. While equation
(\ref{eq:Glauber_master}) captures the statistics of the Glauber
dynamics, we can also simulate the dynamics as they are described
above. To do this most efficiently, we make the following modification:
We use the fact that at each time point only one spin will be updated.
The time between two updates is, by definition of the Glauber dynamics
above, an exponentially distributed random variable. During this time,
spin configurations remain constant. Since we are only interested
in finding the ground state spin configuration, we can speed up the
simulation of the dynamics by imposing that in each time step exactly
one randomly chosen spin is updated. This corresponds to the average
behavior of the stochastic update described above for the choice $\tau=N\delta t$.
To obtain a solution for a specific CSP we let the system equilibrate
for a fixed time interval and subsequently measure the spins' magnetizations
over another fixed time interval. The master equation (\ref{eq:Glauber_master})
allows the theoretical investigation of these simulations which we
will use to gain mechanistic insights on the role of fluctuations
in section \ref{sec:Theoretical-analysis-of}.

\begin{figure}
\centering{}\includegraphics[width=1\linewidth]{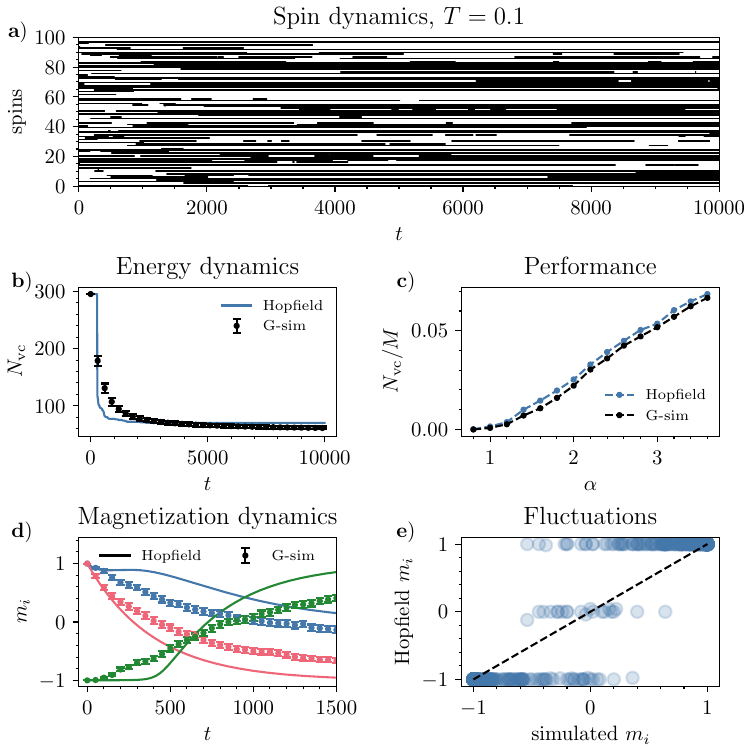}\caption{\protect\label{fig:performance_discrete_Glauber_and_Hopfield}\textbf{Comparison
of simulated Glauber dynamics and Hopfield networks.} $\textbf{a)}$
Glauber spin dynamics in a random CSP with $\alpha=3$ and $N=400$.
The evolution of a subset of $N=100$ spins as a function of time
starting from a random initialization. Black indicates a spin in the
down state $S_{i}=-1$ while white indicates a spin in the up state
$S_{i}=+1$. $\textbf{b)}$ Energy as a function of time for one CSP
instance and a single initial condition. $\textbf{c)}$ Fraction of
violated clauses dependent on clause density $\alpha=M/N$ averaged
over $50$ instances from a single random initial spin configuration.
$\textbf{d)}$ Magnetizations of three randomly chosen spins as a
function of time $t$ for Hopfield and Glauber dynamics (mean and
standard deviation of the mean for $300$ Glauber dynamics realizations)
starting from the same initial condition. $\textbf{e)}$ Magnetizations
of spins in simulated Glauber dynamics and in the Hopfield network
after convergence. All experiments in this figure are done with temperature
$T=0.1$. Details for CSP generation are provided in appendix \ref{app:generation-of-random-csps}.}
\end{figure}

In Figure \ref{fig:performance_discrete_Glauber_and_Hopfield} we
show the result for both simulating the discrete Glauber dynamics
and using the traditional approach of a continuous Hopfield network
at low but non-zero temperature. Panel a) shows the spin evolution
of the simulated Glauber dynamics. Starting from a random configuration,
many spins flip their orientation shortly after the start of the simulation.
At later times spins flip less frequently, but they do not fully freeze.
The qualitatively same behavior can also be seen in panel b), showing
the energy, or equivalently the number of violated clauses, as a function
of time: For both the Glauber dynamics and the Hopfield network the
energy drops rapidly after initialization and seems to converge for
late times $t$ where the energy of the simulation is eventually lower
than that of the Hopfield network. The Glauber dynamics simulation
thus found a better solution than the Hopfield dynamics in this specific
CSP. To investigate the performance with better statistics, we show
the performance of the two solvers for multiple instances with different
clause densities in panel c). Interestingly the discrete Glauber dynamics
outperform the continuous Hopfield network dynamics also here; they
tend to find spin states violating less clauses than the Hopfield
solution. The fraction of violated clauses for a wider range of $\alpha$
is shown in appendix \ref{app:additonal_plots} Figure \ref{fig:app_violated_clauses_wider_range}.
To gain first insights into how these solvers work, we show the magnetizations
of three randomly chosen spins in both the Hopfield network and the
Glauber dynamics in panel d). While the magnetizations are qualitatively
similar, there are clear quantitative differences. In particular,
the magnetizations in the Glauber dynamics are weaker, implying that
there the spins fluctuate more strongly. The magnetization in the
Hopfield network on the other hand quickly freeze after some time
in fully magnetized states. In panel e) we observe these spin magnetizations
after convergence in both the Hopfield network and the Glauber dynamics.
As in panel d) we observe similar magnetizations for most spins but
also clear quantitative differences. Most prominently we notice that
spins in the Hopfield network tend to be fully magnetized, while spins
in the Glauber dynamics may obtain intermediate magnetization values.

The performance plot in panel b) finds the Glauber dynamics to be
more performant, so seemingly the Glauber dynamics captures information
of the underlying CSPs better. As we show in the next section, the
performance difference can be explained by the stochastic spin fluctuations
of the Glauber dynamics, leading to freely fluctuating spins with
intermediate magnetizations which we already observed in panels d)
and e). These fluctuations are absent in the Hopfield network which
thus freezes the spins.

\section{Theoretical analysis of Glauber dynamics\protect\label{sec:Theoretical-analysis-of}}

Next we analyze the Glauber dynamics to better understand how fluctuations
shape the evolution of spin magnetizations and energies and how they
may be beneficial for solving CSPs. Specifically, we perform a systematic
cumulant expansion up to second order of the master equation (\ref{eq:Glauber_master}).
Taking into account all cumulants would result in the full Glauber
dynamics including all random fluctuations, which guarantees the convergence
of this expansion. Note that these cumulant-based approximations,
their quality and relation to previous approximations of Glauber dynamics
have been discussed in detail in a previous study \citep{Dahmen16_031024}
(see Appendix \ref{sec:Derivations-of-solvers} for a complete re-derivation).
In the current work the focus is on setting them in the context of
solving CSPs and to discuss the role of fluctuations in the solution
process. 

First, we apply a mean-field approximation in section \ref{subsec:Mean-field-approximation},
then move on to additionally include the variances of the spins while
still discarding the cross-covariances in section\ref{subsec:Variance-of-spins},
and ultimately we perform the full expansion in the second cumulant
in section \ref{subsec:Covariance-of-spins:}.

\subsection{Mean-field approximation\protect\label{subsec:Mean-field-approximation}}

The simplest approximation of the Glauber dynamics is a mean-field
approximation. In the mean-field approach one replaces the description
of the probability distribution $p(\boldsymbol{S},t)$ by a description
of its mean values of the spins, $\boldsymbol{m}_{i}(t)=\langle\boldsymbol{S}(t)\rangle$,
discarding all fluctuations. The mean-field approximation results
in (for a derivation see appendix \ref{app:mean-field-approximation})
\begin{subequations}
\begin{eqnarray}
\tau\frac{\mathrm{d}m_{i}(t)}{\mathrm{d}t} & = & -m_{i}(t)+\tanh\left(\beta\mu_{i}(t)\right),\label{eq:mean_field_dynamics}\\
\mu_{i}(t) & = & H_{i}+\sum_{j}J_{ij}m_{j}(t)\,.\label{eq:mean_field_local_field}
\end{eqnarray}
\end{subequations}
The dynamics of a mean value $m_{i}$ are described by the right hand
side of equation (\ref{eq:mean_field_dynamics}) including a leak
term and a local field ($\mu_{i}$) term. In the absence of a local
field, the leak term leads to exponentially vanishing magnetizations.
The local field contains contributions from both the external field
$H_{i}$ of the spin $S_{i}$ as well as an effective field generated
by the other spins. These equations are equivalent to the continuous
Hopfield dynamics, when written in the form of equation (\ref{eq:Hopfield_dynamics})
(see appendix \ref{app:equivalence-Hopfield-mf}).

\begin{figure}
\begin{centering}
\includegraphics[width=1\linewidth]{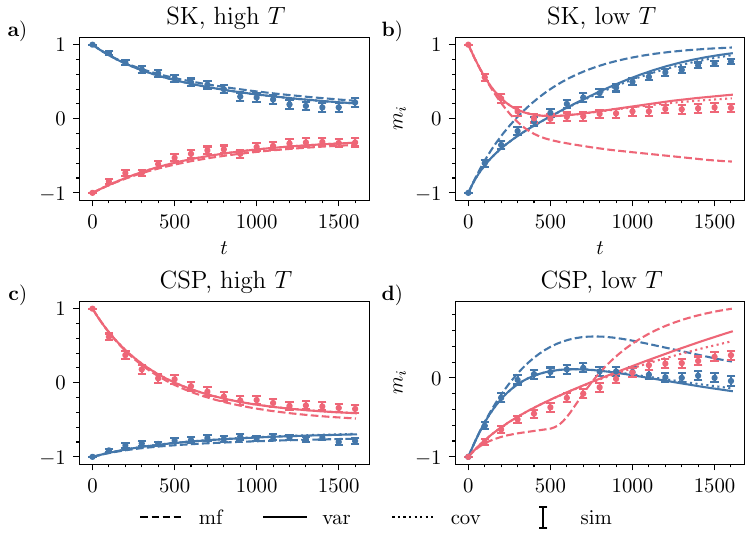}
\par\end{centering}
\caption{\protect\label{fig:mean-field-evaluation}\textbf{Comparison of mean-field,
variance and covariance approximations with simulated Glauber dynamics.}
Magnetization of two spins as a function of time for simulated Glauber
dynamics (mean and standard deviation of the mean over $300$ trials)
and solution for mean-field/variance/covariance equations. \textbf{a,b)}
Sherrington-Kirkpatrick (SK) model instance at temperatures $T=0.1$
and $T=1$, respectively. \textbf{c,d)} CSP instance at temperatures
$T=0.1$ and $T=1$, respectively. The first and second order statistics
of the Gaussian couplings $J_{ij}$ and fields $H_{i}$ in the SK
instance are matched to the corresponding statistics of the CSP instance.
Parameters: $N=400$, $\alpha=3$.}
\end{figure}
This approximation has been very successful in the study of spin systems
\citep{Mezard87}. A prototypical example is the widely studied Sherrington-Kirkpatrick
(SK) spin model \citep{Sherrington75_1792}, where spin couplings
are independently and identically drawn from a Gaussian distribution
$J_{ij}\overset{\text{i.i.d.}}{\sim}\mathcal{N}(J,J^{\prime}/N)$.
Here we additionally impose random external fields $H_{i}\sim\mathcal{N}(H,H^{\prime})$
to mimic the structure of the CSP. The mean-field prediction shows
good agreement with the Glauber dynamics in the SK system for high
temperatures (Figure \ref{fig:mean-field-evaluation}a). For low temperatures,
however, theory and simulation deviate (Figure \ref{fig:mean-field-evaluation}b).
Indeed, the same behavior can be observed for CSPs with their sparse
couplings: At high temperatures, the mean-field prediction and simulation
agree (Figure \ref{fig:mean-field-evaluation}c), but they disagree
for small temperatures (Figure \ref{fig:mean-field-evaluation}d).
This mismatch between mean-field approximation and the full system
for low temperatures, which we show quantitatively in appendix \ref{app:additonal_plots}
Figure \ref{fig:Deviation-between-Glauber-and-approximations}, is
a common phenomenon. Disordered spin systems transition to a spin
glass phase at low temperature with a complex energy landscape consisting
of many local minima \citep{de1978stability}, in which particularly
the fluctuation-ignorant mean-field solver can get stuck. This explains
the differences in performance between the deterministic Hopfield
and stochastic Glauber dynamics in Figure \ref{fig:performance_discrete_Glauber_and_Hopfield},
with the latter obtaining significantly better results at low temperatures.

Therefore we conclude that fluctuations described by higher-order
statistics are important for solving CSPs at the required low temperatures.
These fluctuations are not present in the Hopfield network, being
equivalent to the mean-field approximation, thus explaining the worse
performance. We include the higher-order fluctuations in the next
sections.

\subsection{Variance of spins: effective spin-dependent temperatures\protect\label{subsec:Variance-of-spins}}

The mean-field approximation only describes the first cumulant, the
mean. Now we extend the mean-field description by the second cumulant.
However, instead of taking the full second cumulant, the covariance,
into account, we start by including only the variance of each spin,
and include the full covariance in the next section. Given that the
variance of binary variables is fully determined by their mean, this
approximation still amounts to an equation for the means $m_{i}$
only (see appendix \ref{app:variance-approximation} for details)
\begin{subequations}
\begin{eqnarray}
\tau\frac{\mathrm{d}m_{i}(t)}{\mathrm{d}t} & = & -m_{i}(t)+\langle\tanh(\beta h_{i})\rangle_{h_{i}\sim\mathcal{N}(\mu_{i}(t),\sigma_{i}^{2}(t))}\label{eq:variance_of_spins_dynamics}\\
\mu_{i}(t) & = & H_{i}+\sum_{j}J_{ij}m_{j}(t)\\
\sigma_{i}^{2}(t) & = & \sum_{j}J_{ij}^{2}(1-m_{j}^{2}(t))\,.
\end{eqnarray}
\end{subequations}
The average in equation (\ref{eq:variance_of_spins_dynamics}) is
performed over a Gaussian random variable $h_{i}$ (local field) with
mean $\mu_{i}$ as in equation (\ref{eq:mean_field_local_field})
and with variance $\sigma_{i}^{2}$. The fluctuations contained in
this variance smear out the $\tanh$ non-linearity, while, in contrast,
the mean-field approximation evaluates the non-linearity only at a
single point. One obtains the mean-field approximation in the special
case of vanishing variance $\sigma_{i}^{2}=0$.

\begin{figure}
\centering{}\includegraphics[width=1\linewidth]{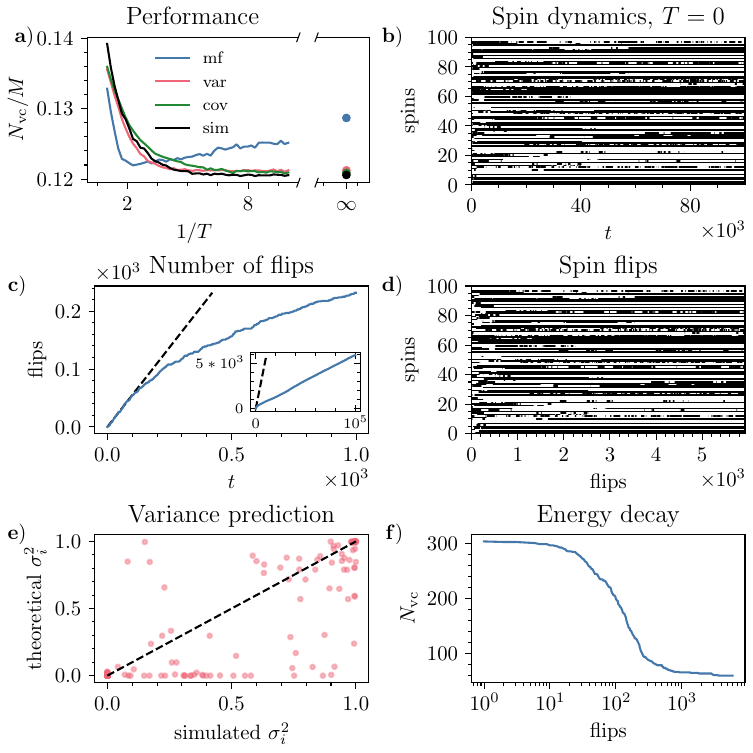}\caption{\protect\label{fig:variance-zero-temperature-panels}\textbf{Analysis
of the variance solver.} $\textbf{a)}$ Fraction of violated clauses
as a function of inverse temperature $1/T$ for the mean-field, variance
and covariance (see section \ref{subsec:Covariance-of-spins:}) solver
averaged over $50$ CSP instances. $\textbf{b)}$ Evolution of spin
values over time (white: spin up, black: spin down) for simulated
Glauber dynamics at $T=0$ for one example CSP instance with $\alpha=3$
and $100$ randomly chosen spins. $\textbf{c)}$ Number of spin flips
as a function of time. The inset additionally shows later times and
the black dashed line shows a linear fit for times shortly after initialization.
$\textbf{d)}$ Same as panel a) but dependent on spin flips instead
of time. $\textbf{e)}$ Simulated and theoretical prediction from
the variance solver for variances $\sigma_{i}^{2}$ of spins. Variances
of simulated dynamics are measured over $5\cdot10^{4}$ unit times
after initialization at theoretical equilibrium state. $\textbf{f)}$
Energy decay for simulated Glauber dynamics as a function of spin
flips. Parameters: $N=400$.}
\end{figure}

Figure \ref{fig:mean-field-evaluation}b and d show that the resulting
prediction for spin magnetizations obtained by including the variance
is accurate also at low temperatures. Moreover the temperature dependent
performance of both the mean-field and the variance approximation
together with the simulated Glauber dynamics in figure \ref{fig:variance-zero-temperature-panels}a
shows two important aspects: 1) The best performance of both the simulation
and the variance solver are found at vanishing temperature $T=0$,
while the mean-field solver is optimal at some intermediate temperature
$T\approx0.5$. This may seem counter intuitive since one may expect
the solver to get stuck in local minima due to missing thermal fluctuations.
We explain this phenomenon later in this section. 2) The best performance
of the variance solver is significantly better than the one of the
mean-field solver. In fact, its performance is comparable to that
of the simulation at all temperatures. Thus by including fluctuations
one indeed obtains an accurate description of the simulated Glauber
dynamics. This suggests that the qualitative difference between the
Hopfield dynamics/mean-field approximation and the simulated Glauber
dynamics are carried by the variance.

We explain these two findings in the following, starting with the
temperature dependence. We observed that decreasing the temperature
monotonically decreases the energy after convergence. Thus, we study
the limit of zero temperature $T\to0$ in the following. In this limit,
the expectation value in (\ref{eq:variance_of_spins_dynamics}) reduces
to an expectation value over the Heaviside step function which can
be analytically calculated and results in

\begin{eqnarray}
m_{i} & = & \text{erf}\left(\frac{\mu_{i}}{\sqrt{2}\sigma_{i}}\right)\,.\label{eq:var-expectation-analytic}
\end{eqnarray}
Interestingly, this result allows for weakly magnetized spins, i.e.
spins with magnetization $m_{i}\neq\pm1$, if the corresponding $\sigma_{i}$
is non-zero. Fluctuations driving a non-zero $\sigma_{i}$ appear
in the Glauber dynamics simulation panel b) even for very large times
and at $T=0$ (Figure \ref{fig:variance-zero-temperature-panels}).
Figure \ref{fig:variance-zero-temperature-panels}c shows the number
of flips as a function of time. We find that shortly after initialization
spins flip frequently. After some equilibration time the frequency
of flips declines, but stays finite, such that continually spins are
flipped. The dynamics of the spin flips after removing time steps
in which no spin was flipped is shown in Figure \ref{fig:variance-zero-temperature-panels}d,
while panel e) confirms that the variances of spins in the simulation
are qualitatively correctly described by the variance solver for most
neurons. Lastly panel f) shows the energy decay of the Glauber simulation
and that in many cases spins are flipped although this does not result
in an energy benefit. The explanation for this a priori puzzling finding
can be found in the energy landscape of the underlying spin system:
The energy function is degenerate, different spin configurations may
have the same energy implying that the effective local field (\ref{eq:mean_field_local_field})
for these spins is vanishing. Thus, no energy (and therefore also
no thermal drive) is needed to cause a spin to flip. In terms of the
master equation (\ref{eq:Glauber_master}) this implies that $F_{i}^{+}(\boldsymbol{S}\backslash S_{i})=F_{i}^{-}(\boldsymbol{S}\backslash S_{i})$
if for a flip of spin $S_{i}$ the energy does not change (also see
appendix \ref{app:gain-function} equation (\ref{eq:gain-function-from-energy})).
If such a spin is chosen for an update, it flips with probability
$1/2$, being purely driven by the probabilistic nature of the Glauber
dynamics.

These results suggest the following hypothesis: The continual flips
allow the system to explore degenerate energy plateaus which helps
finding better solutions with lower energy. While the flips themselves
do not lower the energy, they freeze or unfreeze other spins, until
eventually a spin is chosen that reduces the energy when flipped.
The exploration of states of the same energy are observed as plateaus
in Figure \ref{fig:variance-zero-temperature-panels}f. Since some
spins are frozen at some times and unfrozen at other times they may
have a magnetization which is both different from $\pm1$ and from
$1/2$. This effect and the resulting non-vanishing $\sigma_{i}$
can be understood as an effective, spin dependent temperature. In
the following, we will refer to spins which have such a non-zero temperature
at a given time as free spins.

\begin{figure*}[t]
\centering{}\includegraphics[width=1\linewidth]{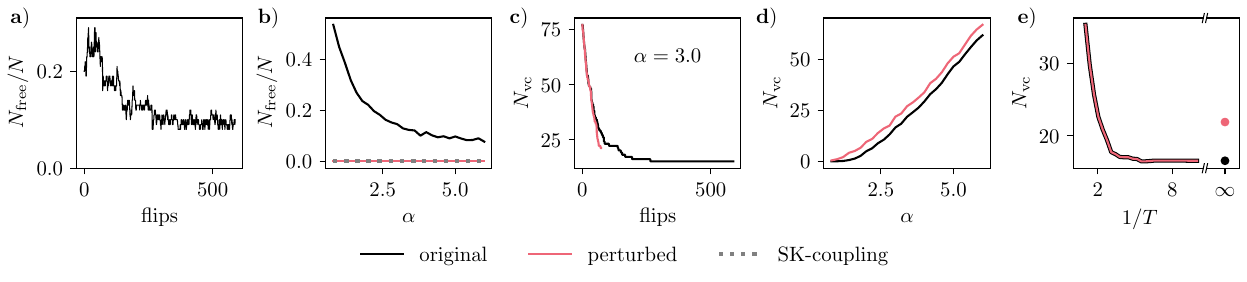}\caption{\protect\label{fig:freespins-perturbed-original}\textbf{Impact of
a small perturbation on Glauber dynamics in a CSP instance.} $\textbf{a)}$
Fraction of free spins in a CSP as a function of spin flips for zero
temperature. $\textbf{b)}$ Fraction of free spins as a function of
clause density. Black line is for original CSP problems, red line
for the corresponding perturbed CSPs with additional external field
$\epsilon_{i}\sim\mathcal{N}(0,10^{-4})$. The dotted gray line shows
the fraction of free spins for an SK-coupling for the original case.
The number of free spins was calculated after $10^{4}$ time steps.
$\textbf{c)}$ Number of violated clauses as a function of spin flips
for the simulated Glauber dynamics for both the original and the perturbed
system of the same CSP as in a) at $T=0$. $\textbf{d)}$ Number of
violated clauses as a function of the clause density both for the
original and the perturbed CSP system at $T=0$, calculated after
$10^{4}$ simulation steps. $\textbf{e)}$ Number of violated clauses
as a function of inverse temperature for the original and the perturbed
system of the CSP from a). Parameters: $N=400$, $\alpha=3$.}
\end{figure*}

To test this hypothesis, we perform a perturbation experiment at temperature
$T=0$. Compared to the original system, the perturbed system has
small additional random external magnetic fields $H_{i}\rightarrow\tilde{H_{i}}=H_{i}+\epsilon_{i}$
with $\epsilon_{i}\sim\mathcal{N}(0,10^{-4})$. Due to this small
perturbation, the energy degeneracy will be destroyed almost surely,
and the effective local field (\ref{eq:mean_field_local_field}) will
be non-zero. In the zero-temperature case, these small fields are
sufficient to freeze the spins, prohibiting the aforementioned exploration
of the energy landscape. Thus we expect the perturbed system to not
exhibit any free spins, thus worsening the performance. The results
of this experiment are shown in Figure \ref{fig:freespins-perturbed-original}.
Panel a) confirms that at all times during the simulation there is
a significant fraction of free spins. This fraction declines after
initialization and converges to a fixed value which depends on the
clause density of the underlying CSP, as shown in panel b). Importantly
there is a significant fraction of free spins at all clause densities
if the system is unperturbed. A small perturbation, however, freezes
all spins for arbitrary clause densities. Panel b) additionally shows
the zero number of free spins for the dense, random SK coupling; free
spins are thus resulting from the discrete and sparse nature of the
CSP couplings, resulting in a relatively large probability that individual
incoming inputs cancel each other and lead to a vanishing effective
field.

Now we study the impact of this property on the solver's performance
by both observing the perturbed and unperturbed CSP system. Figure
\ref{fig:freespins-perturbed-original}c shows the number of violated
clauses (energy) as a function of spins flips. While the perturbed
system lowers the energy with every spin flip and therefore its energy
decays faster in the beginning in this display, all spins eventually
become frozen and no more flips occur. It is thus impossible for the
system to obtain a better solution, it is stuck in a local minimum.
This is qualitatively different for the unperturbed system: At all
times there is a fraction of free spins (see Figure \ref{fig:freespins-perturbed-original}a,b),
allowing spin flips which do not change the energy. This leads to
slower energy decay in the beginning when displayed as a function
of spin flips, but allows exploration of the phase space at later
flips, until eventually a better solution is found. The performance
difference caused by this effect at $T=0$ is shown in Figure \ref{fig:freespins-perturbed-original}d.
Indeed, we observe significantly better performance for the unperturbed
system for all clause densities. The temperature dependent performance
for a fixed clause density is shown in Figure \ref{fig:freespins-perturbed-original}e:
At finite temperature, the performance is comparable. In these cases,
the non-zero thermal fluctuations are enough to overcome the small
energy differences between states of the perturbed system; the perturbation
thus does not qualitatively change the behavior. Only in the $T=0$
case, the energy difference caused by the perturbation freezes the
spins. At this point we thus observe a performance gap. We conclude
that the energy degeneracy allows for spin flips even in the zero
temperature case which help exploring the phase space and thus increase
the performance of the Glauber dynamics solver.

We want to end this section with a short note on the computational
complexity of the above discussed variance solver. In the $T=0$ case
we can use the analytic solution (\ref{eq:var-expectation-analytic})
for the expectation value in (\ref{eq:variance_of_spins_dynamics})
which leads to the same computational cost as the mean-field solver
while obtaining a better performance. Moreover, we can use the insights
from this section to speed up the Glauber dynamics simulation: Instead
of choosing a spin at random for an update, we directly choose one
of the free spins, since the frozen spins have probability zero to
be flipped, even if they are chosen. Lastly we want to point out that
these benefits only hold in the $T=0$ case. At finite temperatures
one would have to evaluate the integral in (\ref{eq:variance_of_spins_dynamics})
numerically.

\subsection{Covariance of spins: expansion up to 2nd cumulant\protect\label{subsec:Covariance-of-spins:}}

Including the variance is only the first expansion term one can include
on top of the mean-field approximation. In this section, we also include
the next correction term, the covariance between spins. The time dynamics
for this case are (derivation in appendix \ref{app:covariance-approximation})
\begin{subequations}
\begin{eqnarray}
\tau\frac{\mathrm{d}m_{i}(t)}{\mathrm{d}t} & = & -m_{i}(t)+\langle\tanh(\beta h_{i})\rangle_{h_{i}\sim\mathcal{N}(\mu_{i}(t),\sigma_{i}^{2}(t))}\label{eq:covvariance_of_spins_dynamics}\\
\tau\frac{\mathrm{d}c_{ij}(t)}{\mathrm{d}t} & = & g(\boldsymbol{m}(t),\boldsymbol{c}(t))\label{eq:h(c)}\\
\sigma_{i}^{2}(t) & = & f(\boldsymbol{m}(t),\boldsymbol{c}(t))\label{eq:f(c)}
\end{eqnarray}
\end{subequations}
where the exact form of the functions $f$ and $g$ is stated in appendix
\ref{app:covariance-approximation}. Note that taking into account
covariances amounts to an enlarged number of degrees of freedom due
to the coupling between mean and covariances via (\ref{eq:f(c)})
and (\ref{eq:h(c)}).

\begin{figure}
\centering{}\includegraphics[width=1\linewidth]{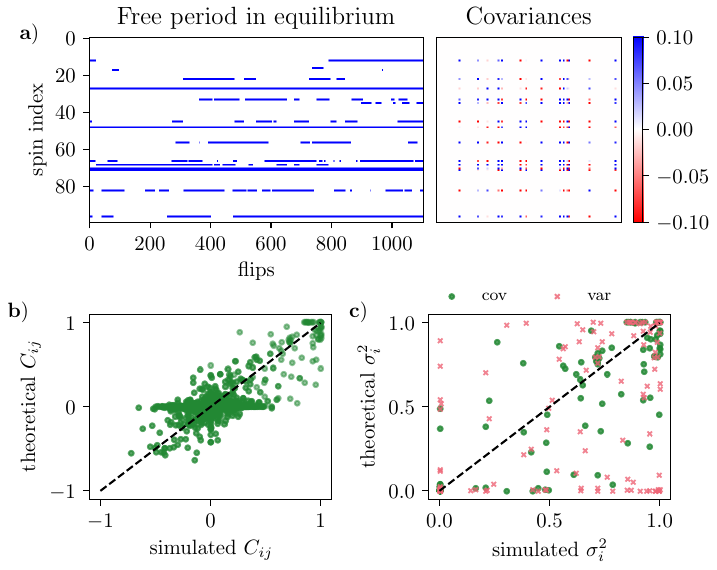}\caption{\protect\label{fig:test-covariance-prediction}\textbf{(Co)variance
comparison of theory and simulation.} \textbf{a)} Left: Frozen and
free spins after equilibration as a function of flips at T=0. Free
spins are marked in blue. Right: Empirical covariance matrix. Example
of $100$ spins. \textbf{b)} Scatter plot of theoretical and simulated
covariance. \textbf{c)} Scatter plot of theoretical and simulated
variance for both the variance solver (red) and covariance solver
(green). Parameters: $N=400$, $\alpha=3$.}
\end{figure}

We test the prediction of this expansion in Figure \ref{fig:test-covariance-prediction}.
In panel a) we show free periods of spins after equilibration in the
Glauber dynamics simulation. These free periods amount to non-trivial
coordinated fluctuations between spins as described by the covariance
matrix of spins. These covariances are, to a good approximation, predicted
by the covariance solver (Figure \ref{fig:test-covariance-prediction}b).
Moreover the prediction of variances is improved by including the
covariances between spins (Figure \ref{fig:test-covariance-prediction}c
and Figure \ref{fig:Covariance-versus-Variance-solver-app} in appendix
\ref{app:additonal_plots}): While the variances are qualitatively
correctly predicted by the variance solver we find quantitatively
better agreement with the covariance solver. This makes sense, since
including all orders would be a complete description of the simulated
Glauber dynamics. In summary, the covariance solver describes the
simulation quantitatively more accurately than the variance solver,
but the variance solver is sufficient to describe the simulation qualitatively.

On the computational side, one may expect an increased performance
from the covariance solver due to the additional degrees of freedom
being the covariances between spins. Graph neural networks like RUN-CSP
\citep{Toenshoff21_580607} profit from their high-dimensional phase
space for solving CSPs. However, observing the performance of the
covariance solver compared to the variance solver in Figure \ref{fig:variance-zero-temperature-panels}a
we only find a minimal benefit for $T\to0$, at the expense of solving
a set of $N+N(N-1)/2$ equations instead of $N$ equations for the
variance solver.

\section{Performance evaluation}

\begin{figure*}
\centering{}\includegraphics[width=1\linewidth]{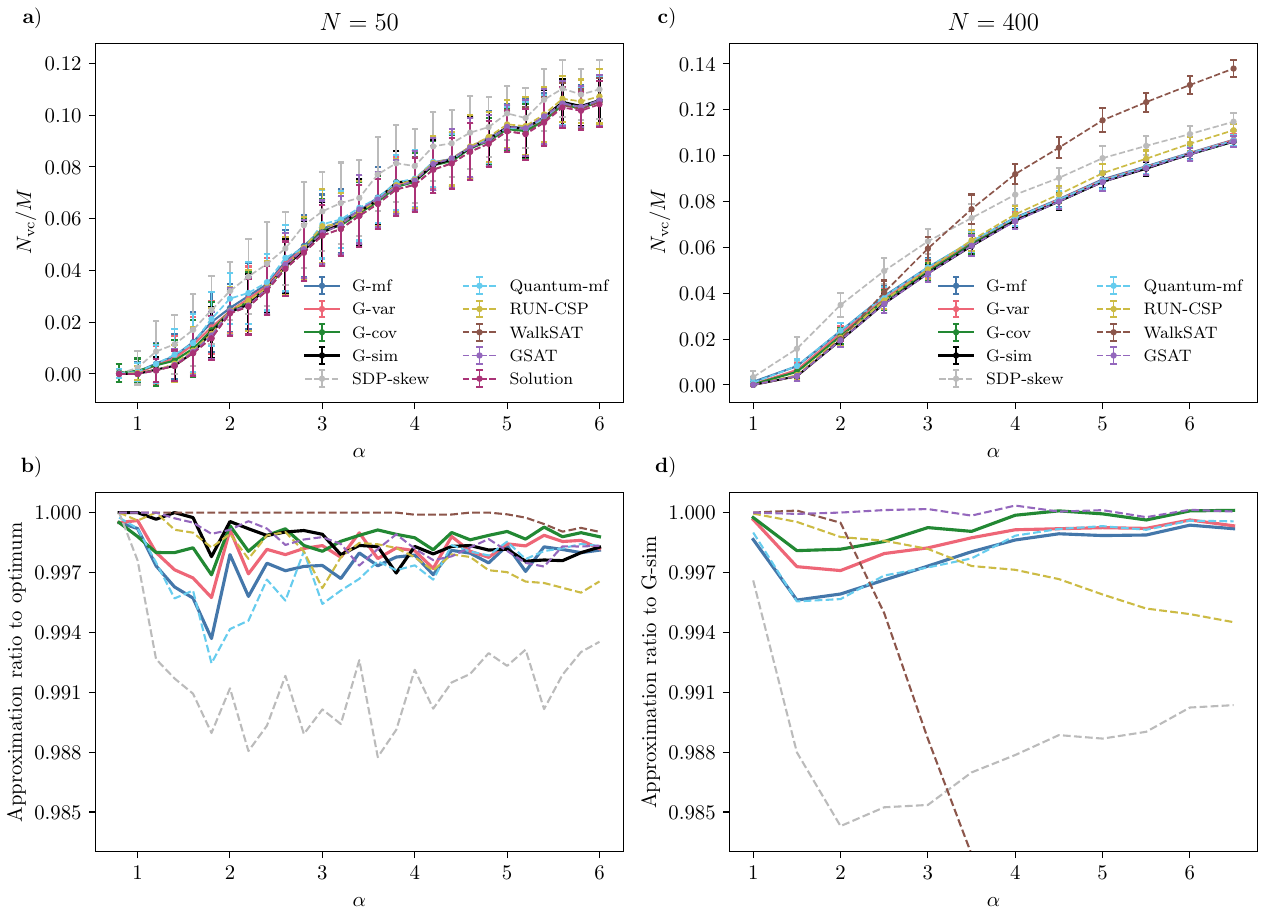}\caption{\protect\label{fig:performance-comparison}\textbf{Performance of
all solvers.} For $N=50$ (left column), the $\alpha$ values are
chosen in the range 0.8 to 6.0 in steps of 0.2. For $N=400$ (right
column), the $\alpha$ values are in the range 1.0 to 6.0 in steps
of 0.5. Markers show the mean and standard deviation across 50 instances
per $\alpha$ and each solver has one random initial condition. The
Glauber-based solvers are running at zero temperature except for the
G-mf solver, which is running at its optimum $T=0.5$ (cf. Figure
\ref{fig:variance-zero-temperature-panels}a).\textbf{ a) }Relative
number of violated clauses as a function of clause density for $N=50$.
\textbf{b) }Mean number of violated clauses relative to the mean optimal
number of violated clauses as a function of clause density.\textbf{
c) }Relative number of violated clauses as a function of clause density
for $N=400$. \textbf{d) }Mean number of violated clauses relative
to the mean violated clauses for the Glauber simulation as a function
of clause density. Solid lines: Glauber-based solvers, dashed lines:
other solvers.}
\end{figure*}
Finally, we test the performance of the solvers investigated in this
work, i.e. the Glauber-dynamics simulation (G--sim) as well as the
solvers inspired by their approximation (G-mf, G-var and G-cov), and
compare to other methods used in the literature, specifically a mean-field
solver inspired by quantum mechanics, semi-definite programming, a
state-of-the-art graph neural network approach, as well as widely
used heuristic solvers.

\textbf{Quantum mean-field (quantum-mf): }The Mean-Field Approximate
Optimization Algorithm (MF-AOA) was introduced in \citet{Bode23_030335}
as a classical limit of the Quantum Approximate Optimization Algorithm
(QAOA) \citep{Farhi15_arXiv}. In its continuous-time formulation
\citep{Bode24_012611}, the MF-AOA leads to equations of motion very
similar to (\ref{eq:mean_field_dynamics}), the most notable difference
being that the quantum mean-field lives on the Bloch sphere, i.e.
instead of the classical $m_{i}(t)$ there are \textit{three} degrees
of freedom $\boldsymbol{n}_{i}(t)=(n_{i}^{x}(t),n_{i}^{y}(t),n_{i}^{z}(t))^{T}$
per spin. Note that it is natural to identify $m_{i}(t)$ with the
$z$-component. The dynamics is given by precession in the effective
magnetic fields $\boldsymbol{B}_{i}(t)$, described by the equations
$\partial_{t}\boldsymbol{n}_{i}(t)=\boldsymbol{n}_{i}(t)\times\boldsymbol{B}_{i}(t)$,
where $\boldsymbol{B}_{i}(t)=2s_{x}(t)\hat{\boldsymbol{e}}_{x}+2s_{z}(t)\mu_{i}(t)\hat{\boldsymbol{e}}_{z}$.
The control fields $s_{x,z}(t)$ obey the boundary conditions $s_{x}(0)=s_{z}(t_{f})=1$
and $s_{x}(t_{f})=s_{z}(0)=0$. For the results shown in Figure \ref{fig:performance-comparison},
we set $t_{f}=2^{10}$. The initial condition for the spins is $\boldsymbol{n}_{i}(t)=(1,0,0)^{T}$,
from which the equations of motion are evolved up to time $t_{f}$.
As above, the (approximate) solution to the optimization problem is
finally obtained from the signs of the $z$-components.

\textbf{Semidefinite programming (SDP-skew): }A classic result of
computer science is the Goemans-Williamson (GW) algorithm \citep{Goemans95_1115}
for the (weighted) MAX-CUT problem. By relaxing the Boolean variables
$x_{i}$ to the $N$-dimensional unit sphere, this problem is turned
into a semidefinite program that can be solved by standard methods
such as MOSEK interfaced in CVXPY \citep{Diamond16_1}. Subsequently,
the binary domain of the problem is recovered by a rounding mechanism,
specifically by cutting the $N$-sphere with a random hyperplane.
One advantage is that this algorithm comes with a performance guarantee
of achieving $\sim0.878$ of the optimal value, which also holds for
MAX-2-SAT. In practice, when applied to any given problem instance,
the algorithm often performs much better than this worst-case guarantee.
For the special case of MAX-2-SAT, later results improve the GW value
to $\sim0.935$ by picking hyperplanes from a skewed distribution
\citep{Lewin02_67}. For the purpose of our work, we compare to this
latter algorithm, which we denote by ``SDP-skew'' in Figure \ref{fig:performance-comparison}.
Note that we show ``single-shot'' results, i.e. we pick a single hyperplane
per random problem instance. This can be considered a fair comparison
in the sense that we proceed similarly for the other algorithms. It
does not take into account runtime considerations, however.

\textbf{Graph neural network (RUN-CSP): }The graph neural network
model RUN-CSP \citep{Toenshoff21_580607} functions by learning a
message passing protocol in an unsupervised manner. Messages are passed
only between literals/spins which participate in the same clause.
The neural network architecture makes use of a high-dimensional state
space (here $d=128$) to store and process information of neighboring
spins. A learnable projection determines the spin values of each spin
from its high-dimensional representation.

\textbf{GSAT:} A classical local search heuristic for solving (MAX)-SAT
problems is G-SAT (Greedy-SAT) \citep{selman1992anew,selmanWalksatWebpage}.
This approach starts from a random variable/spin configuration and
flips in each step the variable that decreases the number of violated
clauses the most (the decrease can be zero or negative). If there
are multiple such variables GSAT picks a random one. Thus this algorithm
only increases the number of violated clauses in a step if it is unavoidable
which is the key difference to WalkSAT.

\textbf{WalkSAT: }WalkSAT \citep{Selman93_521,selmanWalksatWebpage}
is a modification of GSAT which allows spin flips that increase the
energy more frequently. More precisely, in each step WalkSAT performs
a standard GSAT step with probability $1-p$ and flips a randomly
chosen variable of a randomly selected unsatisfied clause with probability
$p$. In the follow-up study \citep{selman1994noise} the authors
found the flips of variables in random unsatisfied clauses to be crucial
for increased performance of WalkSAT over GSAT on 3-SAT, circuit synthesis
and diagnosis problems. For the experiment in Figure \ref{fig:performance-comparison}
we use the default value $p=0.5$ \citet{selmanWalksatWebpage}.

\textbf{Exact solution: }Solving CSPs exactly takes exponential time
and is thus only feasible for small instance sizes; this is the motivation
to develop approximate solvers in the first place. However, it is
valuable to compare the approximate solvers described above with the
exact solution on small instances, where the computing cost is manageable.
To obtain the exact solution we use the RC2 MaxSAT solver from \texttt{PySAT}
\citep{Ignatiev18_428,ignatiev2019rc2,mosek}. For larger system sizes
where finding exact solutions becomes computationally infeasible we
compare the approximate solvers among each other.

We show the result of the performance comparison in Figure \ref{fig:performance-comparison}
for small ($N=50$) and large ($N=400$) CSPs. Panel a) shows the
fraction of violated clauses as a function of the clause density for
small instances. The solution of all solvers, including the exact
solutions, show the same qualitative trend: Below $\alpha=1$, random
MAX-2-SAT problems are almost always satisfiable \citep{Monasson97_1357},
thus the number of violated clauses $N_{\text{vc}}$ is zero. Increasing
$\alpha$, i.e., adding more and more constraints (while having the
same number of spins) naturally increases $N_{\text{vc}}$, implying
that more and more clauses cannot be satisfied, even for the optimal
solution. This trend is thus a property of the underlying CSPs and
not a property of the solvers. The same trend can be observed for
large instances (Figure \ref{fig:performance-comparison}b) where
the exact solution is unfeasible to obtain due to the exponentially
higher computational cost.

To better compare the solution of different solvers we also show the
approximation ratio relative to the optimum for small instances (Figure
\ref{fig:performance-comparison}c), and the approximation ratio relative
to G-sim for large instances (Figure \ref{fig:performance-comparison}d).
For both small and large instances we find the Glauber-based solvers
to match or slightly outperform the state-of-the-art solvers SDP-skew,
quantum-mf and RUN-CSP. In most cases (except for the small instances
and large clause densities) the simulation G-sim thereby performs
best, followed by G-cov, G-var and G-mf (being the Hopfield network).
This is intuitive from the theoretical analysis above: Fluctuations
are important to continue exploring the state space and finding better
solutions. The more fluctuations we include into solvers, the better
the obtained performance.

When comparing Glauber-based solvers to heuristic solvers, we find
differential results for small and large problem sizes: WalkSAT performs
well for small problem sizes $N$ (Figure \ref{fig:performance-comparison}c)
and achieves almost optimal performance for a wide range of clause
densities. For large problem sizes, WalkSAT, however, drastically
drops in performance compared to the other solvers. This is in line
with previous findings in the literature: \citet{Toenshoff21_580607}
showed that RUN-CSP outperforms WalkSAT for large CSPs (see their
Fig. 3). Furthermore, \citet{boulebnane2024_030348} showed that QAOA
outperforms WalkSAT. MF-AOA in turn is similar in performance to QAOA
\citep{Bode23_030335}, and the Glauber-based solvers are similar
to MF-AOA (Figure \ref{fig:performance-comparison}). The drop of
performance in WalkSAT for large $N$ is due to the random choice
of unsatisfied clauses and random flip of spins therein that occurs
with probability $p$ and that, for large $\alpha$, is detrimental
for the solution-finding process of Max-2-SAT. This is evident from
comparing the WalkSAT performance to its predecessor GSAT (Figure
\ref{fig:performance-comparison}), which lacks these random components
and is otherwise almost identical. Throughout, GSAT shows very similar
performance to G-sim. GSAT is a greedy algorithm that always updates
the particular spin whose flip yields the largest possible number
of satisfied clauses or the lowest possible energy. Only if multiple
spins yield the same result, then GSAT chooses one of these spins
randomly. The Glauber dynamics, in contrast, always chooses random
spins for update and contains an update probability which ensures
that the update process at equilibrium exactly samples from the correct
Boltzmann distribution that encodes the CSP. At $T=0$, this makes
sure that the optimal solution is being found asymptotically. In this
case, the update probability is a Heaviside function and G-sim thus
flips only those spins that do not yield an energy increase. Our results
on the equal performance of GSAT and G-sim thus show that there is
no benefit of being greedy for random MAX-2-SAT when it comes to the
final result: whether one chooses the spin with the optimal energy
benefit or just any spin with energy benefit yields similar results.

\section{Discussion}

Solving CSPs is a fundamental problem in computer science and of
practical importance in a wide range of applications. Due to their
computational complexity, often approximate solvers are used in practice
for large instances. The research on approximate solvers is to a large
extent driven by empirical work, resulting in strong solvers that
are, however, often complex to interpret and analyze. Theoretical
investigation of approximate solvers is of fundamental importance
to find key similarities and to guide design choices.

In this study, we contributed to the understanding of approximate
solvers by taking a statistical physics perspective on random MAX-2-SAT
instances. Specifically, we investigated how Glauber dynamics, being
a physical process of Ising-model dynamics \citep{Glauber63_294},
finds approximate solutions and compare this mainly to the traditional
Hopfield-network approach. We found the Glauber dynamics to outperform
the Hopfield network, which corresponds to the mean-field approximation
of the former, and identified statistical fluctuations as a key property
for the high performance of Glauber-based solvers in MAX-2-SAT. Successively
including fluctuations in a systematic manner results in more accurate
descriptions of the Glauber dynamics and thus in stronger solvers.

A large body of previous literature treating CSPs with statistical
physics methodology concentrated on properties of the underlying problems,
e.g. the transition from the solvable to non-solvable regime for random
(MAX-)$K$-SAT problems \citep{Monasson97_1357} or describing how
and why problems near this and other transitions are hard to solve,
even approximately \citep{krzakala2009hiding}. In our work, we use
the Ising description of a CSP to obtain explicit solvers that find
approximate solutions. Specifically, our solvers are able to make
use of the degenerate energy landscape that is typical for disordered
systems such as MAX-2-SAT problems to explore different states and
find better solutions. The main findings of our work are twofold:
First, we show that a series of systematic approximations of the Glauber
dynamics can be used for efficient MAX-2-SAT solving. Each of these
solvers maps to a set of ordinary deterministic differential equations
that hence can be implemented with the same hardware as traditional
Hopfield networks \citep{Mohseni22_363,Cai20_409,Hizzani24_1}, while
reaching considerably higher performance. Second, we identify the
physical mechanism that underlies this increased performance which
is reached in the $T\searrow0$ limit: The appearance of ``free spins'',
which are spin variables that feel a vanishing mean-field from their
peers. Switching such variables does not change energy so they encode
the macroscopically degenerate states in the energy landscape. By
including their statistical fluctuations our algorithms efficiently
explore this landscape, and thus ultimately find a better solution.

Understanding the collective out-of-equilibrium dynamics arising
from the interactions in many-body systems is a challenging subject
of many-particle physics. Dynamic mean-field and replica theory \citep{Sompolinsky81,Sompolinsky82_6860,Sompolinsky88_259,Cugliandolo93_173,Aljadeff15_088101,Kadmon15_041030}
in the $N\to\infty$ limit effectively reduce the many-body problem
of a network comprised of a large number of units to a self-consistency
problem of a single variable, but neglect the cross-covariances of
dynamics between units. \citet{Ginzburg94} introduced pairwise correlations
in the analysis of weakly coupled binary systems and \citet{Renart10_587}
generalized the results to strongly coupled units in the large $N$
limit. Both approaches are, however, limited to averaged pairwise
correlations. Similarly, approximate master equations for binary-state
dynamics obtained from the pair approximation \citep{Gleeson11_068701,Gleeson13_021004}
are restricted to global dynamics in infinite networks. For self-averaging
macroscopic observables such as the mean magnetization or the energy,
Coolen and colleagues presented systematic approaches to derive closed
equations of motion for various network topologies, which, however,
are also only exactly valid in the limit $N\rightarrow\infty$ \citep{Coolen96_8184,Laughton96_763,Mozeika08_115003,Mozeika09_195006}.

Finding specific configurations of variables that provide optimal
solutions for the CSP, however, requires theories that operate on
the microscopic level and can resolve individual units. For out-of-equilibrium
systems with sequential Glauber update in discrete time steps, methods
related to our approach have been proposed by \citet{Mezard11_L07001}.
An alternative approach, based on an expansion of the free energy
in the coupling strength (Plefka-expansion \citep{Plefka82_1971})
leads to the Thouless-Anderson-Palmer mean-field theory \citep{Vasiliev74,Vasiliev75,Thouless77_593,Nakanishi97_8085,Tanaka98_2302,Nishimori01_01,Roudi11_P03031},
which by construction is, however, restricted to weak coupling. Both
theories neglect the influence of cross-covariances on the marginal
statistics, an important finite-size effect shown in \citet{Dahmen16_031024}
to result in a distribution of mean activities due to correlations
alone.

The systematic inclusion of fluctuations in solvers presented here
is based on the cumulant expansion for networks of binary units derived
in \citet{Dahmen16_031024}. At the here considered second order,
it amounts to a Gaussian approximation of the total local field that
each spin experiences. This approximation, which yields coupled self-consistency
equations for all mean magnetizations and cross-covariances between
individual spins, in principle relies on the problems being sufficiently
large, so that even in the sparse coupling case of CSPs each spin
receives a large superposition of signals from other spins. In practice,
we, however, find that even problems of only 50 variables can be addressed
well with our solvers based on the Gaussian approximation (Figure
\ref{fig:performance-comparison}). We furthermore explicitly studied
the effect of sparseness in connectivity on the Gaussian assumption
by comparing the sparse CSP coupling with a fully connected Sherrington-Kirkpatrick
(SK) coupling with matched first and second order statistics (Figures
\ref{fig:mean-field-evaluation} and \ref{fig:Deviation-between-Glauber-and-approximations}),
and find the quality of approximation to be similar in both cases.
For MAX-$K$-SAT problems the entries in the coupling matrix take
on positive and negative values with equal probability, yielding a
zero mean coupling. Therefore there is no net positive input to spins
and no positive feedback loops that would cause on average positive
correlations between spins. Like in the case of balanced networks
\citep{Dahmen16_031024}, correlations between spin variables are
small so that the total local field can be well approximated by a
Gaussian according to the central limit theorem. An inclusion of higher-order
cumulants beyond the Gaussian approximation is, however, also in principle
possible \citep{Dahmen16_031024}. 

The here presented cumulant-based solvers provide a complementary
view on algorithms for CSP solutions compared to previous approaches
that focus on approximations of the joint and conditional probability
distributions of variable configurations. The latter have been employed
to arrive at approximations for the marginal statistics of each variable
\citep{Mezard_Montanari09}. Classical examples of message-passing
algorithms resulting from such approaches are belief propagation \citep{Pearl82_133,Yedidia00,Altarelli13_062115},
and its version that takes into account the effect of clustering of
solutions by including histograms in the messages -- survey propagation
\citep{Mezard02_812,Braunstein05_201}. These methods are based on
the topology of the underlying problem, the factor graph; they operate
under the assumption that the incoming messages into a given factor
node are independent. This is exact for problems whose graph is a
tree. For large $N\,(\rightarrow\infty)$ a random factor graph becomes
locally “tree-like” and cycles become long, constituting a good structure
for the assumed independence of variable nodes and the involved cavity-method
calculations. While belief propagation and survey propagation excel
for big problem sizes $N$, typically above $10^{3}$ \citep{Braunstein05_201},
their theoretical analysis was found to be challenging in certain
cases \citep{Montanari07_arxiv}. There is a variety of extensions
to the classical message-passing algorithms based on out-of-equilibrium
dynamics, graph expansions, reweighing factors, backtracking and decimation
strategies \citep{Montanari07_arxiv,Ricci09_P09001,DelFerraro14_arxiv,Marino16_12996,Gabrie20_223002,Crotti23_e2307935120}.

The here presented approach instead operates on the level of moments
and cumulants and applies approximation schemes for closure directly
on the corresponding time evolution equations, yielding readily interpretable
and theoretically amenable solution algorithms. These solvers differ
from the above mentioned methods in the way they approximate the full
interactions of variables. In particular, the here shown cumulant-based
approximations are agnostic about the factor graph and the particular
arrangement of interactions in the system, and are thus not relying
on large problem sizes. This also qualifies these solvers as candidates
for implementation in neuromorphic hardware \citep{Mohseni22_363,Hizzani24_1,Pedretti25_7}.
The dimensionality of the problem, i.e. the number of coupled equations,
thereby solely depends on the number of variables ($N$ for mean-field
and variance-improved solvers, and $N+N(N-1)/2$ for the covariance-improved
solver). Another recently proposed factor-graph based approach derives
a set of approximate master equations for the joint probability associated
with a specific clause employing the conditioned dynamic approximations
(CDA-1 and CDA-2) \citep{Machado23_123301,Machado25_arxiv}. The complexity
of this approach instead scales with the number of clauses ($M$).
The here presented mean-field and variance-improved solvers ignore
cross-covariances between all variables and thus correspond to an
approximation of a factorized joint probability distribution, similar
to equation (2) in \citet{Machado23_123301}. This is a stronger simplification
than the conditioned dynamic approximations. Our covariance-improved
solver instead takes into account all cross-covariances between variables
self-consistently, irrespective of their arrangement in the factor
graph and without any conditioning on certain shared variables between
clauses. For the here studied MAX-2-SAT problem, the covariance-improved
solver does not yield much increase in performance with respect to
keeping only the variance. Investigating whether this changes when
studying $k\geq3$-SAT close to the clustering/critical clause-to-variable
ratio is an interesting future direction. The recent work \citet{Machado25_arxiv}
furthermore discusses local equations based on the CDA approach for
Focused Metropolis Search (FMS) \citep{Seitz05_P06006} and greedy-WalkSAT
(G-WalkSAT) \citep{Selman93_521}. The transition rates for the master
equation of the corresponding algorithms presented there could be
combined with the here shown cumulant-based approach and replace the
Glauber dynamics transition rates to obtain effective deterministic
solvers for FMS and greedy-WalkSAT.

 Apart from the theoretically related Hopfield network, we compare
the performance of Glauber-based solvers to a number of recently published
state-of-the-art solvers as well as widely used heuristic solvers.
We chose the shown solvers based on the fact that they stem from different
approaches and that all of them have been proven to yield good performance
on random and real CSPs \citep{Toenshoff20_1909,muller2025_023165,boulebnane2024_030348}.
While the MF-AOA solver is based on physics methods, like our solvers,
the other solvers are from the field of computer science (SDP) and
machine learning (RUN-CSP). We also compared the results of our solvers
to exact CSP solutions which were obtained by the RC2 MaxSAT solver.
The SDP algorithm furthermore yields performance guarantees, and MF-AOA
derives from well known quantum algorithms (QAOA), a promising field
of research for solving optimization problems. Additionally, we
compared our solvers to widely used heuristic solvers. In the $T\searrow0$
limit, the Glauber dynamics function similar to GSAT which flips in
each step the spin that yields the lowest energy. If there are multiple
such spins, the algorithm chooses one of them at random. For later
steps, this leads to qualitatively similar behavior to the zero-temperature
Glauber simulation: Spins are flipped continually on an energy plateau
until flipping one spin decreases the energy (compare Figure \ref{fig:variance-zero-temperature-panels}f
to Figure 1 from \citet{Selman93_521}). This similarity may also
explain the similar performance of the two solvers. The ability of
greedy algorithms to navigate the energy landscape along plateaus
was also demonstrated previously \citep{angelini2025algorithmic}.
Here we show that the energy plateaus correspond to free spin spaces,
induced by net zero local fields, which the stochastic Glauber dynamics
and our deterministic variance and covariance solvers can navigate
at $T=0$. The similar behavior of GSAT and Glauber dynamics is particularly
relevant for neuromorphic applications \citep{Mohseni22_363,Hizzani24_1,Pedretti25_7}.
A spin update in the GSAT algorithm requires global information: At
each update step, all local fields and thus potential energy benefits
need to be computed in order to decide which spin is eventually flipped.
The Glauber simulation in contrast can be implemented asynchronously
and in a decentralized manner: when a spin is selected for update
based on its own Poisson process update times, this spin only needs
to compute its own local field to decide whether it is flipped. No
information about the local fields of any of the other spins is needed
to make that decision. Glauber dynamics are thus naturally implementable
as a neural network with private information to each spin. 

We also compare our solvers to WalkSAT. The WalkSAT solver functions
similar to GSAT, however it also repeatedly performs flips which solve
a random unsatisfied clause, even if this increases the total number
of unsatisfied clauses/energy. In \citet{selman1994noise} the authors
found this property to be crucial for solving SAT problems like 3-SAT
or satisfiability problems from circuit design or diagnosis. In contrast,
in this study on MAX-2-SAT problems, we found the greedy steps of
GSAT or the Glauber simulation to be more performant for large problem
sizes. We hypothesize that this is due to the different problem classes,
MAX-2-SAT vs 3-SAT. Nevertheless, the stronger performance of WalkSAT
over GSAT in some SAT problems is evidence for the importance of fluctuations
on CSP solution strategies \citep{Barthel03_066104}. A direction
of future research would thus be to investigate how particular types
of fluctuations help solving some problem classes but are hindering
in others.

The strongest limitation of our work in its current form is that
we only consider MAX-$K$-SAT instances with $K=2$, a specific class
where the decision variant of the problem is solvable in polynomial
time. MAX-$K$-SAT is nevertheless an NP-hard problem even for the
case $K=2$. An extension to other CSPs, such as $K>2$-SAT is an
exciting and promising direction of future research: The key property
of the MAX-2-SAT instances used in this work was the degeneracy of
the energy landscape, which results from the discrete-valued couplings
and magnetic fields in the Ising system. Since CSPs are discrete in
nature, we hypothesize that the degenerate energy landscape is a property
shared by many classes of CSPs, enabling a similar analysis as done
here. Likewise, the employed method to derive effective deterministic
equations of motion from a cumulant hierarchy is completely general
and carries through to $K>2$-SAT problems with minor adaptations
(see Appendix \ref{app:3-SAT} for an application to $3$-SAT). Whether
these cumulant-based solvers show a similarly good performance for
$K>2$ remains to be tested in future studies.

This study focuses on the mechanistic role of fluctuations in the
solution-finding process of Glauber dynamics at fixed temperature.
While we typically found the best performance in the zero-temperature
case, established techniques such as parallel tempering \citep{swendsen1986_2607,geyer1991computing,Chowdhury25_9193}
and simulated annealing \citep{bertsimas1993_10,kirkpatrick1983optimization}
can be combined with the here proposed solvers to potentially further
optimize their performance. 

Another direction of future research is to use the covariance of
spins in the solver more directly. Instead of picking each spin for
update with the same probability, as in the classical Glauber dynamics,
one could focus on those spins $j$ that are correlated to spin $i$
after a flip of spin $i$. The correlation makes it more likely for
spin $j$ to decrease the energy with another flip. Our method obtains
approximations of this covariance, thus opening the door for such
correlation-informed cluster update schemes.

In summary, our analysis provides a formalism for the theoretical
analysis of energy minimization and CSP optimization, and paves the
way to systematically construct physics-based solvers that combine
performance with interpretability.

\subsection*{Acknowledgments}

This work was partially supported by the German Federal Ministry for
Education and Research  through BMBF Grant 01IS19077A to Jülich and
via the funding program quantum technologies -- from basic research
to the market -- under contract number 13N15680 “GeQCoS”. Open access
publication funded by the Deutsche Forschungsgemeinschaft (DFG, German
Research Foundation) -- 491111487.

\appendix
\renewcommand{\thesubsection}{\Alph{subsection}}
\renewcommand\thefigure{A\arabic{figure}}   
\renewcommand{\thetable}{A\arabic{table}}
\setcounter{table}{0}    \newpage{}

\noindent\onecolumngrid

\appendix

\section{Mapping CSPs to Ising systems\protect\label{app:CSP-to-Ising}}

Here we shortly repeat the mapping of MAX-$K$-SAT problems to Ising
systems \citep{Monasson97_1357}, done in a way such that the number
of violated clauses in the CSP matches the energy in the Ising system
(we use $K=2$ throughout this work). We start by identifying spin
$S_{i}$ with variable $x_{i}$, where $x_{i}=0$ and $x_{i}=1$ correspond
to $S_{i}=-1$ and $S_{i}=+1$, respectively. In the CSP, one aims
to minimize the number of violated clauses
\[
C_{l}=\vee_{k=1}^{K}z_{i_{k,l}}\,,
\]
i.e., the number of clauses with value $C_{l}=0$. Here the literals
$z_{i}$ are either $x_{i}$ of their negation $\tilde{x}_{i}=1-x_{i}$.
Requiring the number of violated clauses to match the energy of the
Ising system, we get
\begin{align}
E(\boldsymbol{S}) & =\sum_{l=1}^{M}\delta\Bigg[\sum_{i=1}^{N}\Delta_{l,i}S_{i};-K\Bigg]
\end{align}
as the energy function, where $\delta[a;b]=\delta_{a,b}$ is the Kronecker
delta: If all Boolean variables in one clause $C_{l}$ take the opposite
of the value required to make $C_{l}$ true, one requires $\sum_{i=1}^{N}\Delta_{l,i}S_{i}=-K$.
For this to hold, one needs that
\begin{align}
\Delta_{l,i} & =\begin{cases}
+1 & \text{if }x_{i}\text{ appears}\\
-1 & \text{if }\overline{x}_{i}=1-x_{i}\text{ appears}\\
0 & \text{otherwise}
\end{cases}\,.
\end{align}
Now we notice that
\begin{align}
\prod_{i=1}^{N}(1-\Delta_{l,i}S_{i}) & =\begin{cases}
2^{K} & \text{if }C_{l}=0\\
0 & \text{if }C_{l}=1
\end{cases}
\end{align}
which we use to rewrite the energy
\begin{align}
E(\boldsymbol{S}) & =\frac{1}{2^{K}}\sum_{l=1}^{M}\prod_{i=1}^{N}(1-\Delta_{l,i}S_{i})\,.
\end{align}
The product can be expanded to
\begin{align}
\prod_{i=1}^{N}(1-\Delta_{l,i}S_{i}) & =1+\sum_{R=1}^{K}(-1)^{R}\sum_{i_{1}<i_{2}<\dots<i_{R}}\Delta_{l,i_{1}}\Delta_{l,i_{2}}\dots\Delta_{l,i_{R}}S_{i_{1}}S_{i_{2}}\dots S_{i_{R}}\\
 & =1-\sum_{i}\Delta_{l,i}S_{i}+\frac{1}{2}\sum_{i\neq j}\Delta_{l,i}\Delta_{l,j}S_{i}S_{j}\mp\dots\,.
\end{align}
Restricting to $K=2$ yields the energy function (\ref{eq:Energy_function})
where the parameters are
\begin{align}
H_{i} & =\frac{1}{4}\sum_{l=1}^{M}\Delta_{l,i}\\
J_{ij} & =\frac{1}{4}(\delta_{ij}-1)\sum_{l=1}^{M}\Delta_{l,i}\Delta_{l,j}\,.
\end{align}

\section{Implementation}

Our code will be available at Zenodo upon acceptance of this manuscript.

\subsection{Generation of random CSPs\protect\label{app:generation-of-random-csps}}

Throughout our manuscript, we refer to random MAX-2-SAT problems as
random CSPs. The clauses are generated by picking two random spins
and assigning two random values to the spins (both uniformly) while
respecting the following conditions:
\begin{enumerate}
\item Each variable appears at least once in a clause. If this would not
be the case, the value assigned to these disconnected variables would
be irrelevant and the problem is reduced to a problem with $N-N_{\text{disconnected}}$
variables.
\item Each clause contains two distinct variables. If both variables of
one clause are the same, either the truth value will be independent
of the variable's value (for $x_{i}\lor\tilde{x}_{i}$) which would
have no impact on the difficulty of the CSP, or the clause will be
reducible to length one, violating our prerequisite to only have length
two clauses.
\item All clauses have to be distinct from one another.
\end{enumerate}

\section{Hopfield networks in simplified notation\protect\label{app:Hopfield-renaming}}

In the original work \citep{Hopfield84} the Hopfield network was
written in the form
\begin{align}
C_{i}\frac{\mathrm{d}u_{i}(t)}{\mathrm{d}t} & =-\frac{u_{i}(t)}{R_{i}}+T_{ij}V_{j}(t)+I_{i}\\
u_{i}(t) & =g_{i}^{-1}(V_{i}(t))\,.
\end{align}
The network equations are biologically motivated, $u_{i}$ and $V_{i}$
play the role of input and output to/from neuron $i$, respectively,
and $g_{i}$ denotes the input-output characteristic. Moreover, $C_{i}$
is the input capacitance of the cell membrane and $R_{i}$ the transmembrane
resistance. Finally, $T_{ij}^{-1}$ denotes the impedance between
outputs $V_{j}$ and neuron $i$, and $I_{i}$ is a (time-independent)
input current. For our work, we rename $R_{i}=\beta$, $u_{i}\to\mu_{i}/\beta$,
$C_{i}\to\tau/\beta$, $T_{ij}\to J_{ij}$ and $I_{i}\to H_{i}$ and
choose $g_{i}=\tanh$ which is a standard choice for practical use
of the Hopfield network. In this notation, the Hopfield dynamics take
the form (\ref{eq:Hopfield_dynamics}).

\section{Derivations of Glauber-based solvers\protect\label{sec:Derivations-of-solvers}}

In the following, for completeness, we recapitulate some aspects of
the derivations of the Glauber dynamics \citep{Glauber63_294}, in
particular the choice of gain function $F$ that is required for the
system to sample from the correct Ising system, specified by the energy
in equation (\ref{eq:Energy_function}). Furthermore, we present a
systematic derivation of the different Glauber-based CSP solvers based
on a cumulant expansion \citep{Dahmen16_031024}.

\subsection{Gain function}

\label{app:gain-function}When evolving the Glauber dynamics, due
to the symmetry in couplings, the spin system will eventually reach
equilibrium at which the probability of leaving a state is the same
as entering a state \citep{Coolen00_arxiv_I}
\begin{equation}
p(\boldsymbol{S}_{i-})F_{i}^{+}(\boldsymbol{S}\backslash S_{i})=p(\boldsymbol{S}_{i+})\underbrace{(1-F_{i}^{+}(\boldsymbol{S}\backslash S_{i}))}_{=F_{i}^{-}(\boldsymbol{S}\backslash S_{i})}.\label{eq:detailed_balance}
\end{equation}
In an equilibrium state, the joint probability distribution is proportional
to the Boltzmann factor
\begin{equation}
p(\boldsymbol{S})\propto e^{-\beta E[\boldsymbol{S}]}.
\end{equation}
When we plug in the energy Hamiltonian (\ref{eq:Energy_function})
we can calculate the ratio of $p(\boldsymbol{S}_{i-})$ and $p(\boldsymbol{S}_{i+})$
as 
\begin{align}
\frac{p(\boldsymbol{S}_{i-})}{p(\boldsymbol{S}_{i+})}=\frac{F_{i}^{-}(\boldsymbol{S}\backslash S_{i})}{F_{i}^{+}(\boldsymbol{S}\backslash S_{i})} & =\frac{\exp(-\beta E[\boldsymbol{S}_{i-}])}{\exp(-\beta E[\boldsymbol{S}_{i+}])}\label{eq:gain-function-from-energy}\\
 & =\frac{\exp(-\beta h_{i})}{\exp(\beta h_{i})}\\
 & =\exp(-2\beta h_{i}).
\end{align}
and solve (\ref{eq:detailed_balance}) for $F_{i}^{+}(\boldsymbol{S}\backslash S_{i})$
to obtain the gain function
\[
F_{i}^{+}(\boldsymbol{S}\backslash S_{i})=\frac{1}{1+e^{-2\beta h_{i}}}=\frac{1}{2}\tanh(\beta h_{i})+\frac{1}{2}.
\]
Here $h_{i}=H_{i}+\sum_{j}J_{ij}S_{j}(t)$ denotes the local field,
i.e. the total magnetic field that spin $i$ experiences.

\subsection{Mean-field approximation\protect\label{app:mean-field-approximation}}

The mean values of the spins are defined by
\begin{equation}
m_{i}(t)=\sum_{\{\boldsymbol{S}\}}S_{i}p(\boldsymbol{S},t).
\end{equation}
By employing the master equation (\ref{eq:Glauber_master}) we can
calculate the time derivative of the mean as
\begin{align}
\tau\frac{d}{dt}m_{i}(t) & =\sum_{\{\boldsymbol{S}\}}S_{i}\tau\frac{d}{dt}p(\boldsymbol{S},t)\\
 & =\sum_{\{\boldsymbol{S}\}}\sum_{j=1}^{N}S_{i}S_{j}\left(p(\boldsymbol{S}_{j-})F_{j}^{+}(\boldsymbol{S}\backslash S_{j})-p(\boldsymbol{S}_{j+})F_{j}^{-}(\boldsymbol{S}\backslash S_{j})\right).
\end{align}
For terms with $i\neq j$, the sum over all configurations cancels
every term, leaving
\begin{align}
\tau\frac{d}{dt}m_{i}(t) & =\sum_{\{\boldsymbol{S}\}}\left(p(\boldsymbol{S}_{i-})F_{i}^{+}(\boldsymbol{S}\backslash S_{i})-p(\boldsymbol{S}_{i+})F_{i}^{-}(\boldsymbol{S}\backslash S_{i})\right).
\end{align}
Using $F_{i}^{-}(\boldsymbol{S}\backslash S_{i})=1-F_{i}^{+}(\boldsymbol{S}\backslash S_{i})$
yields
\begin{align}
\tau\frac{d}{dt}m_{i}(t) & =\sum_{\{\boldsymbol{S}\}}p(\boldsymbol{S}_{i-},t)F_{i}^{+}(\boldsymbol{S}\backslash S_{i})+p(\boldsymbol{S}_{i+},t)F_{i}^{+}(\boldsymbol{S}\backslash S_{i})-p(\boldsymbol{S}_{i+},t)\label{eq:append_B_1}\\
 & =\sum_{\{\boldsymbol{S}\}}2\left(p(\boldsymbol{S},t)F_{i}^{+}(\boldsymbol{S}\backslash S_{i})-\frac{1+S_{i}}{2}p(\boldsymbol{S},t)\right)\label{eq:append_B_2}\\
 & =2\langle F_{i}^{+}(\boldsymbol{S}\backslash S_{i})\rangle-1-\langle S_{i}\rangle.
\end{align}
To get from (\ref{eq:append_B_1}) to (\ref{eq:append_B_2}) we used
the fact that 
\begin{equation}
\sum_{\{\boldsymbol{S}\}}f(\boldsymbol{S})\ p(\boldsymbol{S}_{i\pm},t)=\sum_{\{\boldsymbol{S}\}}2f(\boldsymbol{S})\left(\frac{1\pm S_{i}}{2}p(\boldsymbol{S},t)\right),\label{eq:useful_formula}
\end{equation}
for an arbitrary function $f(\boldsymbol{S}).$ Now we can use the
explicit form of $F_{i}(\boldsymbol{S}\backslash S_{i})$ to obtain
\begin{align}
\tau\frac{d}{dt}m_{i}(t)+m_{i}(t) & =\langle2F_{i}^{+}(\boldsymbol{S}\backslash S_{i})-1\rangle\\
 & =\langle\tanh(\beta x_{i})\rangle.
\end{align}
At lowest order, i.e. neglecting all fluctuations in $x_{i}$, this
matches the result of the Hopfield networks (see Appendix \ref{app:equivalence-Hopfield-mf})
\begin{equation}
\tau\frac{d}{dt}m_{i}(t)=-m_{i}+\tanh(\beta\mu_{i}).
\end{equation}

\subsection{Inclusion of spin variances\protect\label{app:variance-approximation}}

Now we want to include also the spin variances. To do so we can make
use of the fact that the variances of total inputs
\begin{align}
\sigma_{i}^{2}=\langle h_{i}^{2}\rangle-\langle h_{i}\rangle^{2} & =\sum_{j,k}J_{ij}J_{ik}\left(\left\langle S_{j}S_{k}\right\rangle -\left\langle S_{j}\right\rangle \left\langle S_{k}\right\rangle \right)\\
 & \approx\sum_{j}J_{ij}J_{ij}\left(1-\left\langle S_{j}\right\rangle \left\langle S_{j}\right\rangle \right)+\sum_{j\neq k}J_{ij}J_{ik}\left(\left\langle S_{j}\right\rangle \left\langle S_{k}\right\rangle -\left\langle S_{j}\right\rangle \left\langle S_{k}\right\rangle \right)\\
 & =\sum_{j}J_{ij}^{2}(1-m_{j}^{2})
\end{align}
are a direct function of the means $m_{i}$ when covariances between
neurons are neglected. At lowest order the local field $h_{i}$ is
by the central limit theorem Gaussian distributed \citep{Dahmen16_031024}
so the equation becomes
\begin{equation}
\tau\frac{d}{dt}m_{i}(t)=-m_{i}+\langle\tanh(\beta h_{i})\rangle_{\mathcal{N}(\mu_{i},\sigma_{i})}.
\end{equation}

\subsection{Including spin covariances}

\label{app:covariance-approximation}To include spin covariances we
first calculate the time derivative of the second moment $\langle S_{i}S_{j}\rangle$
for $i\neq j$ only since $\langle S_{i}^{2}\rangle=1$. To simplify
the notation we introduce the abbreviation $\phi_{i}(\boldsymbol{S}\backslash S_{i})\coloneqq p(\boldsymbol{S}_{i-})F_{i}^{+}(\boldsymbol{S}\backslash S_{i})-p(\boldsymbol{S}_{i+})F_{i}^{-}(\boldsymbol{S}\backslash S_{i})$
for the probability flux that was already mentioned in the introduction
of the Glauber dynamics \citep{Ginzburg94,Renart10_587,Dahmen16_031024}:
\begin{align}
\tau\frac{d}{dt}\langle S_{i}S_{j}\rangle & =\sum_{\{\boldsymbol{S}\}}S_{i}S_{j}\tau\frac{d}{dt}p(\boldsymbol{S},t)\\
 & \stackrel{\text{eq.}(\ref{eq:Glauber_master})}{=}\sum_{\{\boldsymbol{S}\}}\sum_{k=1}^{N}S_{i}S_{j}S_{k}\phi_{k}(\boldsymbol{S}\backslash S_{k},t)\\
 & \stackrel{i\neq j}{=}\sum_{\{\boldsymbol{S}\}}\sum_{k\neq i,j}S_{i}S_{j}S_{k}\phi_{k}(\boldsymbol{S}\backslash S_{k},t)+\underbrace{S_{j}^{2}}_{=1}S_{i}\phi_{j}(\boldsymbol{S}\backslash S_{j},t)+\underbrace{S_{i}^{2}}_{=1}S_{j}\phi_{i}(\boldsymbol{S}\backslash S_{i},t)\\
 & =\sum_{\{\boldsymbol{S}\}}(S_{i}\phi_{j}(\boldsymbol{S}\backslash S_{j},t)+S_{j}\phi_{i}(\boldsymbol{S}\backslash S_{i},t)).
\end{align}
Plugging in the flux yields
\begin{align}
\sum_{\{\boldsymbol{S}\}}S_{i}\phi_{j}(\boldsymbol{S}\backslash S_{j},t) & =\sum_{\{\boldsymbol{S}\}}S_{i}(F_{j}^{+}(\boldsymbol{S}\backslash S_{j})(p(\boldsymbol{S}_{j+},t)+p(\boldsymbol{S}_{j-},t))-p(\boldsymbol{S}_{j+},t))\\
 & \stackrel{\text{eq.}(\ref{eq:useful_formula})}{=}\sum_{\{\boldsymbol{S}\}}S_{i}(2F_{j}^{+}(\boldsymbol{S}\backslash S_{j})p(\boldsymbol{S},t)-(1+S_{j})p(\boldsymbol{S},t))\\
 & =\langle S_{i}(2F_{j}^{+}(\boldsymbol{S}\backslash S_{j})-1)\rangle-\langle S_{i}S_{j}\rangle\\
 & =\langle S_{l}G_{k}(\boldsymbol{S})\rangle-\langle S_{l}S_{k}\rangle,
\end{align}
where $G_{i}(\boldsymbol{S})=2F_{i}^{+}(\boldsymbol{S}\backslash S_{i})-1=\tanh(\beta x_{i})$.
So in total we have
\begin{equation}
\tau\frac{d}{dt}\langle S_{i}S_{j}\rangle=\langle G_{i}(\boldsymbol{S})S_{j}\rangle+\langle G_{j}(\boldsymbol{S})S_{i}\rangle-2\langle S_{i}S_{j}\rangle.
\end{equation}
The time derivative of the covariance then follows as
\begin{align}
\tau\frac{d}{dt}c_{ij} & =\langle G_{i}(\boldsymbol{S})S_{j}\rangle+\langle G_{j}(\boldsymbol{S})S_{i}\rangle-2\langle S_{i}S_{j}\rangle-\langle S_{i}\rangle(\langle G_{j}(\boldsymbol{S})\rangle-\langle S_{j}\rangle)-\langle S_{j}\rangle(\langle G_{i}(\boldsymbol{S})\rangle-\langle S_{i}\rangle)\\
 & =\langle G_{i}(\boldsymbol{S})\delta S_{j}\rangle+\langle G_{j}(\boldsymbol{S})\delta S_{i}\rangle-2c_{ij},
\end{align}
where $\delta S_{i}=S_{i}-\langle S_{i}\rangle$ is the centralized
variable. We can now write $\langle G_{i}(\boldsymbol{S})S_{j}\rangle$
with the Fourier transform of $G_{i}(\boldsymbol{S})=G_{i}(\beta x_{i})$
as \citep{Dahmen16_031024}
\begin{align}
\langle G_{i}(\boldsymbol{S})S_{j}\rangle & =\frac{1}{2\pi}\int d\omega\,\hat{g}(\omega)\langle\exp(i\omega\beta x_{i})S_{j}\rangle\\
 & =\frac{1}{2\pi}\int d\omega\,\hat{g}(\omega)\left\langle \exp\left(i\omega\beta\left(\sum\nolimits_{k}J_{ik}S_{k}+H_{i}\right)\right)S_{j}\right\rangle \\
 & =\frac{1}{2\pi}\int d\omega\,\hat{g}(\omega)e^{i\omega\beta H_{i}}\left.\frac{\partial}{\partial j_{j}}\langle\exp(\boldsymbol{j}^{T}\boldsymbol{S})\rangle\right|_{j_{j}=i\omega\beta J_{ij}}.
\end{align}
Here we introduced the source $j_{k}=i\omega\beta J_{ik}.$ We can
identify the moment generation function $\varphi(\boldsymbol{j})=\langle\exp(\boldsymbol{j}^{T}\boldsymbol{S})\rangle$
and express it by the cumulant generating function $\Phi(\boldsymbol{j})=\ln\varphi(\boldsymbol{j})$
\begin{align}
\langle G_{i}(\boldsymbol{S})S_{j}\rangle & =\frac{1}{2\pi}\int d\omega\hat{g}(\omega)e^{i\omega\beta H_{i}}\left.\frac{\partial}{\partial j_{j}}e^{\Phi(\boldsymbol{j})}\right|_{j_{j}=i\omega\beta J_{ij}}\\
 & =\frac{1}{2\pi}\int d\omega\hat{g}(\omega)e^{i\omega\beta H_{i}}\langle\exp(\boldsymbol{j}^{T}\boldsymbol{S})\rangle\left.\frac{\partial}{\partial j_{j}}\Phi(\boldsymbol{j})\right|_{j_{j}=i\omega\beta J_{ij}}.\label{eq:cumulant_expansion}
\end{align}
Now we do a Gaussian approximation by assuming all cumulants higher
than two are zero, i.e.
\begin{equation}
\Phi(\boldsymbol{j})=\sum_{k}m_{k}j_{k}+\frac{1}{2}\sum_{k,l}c_{kl}j_{k}j_{l}.
\end{equation}
Plugging that into (\ref{eq:cumulant_expansion}) yields \citep{Dahmen16_031024}
\begin{align}
\langle G_{i}(\boldsymbol{S})S_{j}\rangle & \simeq\frac{1}{2\pi}\int d\omega\hat{g}(\omega)\langle\exp(i\omega\beta x_{i})\rangle\left(m_{j}+\sum_{k}i\omega\beta J_{ik}c_{kj}\right),
\end{align}
where we can identify the moment generating function $\langle\exp(i\omega\beta x_{i})\rangle$
of the field $x_{i}.$ As for the case with only spin variances, we
assume $x_{i}$ to be Gaussian distributed so
\begin{equation}
\langle\exp(i\omega\beta x_{i})\rangle=\exp\left(i\omega\beta\mu_{i}+\frac{1}{2}(i\omega\beta\sigma_{i})^{2}\right).
\end{equation}
We can get the factor $i\omega\beta$ in the covariance term by differentiating
this moment generating function by $\mu_{i}.$ Therefore we get
\begin{align}
\langle G_{i}(\boldsymbol{S})S_{j}\rangle & \simeq\frac{1}{2\pi}\int d\omega\hat{g}(\omega)\left(m_{j}+\sum_{k}J_{ik}c_{kj}\frac{\partial}{\partial\mu_{i}}\right)\langle\exp(i\omega\beta x_{i})\rangle\\
 & =\langle G_{i}(\boldsymbol{S})\rangle\langle S_{j}\rangle+\sum_{k}J_{ik}c_{kj}\frac{\partial}{\partial\mu_{i}}\langle G_{i}(\boldsymbol{S})\rangle,
\end{align}
and with the definition of the susceptibility $\chi_{i}=\frac{\partial}{\partial\mu_{i}}\langle G_{i}(\boldsymbol{S})\rangle$
we find
\begin{equation}
\langle G_{i}(\boldsymbol{S})\delta S_{j}\rangle\simeq\chi_{i}\sum_{k}J_{ik}c_{kj}.
\end{equation}

Thereby, the differential equation that describes the time evolution
of the covariance is (\ref{eq:h(c)}) with
\begin{equation}
h_{i}(m(t),c(t))=\sum_{k}(\chi_{i}(t)J_{ik}c_{kj}(t)+\chi_{j}(t)J_{jk}c_{ki}(t))-2c_{ij}(t).
\end{equation}
Here $\chi_{i}(t)$ is a time-dependent susceptibility, which depends
on $m(t)$ and $c(t)$. The variances of the total input are described
by (\ref{eq:f(c)}) with
\begin{align}
f_{i}(m(t),c(t)) & =\langle h_{i}^{2}\rangle-\langle h_{i}\rangle^{2}\\
 & =\sum_{j,k}J_{ij}J_{ik}c_{jk}(t)\\
 & =\sum_{j}J_{ij}^{2}(1-m_{j}^{2}(t))\\
 & +\sum_{j\neq k}J_{ij}J_{ik}c_{jk}(t)
\end{align}

For a stationary state $\left(\frac{d}{dt}c_{ij}=0\right),$the equation
self-consistency equation for the covariances becomes
\begin{equation}
c_{ij}=\frac{1}{2}\sum_{k}(\chi_{i}J_{ik}c_{kj}+\chi_{j}J_{jk}c_{ki}).
\end{equation}

\section{Equivalence of mean-field Glauber dynamics and Hopfield networks\protect\label{app:equivalence-Hopfield-mf}}

Here we show that the Hopfield dynamics stated in (\ref{eq:Hopfield_dynamics})
and the mean field dynamics in (\ref{eq:mean_field_dynamics}) and
(\ref{eq:mean_field_local_field}) are equivalent. We start with taking
the time derivative of (\ref{eq:mean_field_local_field}) (multiplied
with $\tau$) to calculate
\begin{equation}
\tau\frac{\mathrm{d}\mu_{i}}{\mathrm{d}t}=\tau\sum_{j}J_{ij}\frac{\mathrm{d}m_{j}}{\mathrm{d}t}\,.
\end{equation}
Insertion of (\ref{eq:mean_field_dynamics}) on the right hand side
yields
\begin{align}
\frac{\mathrm{d}\mu_{i}}{\mathrm{d}t} & =\sum_{j}J_{ij}(-m_{j}+\tanh(\beta\mu_{j}))\\
 & =H_{i}-\mu_{i}+J_{ij}\tanh(\beta\mu_{j})
\end{align}
where we used (\ref{eq:mean_field_local_field}) again to replace
$\sum_{j}J_{ij}m_{j}=\mu_{i}-H_{i}$. This is exactly equation (\ref{eq:Hopfield_dynamics}).

\section{Additional plots\protect\label{app:additonal_plots}}

\begin{figure}[H]
\centering{}\includegraphics[width=0.5\linewidth]{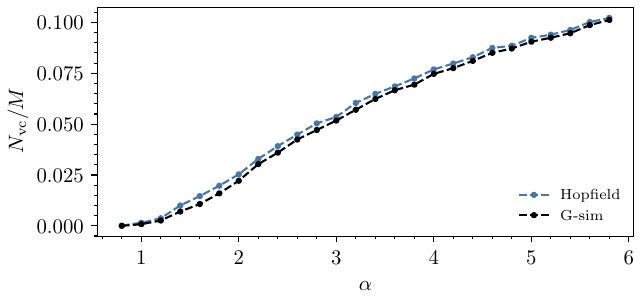}\caption{\protect\label{fig:app_violated_clauses_wider_range}Performance of
Glauber dynamics and Hopfield network for a wider range of $\alpha$
compared to Figure \ref{fig:performance_discrete_Glauber_and_Hopfield}c.
Parameters: $N=400$, $T=0.1$.}
\end{figure}

\begin{figure}[H]
\centering{}\includegraphics[width=1\linewidth]{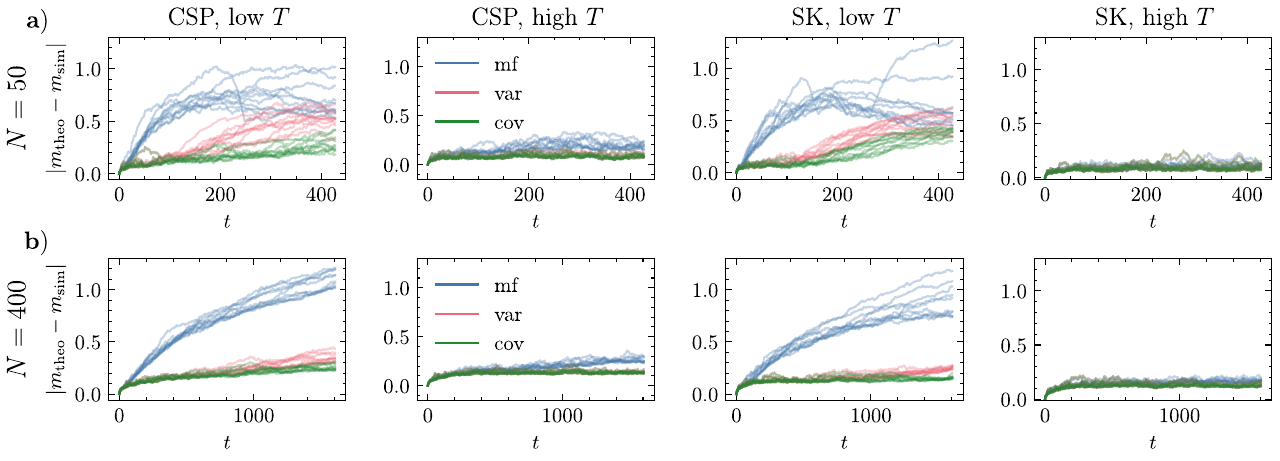}\caption{\protect\label{fig:Deviation-between-Glauber-and-approximations}Deviation
between Glauber simulation magnetizations and the predictions of the
mean-field, variance and covariance solver for CSP and SK couplings
shown for the $10$ spins with the highest deviation for both $N=50$
(panel a) and $N=400$ (panel b). High $T$ is $T=1$ and low $T$
is $T=0.1$. Parameters: $N=400$, $\alpha=3$.}
\end{figure}

\begin{figure}[H]
\centering{}\includegraphics[width=0.5\linewidth]{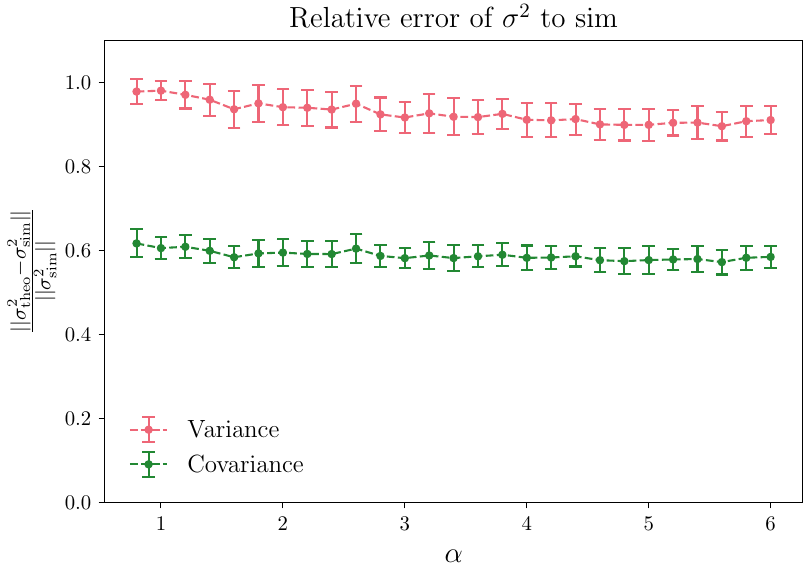}\caption{\protect\label{fig:Covariance-versus-Variance-solver-app}Covariance
versus variance solver. Relative norm of the difference between theoretical
and simulated variance plotted as a function of clause density $\alpha$.
Statistics was calculated over 50 different random CSPs per $\alpha$.}
\end{figure}

Here we show additional plots to complement our analysis in the main
text. Figure \ref{fig:app_violated_clauses_wider_range} shows the
performance comparison of Glauber dynamics and the Hopfield-network
dynamics over a larger range of clause densities $\alpha$ compared
to Figure \ref{fig:performance_discrete_Glauber_and_Hopfield}c).
While the performance difference is most clear for small $\alpha$,
we find the Glauber dynamics to outperform the Hopfield network for
all $\alpha$.
In Figure \ref{fig:mean-field-evaluation} we compared the simulated
Glauber dynamics with the mean-field, variance and covariance predictions
for two randomly chosen spins, from which we conclude that the mean-field
approximation is not sufficient for small temperatures. To corroborate
this conclusion, we show the deviation between Glauber dynamics and
approximations for the ten spins with the strongest deviations in
Figure \ref{fig:Deviation-between-Glauber-and-approximations}; the
spins are individually chosen for each approximation. While the mean-field
deviation is small for high temperatures, we observe strong deviations
for small temperatures and both CSPs and SK couplings for $N=50$
as well as $N=400$. In contrast, the variance and covariance solver
both provide accurate descriptions of the simulated dynamics in all
cases. 
Figure \ref{fig:test-covariance-prediction}c compares the variance
prediction obtained by the different theoretic solvers (G-var and
G-cov) with the results from the simulation for one random CSP instance.
In \ref{fig:Covariance-versus-Variance-solver-app} we show the difference
between the variance prediction and the result from the simulation
averaged over different instances and across different clause densities
to demonstrate that the covariance solver universally matches the
simulation better than the variance solver. Taking into account spin
correlations results in better agreement between theory and simulation.
The magnitude of this improvement is not dependent on $\alpha$, which
shows that correlations play a role for all tested clause densities.

\section{Cumulant-based solvers for 3-SAT \protect\label{app:3-SAT}}

In the 3-SAT case, due to clauses of length 3, the energy is formulated
with the 3-tensor $L_{ijk}$ in addition to the parameters $H_{i}$
and $J_{ij}$ from equation $(\ref{eq:Energy_function})$:

\[
E=\frac{M}{2^{3}}-\sum_{i}H_{i}S_{i}-\frac{1}{2}\sum_{i,j}J_{ij}S_{i}S_{j}-\frac{1}{6}\sum_{i,j,k}L_{ijk}S_{i}S_{j}S_{k}.
\]
The corresponding local field on spin $S_{i}$ reads

\[
h_{i}=H_{i}+\sum_{j}J_{ij}S_{j}+\frac{1}{2}\sum_{j,k}L_{ijk}S_{j}S_{k},
\]
where an additional $S_{j}S_{k}$ term appears relative to section
$\ref{app:gain-function}$.

In the mean-field approximation (compare with $(\ref{eq:mean_field_local_field})$),
the expected value of the local field reads

\[
\langle h_{i}\rangle=H_{i}+\sum_{j}J_{ij}m_{j}+\frac{1}{2}\sum_{j,k}L_{ijk}m_{j}m_{k}.
\]
In a variance solver, in addition to the mean value, the local fields
are endowed with a variance, just as in Section $\ref{app:variance-approximation}$:

\[
\sigma_{i}^{2}=\sum_{j}J_{ij}^{2}\left[1-m_{j}^{2}\right]+2\sum_{j,k}J_{ij}L_{ikj}m_{k}\left[1-m_{j}^{2}\right]+\sum_{j,k,l}(1-\delta_{kl})L_{ijk}L_{ijl}m_{k}m_{l}\left[1-m_{j}^{2}\right]+\frac{1}{2}\sum_{k,l}L_{ikl}^{2}\left[1-m_{k}^{2}m_{l}^{2}\right]
\]


\begin{thebibliography}{103}%
\makeatletter
\providecommand \@ifxundefined [1]{%
 \@ifx{#1\undefined}
}%
\providecommand \@ifnum [1]{%
 \ifnum #1\expandafter \@firstoftwo
 \else \expandafter \@secondoftwo
 \fi
}%
\providecommand \@ifx [1]{%
 \ifx #1\expandafter \@firstoftwo
 \else \expandafter \@secondoftwo
 \fi
}%
\providecommand \natexlab [1]{#1}%
\providecommand \enquote  [1]{``#1''}%
\providecommand \bibnamefont  [1]{#1}%
\providecommand \bibfnamefont [1]{#1}%
\providecommand \citenamefont [1]{#1}%
\providecommand \href@noop [0]{\@secondoftwo}%
\providecommand \href [0]{\begingroup \@sanitize@url \@href}%
\providecommand \@href[1]{\@@startlink{#1}\@@href}%
\providecommand \@@href[1]{\endgroup#1\@@endlink}%
\providecommand \@sanitize@url [0]{\catcode `\\12\catcode `\$12\catcode
  `\&12\catcode `\#12\catcode `\^12\catcode `\_12\catcode `\%12\relax}%
\providecommand \@@startlink[1]{}%
\providecommand \@@endlink[0]{}%
\providecommand \url  [0]{\begingroup\@sanitize@url \@url }%
\providecommand \@url [1]{\endgroup\@href {#1}{\urlprefix }}%
\providecommand \urlprefix  [0]{URL }%
\providecommand \Eprint [0]{\href }%
\providecommand \doibase [0]{https://doi.org/}%
\providecommand \selectlanguage [0]{\@gobble}%
\providecommand \bibinfo  [0]{\@secondoftwo}%
\providecommand \bibfield  [0]{\@secondoftwo}%
\providecommand \translation [1]{[#1]}%
\providecommand \BibitemOpen [0]{}%
\providecommand \bibitemStop [0]{}%
\providecommand \bibitemNoStop [0]{.\EOS\space}%
\providecommand \EOS [0]{\spacefactor3000\relax}%
\providecommand \BibitemShut  [1]{\csname bibitem#1\endcsname}%
\let\auto@bib@innerbib\@empty
\bibitem [{\citenamefont {Pierce}\ and\ \citenamefont
  {Turner}(2000)}]{pierce2000local}%
  \BibitemOpen
  \bibfield  {author} {\bibinfo {author} {\bibfnamefont {B.~C.}\ \bibnamefont
  {Pierce}}\ and\ \bibinfo {author} {\bibfnamefont {D.~N.}\ \bibnamefont
  {Turner}},\ }\bibfield  {title} {\bibinfo {title} {Local type inference},\
  }\href@noop {} {\bibfield  {journal} {\bibinfo  {journal} {Acm transactions
  on programming languages and systems (toplas)}\ }\textbf {\bibinfo {volume}
  {22}},\ \bibinfo {pages} {1} (\bibinfo {year} {2000})}\BibitemShut {NoStop}%
\bibitem [{\citenamefont {Ge}\ \emph {et~al.}(1999)\citenamefont {Ge},
  \citenamefont {Chou},\ and\ \citenamefont {Gao}}]{ge1999geometric}%
  \BibitemOpen
  \bibfield  {author} {\bibinfo {author} {\bibfnamefont {J.-X.}\ \bibnamefont
  {Ge}}, \bibinfo {author} {\bibfnamefont {S.-C.}\ \bibnamefont {Chou}},\ and\
  \bibinfo {author} {\bibfnamefont {X.-S.}\ \bibnamefont {Gao}},\ }\bibfield
  {title} {\bibinfo {title} {Geometric constraint satisfaction using
  optimization methods},\ }\href@noop {} {\bibfield  {journal} {\bibinfo
  {journal} {Computer-Aided Design}\ }\textbf {\bibinfo {volume} {31}},\
  \bibinfo {pages} {867} (\bibinfo {year} {1999})}\BibitemShut {NoStop}%
\bibitem [{\citenamefont {Hell}\ and\ \citenamefont
  {Ne{\v{s}}et{\v{r}}il}(2008)}]{hell2008colouring}%
  \BibitemOpen
  \bibfield  {author} {\bibinfo {author} {\bibfnamefont {P.}~\bibnamefont
  {Hell}}\ and\ \bibinfo {author} {\bibfnamefont {J.}~\bibnamefont
  {Ne{\v{s}}et{\v{r}}il}},\ }\bibfield  {title} {\bibinfo {title} {Colouring,
  constraint satisfaction, and complexity},\ }\href@noop {} {\bibfield
  {journal} {\bibinfo  {journal} {Computer Science Review}\ }\textbf {\bibinfo
  {volume} {2}},\ \bibinfo {pages} {143} (\bibinfo {year} {2008})}\BibitemShut
  {NoStop}%
\bibitem [{\citenamefont {Aceto}\ \emph {et~al.}(2004)\citenamefont {Aceto},
  \citenamefont {Hansen}, \citenamefont {Ing{\'o}lfsd{\'o}ttir}, \citenamefont
  {Johnsen},\ and\ \citenamefont {Knudsen}}]{aceto2004complexity}%
  \BibitemOpen
  \bibfield  {author} {\bibinfo {author} {\bibfnamefont {L.}~\bibnamefont
  {Aceto}}, \bibinfo {author} {\bibfnamefont {J.~A.}\ \bibnamefont {Hansen}},
  \bibinfo {author} {\bibfnamefont {A.}~\bibnamefont {Ing{\'o}lfsd{\'o}ttir}},
  \bibinfo {author} {\bibfnamefont {J.}~\bibnamefont {Johnsen}},\ and\ \bibinfo
  {author} {\bibfnamefont {J.}~\bibnamefont {Knudsen}},\ }\bibfield  {title}
  {\bibinfo {title} {The complexity of checking consistency of pedigree
  information and related problems},\ }\href@noop {} {\bibfield  {journal}
  {\bibinfo  {journal} {Journal of Computer Science and Technology}\ }\textbf
  {\bibinfo {volume} {19}},\ \bibinfo {pages} {42} (\bibinfo {year}
  {2004})}\BibitemShut {NoStop}%
\bibitem [{\citenamefont {Ramamoorthy}\ and\ \citenamefont
  {Jayagowri}(2023)}]{RAMAMOORTHY20232539}%
  \BibitemOpen
  \bibfield  {author} {\bibinfo {author} {\bibfnamefont {A.}~\bibnamefont
  {Ramamoorthy}}\ and\ \bibinfo {author} {\bibfnamefont {P.}~\bibnamefont
  {Jayagowri}},\ }\bibfield  {title} {\bibinfo {title} {The state-of-the-art
  boolean satisfiability based cryptanalysis},\ }\href
  {https://doi.org/https://doi.org/10.1016/j.matpr.2021.06.404} {\bibfield
  {journal} {\bibinfo  {journal} {Materials Today: Proceedings}\ }\textbf
  {\bibinfo {volume} {80}},\ \bibinfo {pages} {2539} (\bibinfo {year}
  {2023})},\ \bibinfo {note} {sI:5 NANO 2021}\BibitemShut {NoStop}%
\bibitem [{\citenamefont {Brailsford}\ \emph {et~al.}(1999)\citenamefont
  {Brailsford}, \citenamefont {Potts},\ and\ \citenamefont
  {Smith}}]{brailsford1999constraint}%
  \BibitemOpen
  \bibfield  {author} {\bibinfo {author} {\bibfnamefont {S.~C.}\ \bibnamefont
  {Brailsford}}, \bibinfo {author} {\bibfnamefont {C.~N.}\ \bibnamefont
  {Potts}},\ and\ \bibinfo {author} {\bibfnamefont {B.~M.}\ \bibnamefont
  {Smith}},\ }\bibfield  {title} {\bibinfo {title} {Constraint satisfaction
  problems: Algorithms and applications},\ }\href@noop {} {\bibfield  {journal}
  {\bibinfo  {journal} {European journal of operational research}\ }\textbf
  {\bibinfo {volume} {119}},\ \bibinfo {pages} {557} (\bibinfo {year}
  {1999})}\BibitemShut {NoStop}%
\bibitem [{\citenamefont {Karp}(1972)}]{Karp72_85}%
  \BibitemOpen
  \bibfield  {author} {\bibinfo {author} {\bibfnamefont {R.~M.}\ \bibnamefont
  {Karp}},\ }\bibinfo {title} {Reducibility among combinatorial problems},\ in\
  \href {https://doi.org/10.1007/978-1-4684-2001-2_9} {\emph {\bibinfo
  {booktitle} {Complexity of Computer Computations: Proceedings of a symposium
  on the Complexity of Computer Computations, held March 20--22, 1972, at the
  IBM Thomas J. Watson Research Center, Yorktown Heights, New York, and
  sponsored by the Office of Naval Research, Mathematics Program, IBM World
  Trade Corporation, and the IBM Research Mathematical Sciences Department}}},\
  \bibinfo {editor} {edited by\ \bibinfo {editor} {\bibfnamefont {R.~E.}\
  \bibnamefont {Miller}}, \bibinfo {editor} {\bibfnamefont {J.~W.}\
  \bibnamefont {Thatcher}},\ and\ \bibinfo {editor} {\bibfnamefont {J.~D.}\
  \bibnamefont {Bohlinger}}}\ (\bibinfo  {publisher} {Springer US},\ \bibinfo
  {address} {Boston, MA},\ \bibinfo {year} {1972})\ pp.\ \bibinfo {pages}
  {85--103}\BibitemShut {NoStop}%
\bibitem [{\citenamefont {M{\'e}zard}\ and\ \citenamefont
  {Montanari}(2009)}]{Mezard_Montanari09}%
  \BibitemOpen
  \bibfield  {author} {\bibinfo {author} {\bibfnamefont {M.}~\bibnamefont
  {M{\'e}zard}}\ and\ \bibinfo {author} {\bibfnamefont {A.}~\bibnamefont
  {Montanari}},\ }\href@noop {} {\emph {\bibinfo {title} {Information, Physics
  and Computation}}}\ (\bibinfo  {publisher} {Oxford University Press},\
  \bibinfo {address} {Oxford},\ \bibinfo {year} {2009})\BibitemShut {NoStop}%
\bibitem [{\citenamefont {Goldreich}(2010)}]{goldreich2010p}%
  \BibitemOpen
  \bibfield  {author} {\bibinfo {author} {\bibfnamefont {O.}~\bibnamefont
  {Goldreich}},\ }\href@noop {} {\emph {\bibinfo {title} {P, NP, and
  NP-Completeness: The basics of computational complexity}}}\ (\bibinfo
  {publisher} {Cambridge University Press},\ \bibinfo {year}
  {2010})\BibitemShut {NoStop}%
\bibitem [{\citenamefont {Garey~Michael}\ and\ \citenamefont
  {Johnson~David}(1979)}]{garey1979computers}%
  \BibitemOpen
  \bibfield  {author} {\bibinfo {author} {\bibfnamefont {R.}~\bibnamefont
  {Garey~Michael}}\ and\ \bibinfo {author} {\bibfnamefont {S.}~\bibnamefont
  {Johnson~David}},\ }\href@noop {} {\bibinfo {title} {Computers and
  intractability: A guide to the theory of np-completeness}} (\bibinfo {year}
  {1979})\BibitemShut {NoStop}%
\bibitem [{\citenamefont {Hopfield}(1984)}]{Hopfield84}%
  \BibitemOpen
  \bibfield  {author} {\bibinfo {author} {\bibfnamefont {J.~J.}\ \bibnamefont
  {Hopfield}},\ }\bibfield  {title} {\bibinfo {title} {Neurons with graded
  response have collective computational properties like those of two-state
  neurons},\ }\href@noop {} {\bibfield  {journal} {\bibinfo  {journal} {Proc.
  Natl. Acad. Sci. USA}\ }\textbf {\bibinfo {volume} {81}},\ \bibinfo {pages}
  {3088} (\bibinfo {year} {1984})}\BibitemShut {NoStop}%
\bibitem [{\citenamefont {Looi}(1992)}]{looi_neural_1992}%
  \BibitemOpen
  \bibfield  {author} {\bibinfo {author} {\bibfnamefont {C.-K.}\ \bibnamefont
  {Looi}},\ }\bibfield  {title} {\bibinfo {title} {Neural network methods in
  combinatorial optimization},\ }\href
  {https://doi.org/10.1016/0305-0548(92)90044-6} {\bibfield  {journal}
  {\bibinfo  {journal} {Computers \& Operations Research}\ }\textbf {\bibinfo
  {volume} {19}},\ \bibinfo {pages} {191} (\bibinfo {year} {1992})}\BibitemShut
  {NoStop}%
\bibitem [{\citenamefont {Hopfield}\ and\ \citenamefont
  {Tank}(1985)}]{hopfield85_141}%
  \BibitemOpen
  \bibfield  {author} {\bibinfo {author} {\bibfnamefont {J.~J.}\ \bibnamefont
  {Hopfield}}\ and\ \bibinfo {author} {\bibfnamefont {D.~W.}\ \bibnamefont
  {Tank}},\ }\bibfield  {title} {\bibinfo {title} {Neural computation of
  decisions in optimization problems},\ }\href
  {https://doi.org/10.1007/BF00339943} {\bibfield  {journal} {\bibinfo
  {journal} {Biological Cybernetics}\ }\textbf {\bibinfo {volume} {52}},\
  \bibinfo {pages} {141} (\bibinfo {year} {1985})}\BibitemShut {NoStop}%
\bibitem [{\citenamefont {Peterson}\ and\ \citenamefont
  {Anderson}(1987)}]{Peterson87_995}%
  \BibitemOpen
  \bibfield  {author} {\bibinfo {author} {\bibfnamefont {C.}~\bibnamefont
  {Peterson}}\ and\ \bibinfo {author} {\bibfnamefont {J.~R.}\ \bibnamefont
  {Anderson}},\ }\bibfield  {title} {\bibinfo {title} {A mean field theory
  learning algorithm for neural networks},\ }\href@noop {} {\bibfield
  {journal} {\bibinfo  {journal} {Complex Syst.}\ ,\ \bibinfo {pages} {995 }}
  (\bibinfo {year} {1987})}\BibitemShut {NoStop}%
\bibitem [{\citenamefont {Hopfield}(1982)}]{Hopfield82}%
  \BibitemOpen
  \bibfield  {author} {\bibinfo {author} {\bibfnamefont {J.~J.}\ \bibnamefont
  {Hopfield}},\ }\bibfield  {title} {\bibinfo {title} {Neural networks and
  physical systems with emergent collective computational abilities},\
  }\href@noop {} {\bibfield  {journal} {\bibinfo  {journal} {Proc. Natl. Acad.
  Sci. USA}\ }\textbf {\bibinfo {volume} {79}},\ \bibinfo {pages} {2554}
  (\bibinfo {year} {1982})}\BibitemShut {NoStop}%
\bibitem [{\citenamefont {Peterson}\ and\ \citenamefont
  {Anderson}(1988)}]{peterson_neural_1988}%
  \BibitemOpen
  \bibfield  {author} {\bibinfo {author} {\bibfnamefont {C.}~\bibnamefont
  {Peterson}}\ and\ \bibinfo {author} {\bibfnamefont {J.~R.}\ \bibnamefont
  {Anderson}},\ }\bibfield  {title} {\bibinfo {title} {Neural {Networks} and
  {NP}-complete {Optimization} {Problems}; {A} {Performance} {Study} on the
  {Graph} {Bisection} {Problem}},\ }\href@noop {} {\bibfield  {journal}
  {\bibinfo  {journal} {Complex Systems}\ } (\bibinfo {year}
  {1988})}\BibitemShut {NoStop}%
\bibitem [{\citenamefont {Bilbro}\ \emph {et~al.}(1988)\citenamefont {Bilbro},
  \citenamefont {Mann}, \citenamefont {Miller}, \citenamefont {Snyder},
  \citenamefont {van~den Bout},\ and\ \citenamefont
  {White}}]{bilbro_optimization_1988}%
  \BibitemOpen
  \bibfield  {author} {\bibinfo {author} {\bibfnamefont {G.}~\bibnamefont
  {Bilbro}}, \bibinfo {author} {\bibfnamefont {R.}~\bibnamefont {Mann}},
  \bibinfo {author} {\bibfnamefont {T.}~\bibnamefont {Miller}}, \bibinfo
  {author} {\bibfnamefont {W.}~\bibnamefont {Snyder}}, \bibinfo {author}
  {\bibfnamefont {D.}~\bibnamefont {van~den Bout}},\ and\ \bibinfo {author}
  {\bibfnamefont {M.}~\bibnamefont {White}},\ }\bibfield  {title} {\bibinfo
  {title} {Optimization by {Mean} {Field} {Annealing}},\ }in\ \href
  {https://proceedings.neurips.cc/paper_files/paper/1988/hash/ec5decca5ed3d6b8079e2e7e7bacc9f2-Abstract.html}
  {\emph {\bibinfo {booktitle} {Advances in {Neural} {Information} {Processing}
  {Systems}}}},\ Vol.~\bibinfo {volume} {1}\ (\bibinfo  {publisher}
  {Morgan-Kaufmann},\ \bibinfo {year} {1988})\BibitemShut {NoStop}%
\bibitem [{\citenamefont {Wen}\ \emph {et~al.}(2009{\natexlab{a}})\citenamefont
  {Wen}, \citenamefont {Lan},\ and\ \citenamefont {Shih}}]{wen_review_2009}%
  \BibitemOpen
  \bibfield  {author} {\bibinfo {author} {\bibfnamefont {U.-P.}\ \bibnamefont
  {Wen}}, \bibinfo {author} {\bibfnamefont {K.-M.}\ \bibnamefont {Lan}},\ and\
  \bibinfo {author} {\bibfnamefont {H.-S.}\ \bibnamefont {Shih}},\ }\bibfield
  {title} {\bibinfo {title} {A review of hopfield neural networks for solving
  mathematical programming problems},\ }\href
  {https://doi.org/10.1016/j.ejor.2008.11.002} {\bibfield  {journal} {\bibinfo
  {journal} {European Journal of Operational Research}\ }\textbf {\bibinfo
  {volume} {198}},\ \bibinfo {pages} {675} (\bibinfo {year}
  {2009}{\natexlab{a}})}\BibitemShut {NoStop}%
\bibitem [{\citenamefont {Sathasivam}\ \emph {et~al.}(2020)\citenamefont
  {Sathasivam}, \citenamefont {Mansor}, \citenamefont {Kasihmuddin},\ and\
  \citenamefont {Abubakar}}]{sathasivam_election_2020}%
  \BibitemOpen
  \bibfield  {author} {\bibinfo {author} {\bibfnamefont {S.}~\bibnamefont
  {Sathasivam}}, \bibinfo {author} {\bibfnamefont {M.~A.}\ \bibnamefont
  {Mansor}}, \bibinfo {author} {\bibfnamefont {M.~S.~M.}\ \bibnamefont
  {Kasihmuddin}},\ and\ \bibinfo {author} {\bibfnamefont {H.}~\bibnamefont
  {Abubakar}},\ }\bibfield  {title} {\bibinfo {title} {Election algorithm for
  random k satisfiability in the hopfield neural network},\ }\href
  {https://doi.org/10.3390/pr8050568} {\bibfield  {journal} {\bibinfo
  {journal} {Processes}\ }\textbf {\bibinfo {volume} {8}},\ \bibinfo {pages}
  {568} (\bibinfo {year} {2020})}\BibitemShut {NoStop}%
\bibitem [{\citenamefont {Kasihmuddin}\ \emph {et~al.}(2017)\citenamefont
  {Kasihmuddin}, \citenamefont {Sathasivam},\ and\ \citenamefont
  {Mansor}}]{kasihmuddin_hybrid_2017}%
  \BibitemOpen
  \bibfield  {author} {\bibinfo {author} {\bibfnamefont {M.~S.~M.}\
  \bibnamefont {Kasihmuddin}}, \bibinfo {author} {\bibfnamefont
  {S.}~\bibnamefont {Sathasivam}},\ and\ \bibinfo {author} {\bibfnamefont
  {M.~A.}\ \bibnamefont {Mansor}},\ }\bibfield  {title} {\bibinfo {title}
  {Hybrid genetic algorithm in the hopfield network for maximum
  2-satisfiability problem},\ }\href {https://doi.org/10.1063/1.4995911}
  {\bibfield  {journal} {\bibinfo  {journal} {AIP Conference Proceedings}\
  }\textbf {\bibinfo {volume} {1870}},\ \bibinfo {pages} {050001} (\bibinfo
  {year} {2017})}\BibitemShut {NoStop}%
\bibitem [{\citenamefont {Selsam}\ \emph {et~al.}(2019)\citenamefont {Selsam},
  \citenamefont {Lamm}, \citenamefont {B\"{u}nz}, \citenamefont {Liang},
  \citenamefont {de~Moura},\ and\ \citenamefont {Dill}}]{selsam2018learning}%
  \BibitemOpen
  \bibfield  {author} {\bibinfo {author} {\bibfnamefont {D.}~\bibnamefont
  {Selsam}}, \bibinfo {author} {\bibfnamefont {M.}~\bibnamefont {Lamm}},
  \bibinfo {author} {\bibfnamefont {B.}~\bibnamefont {B\"{u}nz}}, \bibinfo
  {author} {\bibfnamefont {P.}~\bibnamefont {Liang}}, \bibinfo {author}
  {\bibfnamefont {L.}~\bibnamefont {de~Moura}},\ and\ \bibinfo {author}
  {\bibfnamefont {D.~L.}\ \bibnamefont {Dill}},\ }\bibfield  {title} {\bibinfo
  {title} {Learning a {SAT} solver from single-bit supervision},\ }in\ \href
  {https://openreview.net/forum?id=HJMC_iA5tm} {\emph {\bibinfo {booktitle}
  {International Conference on Learning Representations}}}\ (\bibinfo {year}
  {2019})\BibitemShut {NoStop}%
\bibitem [{\citenamefont {Toenshoff}\ \emph {et~al.}(2021)\citenamefont
  {Toenshoff}, \citenamefont {Ritzert}, \citenamefont {Wolf},\ and\
  \citenamefont {Grohe}}]{Toenshoff21_580607}%
  \BibitemOpen
  \bibfield  {author} {\bibinfo {author} {\bibfnamefont {J.}~\bibnamefont
  {Toenshoff}}, \bibinfo {author} {\bibfnamefont {M.}~\bibnamefont {Ritzert}},
  \bibinfo {author} {\bibfnamefont {H.}~\bibnamefont {Wolf}},\ and\ \bibinfo
  {author} {\bibfnamefont {M.}~\bibnamefont {Grohe}},\ }\bibfield  {title}
  {\bibinfo {title} {Graph neural networks for maximum constraint
  satisfaction},\ }\href@noop {} {\bibfield  {journal} {\bibinfo  {journal}
  {Frontiers in artificial intelligence}\ }\textbf {\bibinfo {volume} {3}},\
  \bibinfo {pages} {580607} (\bibinfo {year} {2021})}\BibitemShut {NoStop}%
\bibitem [{\citenamefont {M{\'e}zard}\ \emph {et~al.}(2002)\citenamefont
  {M{\'e}zard}, \citenamefont {Parisi},\ and\ \citenamefont
  {Zecchina}}]{Mezard02_812}%
  \BibitemOpen
  \bibfield  {author} {\bibinfo {author} {\bibfnamefont {M.}~\bibnamefont
  {M{\'e}zard}}, \bibinfo {author} {\bibfnamefont {G.}~\bibnamefont {Parisi}},\
  and\ \bibinfo {author} {\bibfnamefont {R.}~\bibnamefont {Zecchina}},\
  }\bibfield  {title} {\bibinfo {title} {Analytic and algorithmic solution of
  random satisfiability problems},\ }\href
  {https://doi.org/10.1126/science.1073287} {\bibfield  {journal} {\bibinfo
  {journal} {Science}\ }\textbf {\bibinfo {volume} {297}},\ \bibinfo {pages}
  {812} (\bibinfo {year} {2002})},\ \Eprint
  {https://arxiv.org/abs/https://www.science.org/doi/pdf/10.1126/science.1073287}
  {https://www.science.org/doi/pdf/10.1126/science.1073287} \BibitemShut
  {NoStop}%
\bibitem [{\citenamefont {Monasson}\ and\ \citenamefont
  {Zecchina}(1996)}]{monasson1996entropy}%
  \BibitemOpen
  \bibfield  {author} {\bibinfo {author} {\bibfnamefont {R.}~\bibnamefont
  {Monasson}}\ and\ \bibinfo {author} {\bibfnamefont {R.}~\bibnamefont
  {Zecchina}},\ }\bibfield  {title} {\bibinfo {title} {Entropy of the
  k-satisfiability problem},\ }\href@noop {} {\bibfield  {journal} {\bibinfo
  {journal} {Physical review letters}\ }\textbf {\bibinfo {volume} {76}},\
  \bibinfo {pages} {3881} (\bibinfo {year} {1996})}\BibitemShut {NoStop}%
\bibitem [{\citenamefont {Kirkpatrick}\ \emph {et~al.}(1983)\citenamefont
  {Kirkpatrick}, \citenamefont {Gelatt},\ and\ \citenamefont
  {Vecchi}}]{kirkpatrick1983optimization}%
  \BibitemOpen
  \bibfield  {author} {\bibinfo {author} {\bibfnamefont {S.}~\bibnamefont
  {Kirkpatrick}}, \bibinfo {author} {\bibfnamefont {C.~D.}\ \bibnamefont
  {Gelatt}},\ and\ \bibinfo {author} {\bibfnamefont {M.~P.}\ \bibnamefont
  {Vecchi}},\ }\bibfield  {title} {\bibinfo {title} {Optimization by simulated
  annealing},\ }\href {https://doi.org/10.1126/science.220.4598.671} {\bibfield
   {journal} {\bibinfo  {journal} {Science}\ }\textbf {\bibinfo {volume}
  {220}},\ \bibinfo {pages} {671} (\bibinfo {year} {1983})}\BibitemShut
  {NoStop}%
\bibitem [{\citenamefont {Krzaka{\l}a}\ \emph {et~al.}(2007)\citenamefont
  {Krzaka{\l}a}, \citenamefont {Montanari}, \citenamefont {Ricci-Tersenghi},
  \citenamefont {Semerjian},\ and\ \citenamefont
  {Zdeborov{\'a}}}]{Krzakala07_10318}%
  \BibitemOpen
  \bibfield  {author} {\bibinfo {author} {\bibfnamefont {F.}~\bibnamefont
  {Krzaka{\l}a}}, \bibinfo {author} {\bibfnamefont {A.}~\bibnamefont
  {Montanari}}, \bibinfo {author} {\bibfnamefont {F.}~\bibnamefont
  {Ricci-Tersenghi}}, \bibinfo {author} {\bibfnamefont {G.}~\bibnamefont
  {Semerjian}},\ and\ \bibinfo {author} {\bibfnamefont {L.}~\bibnamefont
  {Zdeborov{\'a}}},\ }\bibfield  {title} {\bibinfo {title} {Gibbs states and
  the set of solutions of random constraint satisfaction problems},\
  }\href@noop {} {\bibfield  {journal} {\bibinfo  {journal} {Proceedings of the
  National Academy of Sciences}\ }\textbf {\bibinfo {volume} {104}},\ \bibinfo
  {pages} {10318} (\bibinfo {year} {2007})}\BibitemShut {NoStop}%
\bibitem [{\citenamefont {Monasson}\ and\ \citenamefont
  {Zecchina}(1997)}]{Monasson97_1357}%
  \BibitemOpen
  \bibfield  {author} {\bibinfo {author} {\bibfnamefont {R.}~\bibnamefont
  {Monasson}}\ and\ \bibinfo {author} {\bibfnamefont {R.}~\bibnamefont
  {Zecchina}},\ }\bibfield  {title} {\bibinfo {title} {Statistical mechanics of
  the random k-satisfiability model},\ }\href
  {https://doi.org/10.1103/physreve.56.1357} {\bibfield  {journal} {\bibinfo
  {journal} {Phys. Rev. E}\ }\textbf {\bibinfo {volume} {56}},\ \bibinfo
  {pages} {1357} (\bibinfo {year} {1997})}\BibitemShut {NoStop}%
\bibitem [{\citenamefont {M{\'e}zard}\ \emph {et~al.}(2005)\citenamefont
  {M{\'e}zard}, \citenamefont {Mora},\ and\ \citenamefont
  {Zecchina}}]{PhysRevLett.94.197205}%
  \BibitemOpen
  \bibfield  {author} {\bibinfo {author} {\bibfnamefont {M.}~\bibnamefont
  {M{\'e}zard}}, \bibinfo {author} {\bibfnamefont {T.}~\bibnamefont {Mora}},\
  and\ \bibinfo {author} {\bibfnamefont {R.}~\bibnamefont {Zecchina}},\
  }\bibfield  {title} {\bibinfo {title} {Clustering of solutions in the random
  satisfiability problem},\ }\href
  {https://doi.org/10.1103/PhysRevLett.94.197205} {\bibfield  {journal}
  {\bibinfo  {journal} {Phys. Rev. Lett.}\ }\textbf {\bibinfo {volume} {94}},\
  \bibinfo {pages} {197205} (\bibinfo {year} {2005})}\BibitemShut {NoStop}%
\bibitem [{\citenamefont {M{\'e}zard}\ and\ \citenamefont
  {Parisi}(1985)}]{Mezard1985}%
  \BibitemOpen
  \bibfield  {author} {\bibinfo {author} {\bibfnamefont {M.}~\bibnamefont
  {M{\'e}zard}}\ and\ \bibinfo {author} {\bibfnamefont {G.}~\bibnamefont
  {Parisi}},\ }\bibfield  {title} {\bibinfo {title} {Replicas and
  optimization},\ }\href {https://doi.org/10.1051/jphyslet:019850046017077100}
  {\bibfield  {journal} {\bibinfo  {journal} {J. Phys. Lett.}\ }\textbf
  {\bibinfo {volume} {46}},\ \bibinfo {pages} {771} (\bibinfo {year}
  {1985})}\BibitemShut {NoStop}%
\bibitem [{\citenamefont {M{\'e}zard}\ and\ \citenamefont
  {Zecchina}(2002)}]{PhysRevE.66.056126}%
  \BibitemOpen
  \bibfield  {author} {\bibinfo {author} {\bibfnamefont {M.}~\bibnamefont
  {M{\'e}zard}}\ and\ \bibinfo {author} {\bibfnamefont {R.}~\bibnamefont
  {Zecchina}},\ }\bibfield  {title} {\bibinfo {title} {Random
  \$k\$-satisfiability problem: From an analytic solution to an efficient
  algorithm},\ }\href {https://doi.org/10.1103/PhysRevE.66.056126} {\bibfield
  {journal} {\bibinfo  {journal} {Phys. Rev. E}\ }\textbf {\bibinfo {volume}
  {66}},\ \bibinfo {pages} {056126} (\bibinfo {year} {2002})}\BibitemShut
  {NoStop}%
\bibitem [{\citenamefont {Glauber}(1963)}]{Glauber63_294}%
  \BibitemOpen
  \bibfield  {author} {\bibinfo {author} {\bibfnamefont {R.}~\bibnamefont
  {Glauber}},\ }\bibfield  {title} {\bibinfo {title} {Time-dependent statistics
  of the {I}sing model},\ }\href@noop {} {\bibfield  {journal} {\bibinfo
  {journal} {J. Math. Phys.}\ }\textbf {\bibinfo {volume} {4}},\ \bibinfo
  {pages} {294} (\bibinfo {year} {1963})}\BibitemShut {NoStop}%
\bibitem [{\citenamefont {Dahmen}\ \emph {et~al.}(2016)\citenamefont {Dahmen},
  \citenamefont {Bos},\ and\ \citenamefont {Helias}}]{Dahmen16_031024}%
  \BibitemOpen
  \bibfield  {author} {\bibinfo {author} {\bibfnamefont {D.}~\bibnamefont
  {Dahmen}}, \bibinfo {author} {\bibfnamefont {H.}~\bibnamefont {Bos}},\ and\
  \bibinfo {author} {\bibfnamefont {M.}~\bibnamefont {Helias}},\ }\bibfield
  {title} {\bibinfo {title} {Correlated fluctuations in strongly coupled binary
  networks beyond equilibrium},\ }\href
  {https://doi.org/10.1103/PhysRevX.6.031024} {\bibfield  {journal} {\bibinfo
  {journal} {Phys. Rev. X}\ }\textbf {\bibinfo {volume} {6}},\ \bibinfo {pages}
  {031024} (\bibinfo {year} {2016})}\BibitemShut {NoStop}%
\bibitem [{\citenamefont {M\'{e}zard}\ \emph {et~al.}(1987)\citenamefont
  {M\'{e}zard}, \citenamefont {Parisi},\ and\ \citenamefont
  {Virasoro}}]{Mezard87}%
  \BibitemOpen
  \bibfield  {author} {\bibinfo {author} {\bibfnamefont {M.}~\bibnamefont
  {M\'{e}zard}}, \bibinfo {author} {\bibfnamefont {G.}~\bibnamefont {Parisi}},\
  and\ \bibinfo {author} {\bibfnamefont {M.}~\bibnamefont {Virasoro}},\ }\href
  {http://www.worldcat.org/isbn/9971501163} {\emph {\bibinfo {title} {{Spin
  Glass Theory and Beyond (World Scientific Lecture Notes in Physics, Vol
  9)}}}}\ (\bibinfo  {publisher} {{World Scientific Publishing Company}},\
  \bibinfo {year} {1987})\BibitemShut {NoStop}%
\bibitem [{\citenamefont {Wen}\ \emph {et~al.}(2009{\natexlab{b}})\citenamefont
  {Wen}, \citenamefont {Lan},\ and\ \citenamefont {Shih}}]{wen2009review}%
  \BibitemOpen
  \bibfield  {author} {\bibinfo {author} {\bibfnamefont {U.-P.}\ \bibnamefont
  {Wen}}, \bibinfo {author} {\bibfnamefont {K.-M.}\ \bibnamefont {Lan}},\ and\
  \bibinfo {author} {\bibfnamefont {H.-S.}\ \bibnamefont {Shih}},\ }\bibfield
  {title} {\bibinfo {title} {A review of hopfield neural networks for solving
  mathematical programming problems},\ }\href@noop {} {\bibfield  {journal}
  {\bibinfo  {journal} {European Journal of Operational Research}\ }\textbf
  {\bibinfo {volume} {198}},\ \bibinfo {pages} {675} (\bibinfo {year}
  {2009}{\natexlab{b}})}\BibitemShut {NoStop}%
\bibitem [{\citenamefont {Talav{\'a}n}\ and\ \citenamefont
  {Y{\'a}{\~n}ez}(2002)}]{talavan2002parameter}%
  \BibitemOpen
  \bibfield  {author} {\bibinfo {author} {\bibfnamefont {P.~M.}\ \bibnamefont
  {Talav{\'a}n}}\ and\ \bibinfo {author} {\bibfnamefont {J.}~\bibnamefont
  {Y{\'a}{\~n}ez}},\ }\bibfield  {title} {\bibinfo {title} {Parameter setting
  of the hopfield network applied to tsp},\ }\href@noop {} {\bibfield
  {journal} {\bibinfo  {journal} {Neural Networks}\ }\textbf {\bibinfo {volume}
  {15}},\ \bibinfo {pages} {363} (\bibinfo {year} {2002})}\BibitemShut
  {NoStop}%
\bibitem [{\citenamefont {Wang}\ \emph {et~al.}(1995)\citenamefont {Wang},
  \citenamefont {Sun}, \citenamefont {Golden},\ and\ \citenamefont
  {Jia}}]{wang1995using}%
  \BibitemOpen
  \bibfield  {author} {\bibinfo {author} {\bibfnamefont {Q.}~\bibnamefont
  {Wang}}, \bibinfo {author} {\bibfnamefont {X.}~\bibnamefont {Sun}}, \bibinfo
  {author} {\bibfnamefont {B.~L.}\ \bibnamefont {Golden}},\ and\ \bibinfo
  {author} {\bibfnamefont {J.}~\bibnamefont {Jia}},\ }\bibfield  {title}
  {\bibinfo {title} {Using artificial neural networks to solve the orienteering
  problem},\ }\href@noop {} {\bibfield  {journal} {\bibinfo  {journal} {Annals
  of Operations Research}\ }\textbf {\bibinfo {volume} {61}},\ \bibinfo {pages}
  {111} (\bibinfo {year} {1995})}\BibitemShut {NoStop}%
\bibitem [{\citenamefont {Feng}\ and\ \citenamefont
  {Douligeris}(2000)}]{feng2000using}%
  \BibitemOpen
  \bibfield  {author} {\bibinfo {author} {\bibfnamefont {G.}~\bibnamefont
  {Feng}}\ and\ \bibinfo {author} {\bibfnamefont {C.}~\bibnamefont
  {Douligeris}},\ }\bibfield  {title} {\bibinfo {title} {Using hopfield
  networks to solve traveling salesman problems based on stable state analysis
  technique},\ }in\ \href@noop {} {\emph {\bibinfo {booktitle} {Proceedings of
  the IEEE-INNS-ENNS International Joint Conference on Neural Networks. IJCNN
  2000. Neural Computing: New Challenges and Perspectives for the New
  Millennium}}},\ Vol.~\bibinfo {volume} {6}\ (\bibinfo {organization} {IEEE},\
  \bibinfo {year} {2000})\ pp.\ \bibinfo {pages} {521--526}\BibitemShut
  {NoStop}%
\bibitem [{\citenamefont {Sherrington}\ and\ \citenamefont
  {Kirkpatrick}(1975)}]{Sherrington75_1792}%
  \BibitemOpen
  \bibfield  {author} {\bibinfo {author} {\bibfnamefont {D.}~\bibnamefont
  {Sherrington}}\ and\ \bibinfo {author} {\bibfnamefont {S.}~\bibnamefont
  {Kirkpatrick}},\ }\bibfield  {title} {\bibinfo {title} {Solvable model of a
  spin-glass},\ }\href@noop {} {\bibfield  {journal} {\bibinfo  {journal}
  {Phys. Rev. Lett.}\ }\textbf {\bibinfo {volume} {35}},\ \bibinfo {pages}
  {1792} (\bibinfo {year} {1975})}\BibitemShut {NoStop}%
\bibitem [{\citenamefont {de~Almeida}\ and\ \citenamefont
  {Thouless}(1978)}]{de1978stability}%
  \BibitemOpen
  \bibfield  {author} {\bibinfo {author} {\bibfnamefont {J.~R.}\ \bibnamefont
  {de~Almeida}}\ and\ \bibinfo {author} {\bibfnamefont {D.~J.}\ \bibnamefont
  {Thouless}},\ }\bibfield  {title} {\bibinfo {title} {Stability of the
  sherrington-kirkpatrick solution of a spin glass model},\ }\href@noop {}
  {\bibfield  {journal} {\bibinfo  {journal} {Journal of Physics A:
  Mathematical and General}\ }\textbf {\bibinfo {volume} {11}},\ \bibinfo
  {pages} {983} (\bibinfo {year} {1978})}\BibitemShut {NoStop}%
\bibitem [{\citenamefont {Misra-Spieldenner}\ \emph {et~al.}(2023)\citenamefont
  {Misra-Spieldenner}, \citenamefont {Bode}, \citenamefont {Schuhmacher},
  \citenamefont {Stollenwerk}, \citenamefont {Bagrets},\ and\ \citenamefont
  {Wilhelm}}]{Bode23_030335}%
  \BibitemOpen
  \bibfield  {author} {\bibinfo {author} {\bibfnamefont {A.}~\bibnamefont
  {Misra-Spieldenner}}, \bibinfo {author} {\bibfnamefont {T.}~\bibnamefont
  {Bode}}, \bibinfo {author} {\bibfnamefont {P.~K.}\ \bibnamefont
  {Schuhmacher}}, \bibinfo {author} {\bibfnamefont {T.}~\bibnamefont
  {Stollenwerk}}, \bibinfo {author} {\bibfnamefont {D.}~\bibnamefont
  {Bagrets}},\ and\ \bibinfo {author} {\bibfnamefont {F.~K.}\ \bibnamefont
  {Wilhelm}},\ }\bibfield  {title} {\bibinfo {title} {Mean-field approximate
  optimization algorithm},\ }\href
  {https://doi.org/10.1103/PRXQuantum.4.030335} {\bibfield  {journal} {\bibinfo
   {journal} {PRX Quantum}\ }\textbf {\bibinfo {volume} {4}},\ \bibinfo {pages}
  {030335} (\bibinfo {year} {2023})}\BibitemShut {NoStop}%
\bibitem [{\citenamefont {Farhi}\ \emph {et~al.}(2015)\citenamefont {Farhi},
  \citenamefont {Goldstone},\ and\ \citenamefont {Gutmann}}]{Farhi15_arXiv}%
  \BibitemOpen
  \bibfield  {author} {\bibinfo {author} {\bibfnamefont {E.}~\bibnamefont
  {Farhi}}, \bibinfo {author} {\bibfnamefont {J.}~\bibnamefont {Goldstone}},\
  and\ \bibinfo {author} {\bibfnamefont {S.}~\bibnamefont {Gutmann}},\ }\href
  {https://arxiv.org/abs/1412.6062} {\bibinfo {title} {A quantum approximate
  optimization algorithm applied to a bounded occurrence constraint problem}}
  (\bibinfo {year} {2015}),\ \Eprint {https://arxiv.org/abs/1412.6062}
  {arXiv:1412.6062 [quant-ph]} \BibitemShut {NoStop}%
\bibitem [{\citenamefont {Bode}\ and\ \citenamefont
  {Wilhelm}(2024)}]{Bode24_012611}%
  \BibitemOpen
  \bibfield  {author} {\bibinfo {author} {\bibfnamefont {T.}~\bibnamefont
  {Bode}}\ and\ \bibinfo {author} {\bibfnamefont {F.~K.}\ \bibnamefont
  {Wilhelm}},\ }\bibfield  {title} {\bibinfo {title} {Adiabatic bottlenecks in
  quantum annealing and nonequilibrium dynamics of paramagnons},\ }\href
  {https://doi.org/10.1103/PhysRevA.110.012611} {\bibfield  {journal} {\bibinfo
   {journal} {Phys. Rev. A}\ }\textbf {\bibinfo {volume} {110}},\ \bibinfo
  {pages} {012611} (\bibinfo {year} {2024})}\BibitemShut {NoStop}%
\bibitem [{\citenamefont {Goemans}\ and\ \citenamefont
  {Williamson}(1995)}]{Goemans95_1115}%
  \BibitemOpen
  \bibfield  {author} {\bibinfo {author} {\bibfnamefont {M.~X.}\ \bibnamefont
  {Goemans}}\ and\ \bibinfo {author} {\bibfnamefont {D.~P.}\ \bibnamefont
  {Williamson}},\ }\bibfield  {title} {\bibinfo {title} {Improved approximation
  algorithms for maximum cut and satisfiability problems using semidefinite
  programming},\ }\href {https://doi.org/10.1145/227683.227684} {\bibfield
  {journal} {\bibinfo  {journal} {J. ACM}\ }\textbf {\bibinfo {volume} {42}},\
  \bibinfo {pages} {1115} (\bibinfo {year} {1995})}\BibitemShut {NoStop}%
\bibitem [{\citenamefont {Diamond}\ and\ \citenamefont
  {Boyd}(2016)}]{Diamond16_1}%
  \BibitemOpen
  \bibfield  {author} {\bibinfo {author} {\bibfnamefont {S.}~\bibnamefont
  {Diamond}}\ and\ \bibinfo {author} {\bibfnamefont {S.}~\bibnamefont {Boyd}},\
  }\bibfield  {title} {\bibinfo {title} {{CVXPY}: A {P}ython-embedded modeling
  language for convex optimization},\ }\href@noop {} {\bibfield  {journal}
  {\bibinfo  {journal} {J. Mach. Learn. Res.}\ }\textbf {\bibinfo {volume}
  {17}},\ \bibinfo {pages} {1} (\bibinfo {year} {2016})}\BibitemShut {NoStop}%
\bibitem [{\citenamefont {Lewin}\ \emph {et~al.}(2002)\citenamefont {Lewin},
  \citenamefont {Livnat},\ and\ \citenamefont {Zwick}}]{Lewin02_67}%
  \BibitemOpen
  \bibfield  {author} {\bibinfo {author} {\bibfnamefont {M.}~\bibnamefont
  {Lewin}}, \bibinfo {author} {\bibfnamefont {D.}~\bibnamefont {Livnat}},\ and\
  \bibinfo {author} {\bibfnamefont {U.}~\bibnamefont {Zwick}},\ }\bibfield
  {title} {\bibinfo {title} {Improved rounding techniques for the max 2-sat and
  max di-cut problems},\ }in\ \href@noop {} {\emph {\bibinfo {booktitle}
  {Integer Programming and Combinatorial Optimization}}},\ \bibinfo {editor}
  {edited by\ \bibinfo {editor} {\bibfnamefont {W.~J.}\ \bibnamefont {Cook}}\
  and\ \bibinfo {editor} {\bibfnamefont {A.~S.}\ \bibnamefont {Schulz}}}\
  (\bibinfo  {publisher} {Springer Berlin Heidelberg},\ \bibinfo {address}
  {Berlin, Heidelberg},\ \bibinfo {year} {2002})\ pp.\ \bibinfo {pages}
  {67--82}\BibitemShut {NoStop}%
\bibitem [{\citenamefont {Selman}\ \emph {et~al.}(1992)\citenamefont {Selman},
  \citenamefont {Levesque},\ and\ \citenamefont {Mitchell}}]{selman1992anew}%
  \BibitemOpen
  \bibfield  {author} {\bibinfo {author} {\bibfnamefont {B.}~\bibnamefont
  {Selman}}, \bibinfo {author} {\bibfnamefont {H.}~\bibnamefont {Levesque}},\
  and\ \bibinfo {author} {\bibfnamefont {D.}~\bibnamefont {Mitchell}},\
  }\bibfield  {title} {\bibinfo {title} {A new method for solving hard
  satisfiability problems},\ }\href
  {https://aaai.org/papers/00440-aaai92-068-a-new-method-for-solving-hard-satisfiability-problems/}
  {\bibfield  {journal} {\bibinfo  {journal} {Proceedings of the AAAI
  Conference on Artificial Intelligence, 10}\ } (\bibinfo {year}
  {1992})}\BibitemShut {NoStop}%
\bibitem [{\citenamefont {Selman}\ and\ \citenamefont
  {Kautz}()}]{selmanWalksatWebpage}%
  \BibitemOpen
  \bibfield  {author} {\bibinfo {author} {\bibfnamefont {B.}~\bibnamefont
  {Selman}}\ and\ \bibinfo {author} {\bibfnamefont {H.}~\bibnamefont {Kautz}},\
  }\href {https://www.cs.virginia.edu/~rmw7my/walksat/index.html} {\bibinfo
  {title} {Walksat home page}}\BibitemShut {NoStop}%
\bibitem [{\citenamefont {Selman}\ \emph {et~al.}(1993)\citenamefont {Selman},
  \citenamefont {Kautz}, \citenamefont {Cohen} \emph {et~al.}}]{Selman93_521}%
  \BibitemOpen
  \bibfield  {author} {\bibinfo {author} {\bibfnamefont {B.}~\bibnamefont
  {Selman}}, \bibinfo {author} {\bibfnamefont {H.~A.}\ \bibnamefont {Kautz}},
  \bibinfo {author} {\bibfnamefont {B.}~\bibnamefont {Cohen}}, \emph {et~al.},\
  }\bibfield  {title} {\bibinfo {title} {Local search strategies for
  satisfiability testing.},\ }\href@noop {} {\bibfield  {journal} {\bibinfo
  {journal} {Cliques, coloring, and satisfiability}\ }\textbf {\bibinfo
  {volume} {26}},\ \bibinfo {pages} {521} (\bibinfo {year} {1993})}\BibitemShut
  {NoStop}%
\bibitem [{\citenamefont {Selman}\ \emph {et~al.}(1994)\citenamefont {Selman},
  \citenamefont {Kautz}, \citenamefont {Cohen} \emph
  {et~al.}}]{selman1994noise}%
  \BibitemOpen
  \bibfield  {author} {\bibinfo {author} {\bibfnamefont {B.}~\bibnamefont
  {Selman}}, \bibinfo {author} {\bibfnamefont {H.~A.}\ \bibnamefont {Kautz}},
  \bibinfo {author} {\bibfnamefont {B.}~\bibnamefont {Cohen}}, \emph {et~al.},\
  }\bibfield  {title} {\bibinfo {title} {Noise strategies for improving local
  search},\ }in\ \href@noop {} {\emph {\bibinfo {booktitle} {AAAI}}},\
  Vol.~\bibinfo {volume} {94}\ (\bibinfo {year} {1994})\ pp.\ \bibinfo {pages}
  {337--343}\BibitemShut {NoStop}%
\bibitem [{\citenamefont {Ignatiev}\ \emph {et~al.}(2018)\citenamefont
  {Ignatiev}, \citenamefont {Morgado},\ and\ \citenamefont
  {Marques{-}Silva}}]{Ignatiev18_428}%
  \BibitemOpen
  \bibfield  {author} {\bibinfo {author} {\bibfnamefont {A.}~\bibnamefont
  {Ignatiev}}, \bibinfo {author} {\bibfnamefont {A.}~\bibnamefont {Morgado}},\
  and\ \bibinfo {author} {\bibfnamefont {J.}~\bibnamefont {Marques{-}Silva}},\
  }\bibfield  {title} {\bibinfo {title} {{PySAT:} {A} {Python} toolkit for
  prototyping with {SAT} oracles},\ }in\ \href
  {https://doi.org/10.1007/978-3-319-94144-8_26} {\emph {\bibinfo {booktitle}
  {SAT}}}\ (\bibinfo {year} {2018})\ pp.\ \bibinfo {pages}
  {428--437}\BibitemShut {NoStop}%
\bibitem [{\citenamefont {Ignatiev}\ \emph {et~al.}(2019)\citenamefont
  {Ignatiev}, \citenamefont {Morgado},\ and\ \citenamefont
  {Marques-Silva}}]{ignatiev2019rc2}%
  \BibitemOpen
  \bibfield  {author} {\bibinfo {author} {\bibfnamefont {A.}~\bibnamefont
  {Ignatiev}}, \bibinfo {author} {\bibfnamefont {A.}~\bibnamefont {Morgado}},\
  and\ \bibinfo {author} {\bibfnamefont {J.}~\bibnamefont {Marques-Silva}},\
  }\bibfield  {title} {\bibinfo {title} {Rc2: an efficient maxsat solver},\
  }\href@noop {} {\bibfield  {journal} {\bibinfo  {journal} {Journal on
  Satisfiability, Boolean Modelling and Computation}\ }\textbf {\bibinfo
  {volume} {11}},\ \bibinfo {pages} {53} (\bibinfo {year} {2019})}\BibitemShut
  {NoStop}%
\bibitem [{\citenamefont {ApS}(2024)}]{mosek}%
  \BibitemOpen
  \bibfield  {author} {\bibinfo {author} {\bibfnamefont {M.}~\bibnamefont
  {ApS}},\ }\href {https://docs.mosek.com/10.2/pythonapi/index.html} {\emph
  {\bibinfo {title} {MOSEK Optimizer API for Python. Version 10.2.8}}}
  (\bibinfo {year} {2024})\BibitemShut {NoStop}%
\bibitem [{\citenamefont {Boulebnane}\ and\ \citenamefont
  {Montanaro}(2024)}]{boulebnane2024_030348}%
  \BibitemOpen
  \bibfield  {author} {\bibinfo {author} {\bibfnamefont {S.}~\bibnamefont
  {Boulebnane}}\ and\ \bibinfo {author} {\bibfnamefont {A.}~\bibnamefont
  {Montanaro}},\ }\bibfield  {title} {\bibinfo {title} {Solving boolean
  satisfiability problems with the quantum approximate optimization
  algorithm},\ }\href@noop {} {\bibfield  {journal} {\bibinfo  {journal} {PRX
  Quantum}\ }\textbf {\bibinfo {volume} {5}},\ \bibinfo {pages} {030348}
  (\bibinfo {year} {2024})}\BibitemShut {NoStop}%
\bibitem [{\citenamefont {Krzakala}\ and\ \citenamefont
  {Zdeborov{\'a}}(2009)}]{krzakala2009hiding}%
  \BibitemOpen
  \bibfield  {author} {\bibinfo {author} {\bibfnamefont {F.}~\bibnamefont
  {Krzakala}}\ and\ \bibinfo {author} {\bibfnamefont {L.}~\bibnamefont
  {Zdeborov{\'a}}},\ }\bibfield  {title} {\bibinfo {title} {Hiding quiet
  solutions in random constraint satisfaction problems},\ }\href@noop {}
  {\bibfield  {journal} {\bibinfo  {journal} {Physical review letters}\
  }\textbf {\bibinfo {volume} {102}},\ \bibinfo {pages} {238701} (\bibinfo
  {year} {2009})}\BibitemShut {NoStop}%
\bibitem [{\citenamefont {Mohseni}\ \emph {et~al.}(2022)\citenamefont
  {Mohseni}, \citenamefont {McMahon},\ and\ \citenamefont
  {Byrnes}}]{Mohseni22_363}%
  \BibitemOpen
  \bibfield  {author} {\bibinfo {author} {\bibfnamefont {N.}~\bibnamefont
  {Mohseni}}, \bibinfo {author} {\bibfnamefont {P.~L.}\ \bibnamefont
  {McMahon}},\ and\ \bibinfo {author} {\bibfnamefont {T.}~\bibnamefont
  {Byrnes}},\ }\bibfield  {title} {\bibinfo {title} {Ising machines as hardware
  solvers of combinatorial optimization problems},\ }\href@noop {} {\bibfield
  {journal} {\bibinfo  {journal} {Nature Reviews Physics}\ }\textbf {\bibinfo
  {volume} {4}},\ \bibinfo {pages} {363} (\bibinfo {year} {2022})}\BibitemShut
  {NoStop}%
\bibitem [{\citenamefont {Cai}\ \emph {et~al.}(2020)\citenamefont {Cai},
  \citenamefont {Kumar}, \citenamefont {Van~Vaerenbergh}, \citenamefont
  {Sheng}, \citenamefont {Liu}, \citenamefont {Li}, \citenamefont {Liu},
  \citenamefont {Foltin}, \citenamefont {Yu}, \citenamefont {Xia} \emph
  {et~al.}}]{Cai20_409}%
  \BibitemOpen
  \bibfield  {author} {\bibinfo {author} {\bibfnamefont {F.}~\bibnamefont
  {Cai}}, \bibinfo {author} {\bibfnamefont {S.}~\bibnamefont {Kumar}}, \bibinfo
  {author} {\bibfnamefont {T.}~\bibnamefont {Van~Vaerenbergh}}, \bibinfo
  {author} {\bibfnamefont {X.}~\bibnamefont {Sheng}}, \bibinfo {author}
  {\bibfnamefont {R.}~\bibnamefont {Liu}}, \bibinfo {author} {\bibfnamefont
  {C.}~\bibnamefont {Li}}, \bibinfo {author} {\bibfnamefont {Z.}~\bibnamefont
  {Liu}}, \bibinfo {author} {\bibfnamefont {M.}~\bibnamefont {Foltin}},
  \bibinfo {author} {\bibfnamefont {S.}~\bibnamefont {Yu}}, \bibinfo {author}
  {\bibfnamefont {Q.}~\bibnamefont {Xia}}, \emph {et~al.},\ }\bibfield  {title}
  {\bibinfo {title} {Power-efficient combinatorial optimization using intrinsic
  noise in memristor hopfield neural networks},\ }\href@noop {} {\bibfield
  {journal} {\bibinfo  {journal} {Nature Electronics}\ }\textbf {\bibinfo
  {volume} {3}},\ \bibinfo {pages} {409} (\bibinfo {year} {2020})}\BibitemShut
  {NoStop}%
\bibitem [{\citenamefont {Hizzani}\ \emph {et~al.}(2024)\citenamefont
  {Hizzani}, \citenamefont {Heittmann}, \citenamefont {Hutchinson},
  \citenamefont {Dobrynin}, \citenamefont {Van~Vaerenbergh}, \citenamefont
  {Bhattacharya}, \citenamefont {Renaudineau}, \citenamefont {Strukov},\ and\
  \citenamefont {Strachan}}]{Hizzani24_1}%
  \BibitemOpen
  \bibfield  {author} {\bibinfo {author} {\bibfnamefont {M.}~\bibnamefont
  {Hizzani}}, \bibinfo {author} {\bibfnamefont {A.}~\bibnamefont {Heittmann}},
  \bibinfo {author} {\bibfnamefont {G.}~\bibnamefont {Hutchinson}}, \bibinfo
  {author} {\bibfnamefont {D.}~\bibnamefont {Dobrynin}}, \bibinfo {author}
  {\bibfnamefont {T.}~\bibnamefont {Van~Vaerenbergh}}, \bibinfo {author}
  {\bibfnamefont {T.}~\bibnamefont {Bhattacharya}}, \bibinfo {author}
  {\bibfnamefont {A.}~\bibnamefont {Renaudineau}}, \bibinfo {author}
  {\bibfnamefont {D.}~\bibnamefont {Strukov}},\ and\ \bibinfo {author}
  {\bibfnamefont {J.~P.}\ \bibnamefont {Strachan}},\ }\bibfield  {title}
  {\bibinfo {title} {Memristor-based hardware and algorithms for higher-order
  hopfield optimization solver outperforming quadratic ising machines},\ }in\
  \href@noop {} {\emph {\bibinfo {booktitle} {2024 IEEE International Symposium
  on Circuits and Systems (ISCAS)}}}\ (\bibinfo {organization} {IEEE},\
  \bibinfo {year} {2024})\ pp.\ \bibinfo {pages} {1--5}\BibitemShut {NoStop}%
\bibitem [{\citenamefont {Sompolinsky}\ and\ \citenamefont
  {Zippelius}(1981)}]{Sompolinsky81}%
  \BibitemOpen
  \bibfield  {author} {\bibinfo {author} {\bibfnamefont {H.}~\bibnamefont
  {Sompolinsky}}\ and\ \bibinfo {author} {\bibfnamefont {A.}~\bibnamefont
  {Zippelius}},\ }\bibfield  {title} {\bibinfo {title} {Dynamic theory of the
  spin-glass phase},\ }\href {https://doi.org/10.1103/PhysRevLett.47.359}
  {\bibfield  {journal} {\bibinfo  {journal} {Phys. Rev. Lett.}\ }\textbf
  {\bibinfo {volume} {47}},\ \bibinfo {pages} {359} (\bibinfo {year}
  {1981})}\BibitemShut {NoStop}%
\bibitem [{\citenamefont {Sompolinsky}\ and\ \citenamefont
  {Zippelius}(1982)}]{Sompolinsky82_6860}%
  \BibitemOpen
  \bibfield  {author} {\bibinfo {author} {\bibfnamefont {H.}~\bibnamefont
  {Sompolinsky}}\ and\ \bibinfo {author} {\bibfnamefont {A.}~\bibnamefont
  {Zippelius}},\ }\bibfield  {title} {\bibinfo {title} {Relaxational dynamics
  of the edwards-anderson model and the mean-field theory of spin-glasses},\
  }\href@noop {} {\bibfield  {journal} {\bibinfo  {journal} {Phys. Rev. B}\
  }\textbf {\bibinfo {volume} {25}},\ \bibinfo {pages} {6860} (\bibinfo {year}
  {1982})}\BibitemShut {NoStop}%
\bibitem [{\citenamefont {Sompolinsky}\ \emph {et~al.}(1988)\citenamefont
  {Sompolinsky}, \citenamefont {Crisanti},\ and\ \citenamefont
  {Sommers}}]{Sompolinsky88_259}%
  \BibitemOpen
  \bibfield  {author} {\bibinfo {author} {\bibfnamefont {H.}~\bibnamefont
  {Sompolinsky}}, \bibinfo {author} {\bibfnamefont {A.}~\bibnamefont
  {Crisanti}},\ and\ \bibinfo {author} {\bibfnamefont {H.~J.}\ \bibnamefont
  {Sommers}},\ }\bibfield  {title} {\bibinfo {title} {Chaos in random neural
  networks},\ }\href {https://doi.org/10.1103/PhysRevLett.61.259} {\bibfield
  {journal} {\bibinfo  {journal} {Phys. Rev. Lett.}\ }\textbf {\bibinfo
  {volume} {61}},\ \bibinfo {pages} {259} (\bibinfo {year} {1988})}\BibitemShut
  {NoStop}%
\bibitem [{\citenamefont {Cugliandolo}\ and\ \citenamefont
  {Kurchan}(1993)}]{Cugliandolo93_173}%
  \BibitemOpen
  \bibfield  {author} {\bibinfo {author} {\bibfnamefont {L.~F.}\ \bibnamefont
  {Cugliandolo}}\ and\ \bibinfo {author} {\bibfnamefont {J.}~\bibnamefont
  {Kurchan}},\ }\bibfield  {title} {\bibinfo {title} {Analytical solution of
  the off-equilibrium dynamics of a long-range spin-glass model},\ }\href
  {https://doi.org/10.1103/PhysRevLett.71.173} {\bibfield  {journal} {\bibinfo
  {journal} {Phys. Rev. Lett.}\ }\textbf {\bibinfo {volume} {71}},\ \bibinfo
  {pages} {173} (\bibinfo {year} {1993})}\BibitemShut {NoStop}%
\bibitem [{\citenamefont {Aljadeff}\ \emph {et~al.}(2015)\citenamefont
  {Aljadeff}, \citenamefont {Stern},\ and\ \citenamefont
  {Sharpee}}]{Aljadeff15_088101}%
  \BibitemOpen
  \bibfield  {author} {\bibinfo {author} {\bibfnamefont {J.}~\bibnamefont
  {Aljadeff}}, \bibinfo {author} {\bibfnamefont {M.}~\bibnamefont {Stern}},\
  and\ \bibinfo {author} {\bibfnamefont {T.}~\bibnamefont {Sharpee}},\
  }\bibfield  {title} {\bibinfo {title} {Transition to chaos in random networks
  with cell-type-specific connectivity},\ }\href
  {https://doi.org/10.1103/PhysRevLett.114.088101} {\bibfield  {journal}
  {\bibinfo  {journal} {Phys. Rev. Lett.}\ }\textbf {\bibinfo {volume} {114}},\
  \bibinfo {pages} {088101} (\bibinfo {year} {2015})}\BibitemShut {NoStop}%
\bibitem [{\citenamefont {Kadmon}\ and\ \citenamefont
  {Sompolinsky}(2015)}]{Kadmon15_041030}%
  \BibitemOpen
  \bibfield  {author} {\bibinfo {author} {\bibfnamefont {J.}~\bibnamefont
  {Kadmon}}\ and\ \bibinfo {author} {\bibfnamefont {H.}~\bibnamefont
  {Sompolinsky}},\ }\bibfield  {title} {\bibinfo {title} {Transition to chaos
  in random neuronal networks},\ }\href
  {https://doi.org/10.1103/PhysRevX.5.041030} {\bibfield  {journal} {\bibinfo
  {journal} {Phys. Rev. X}\ }\textbf {\bibinfo {volume} {5}},\ \bibinfo {pages}
  {041030} (\bibinfo {year} {2015})}\BibitemShut {NoStop}%
\bibitem [{\citenamefont {Ginzburg}\ and\ \citenamefont
  {Sompolinsky}(1994)}]{Ginzburg94}%
  \BibitemOpen
  \bibfield  {author} {\bibinfo {author} {\bibfnamefont {I.}~\bibnamefont
  {Ginzburg}}\ and\ \bibinfo {author} {\bibfnamefont {H.}~\bibnamefont
  {Sompolinsky}},\ }\bibfield  {title} {\bibinfo {title} {Theory of
  correlations in stochastic neural networks},\ }\href
  {https://doi.org/10.1103/PhysRevE.50.3171} {\bibfield  {journal} {\bibinfo
  {journal} {Phys. Rev. E}\ }\textbf {\bibinfo {volume} {50}},\ \bibinfo
  {pages} {3171} (\bibinfo {year} {1994})}\BibitemShut {NoStop}%
\bibitem [{\citenamefont {Renart}\ \emph {et~al.}(2010)\citenamefont {Renart},
  \citenamefont {{De La Rocha}}, \citenamefont {Bartho}, \citenamefont
  {Hollender}, \citenamefont {Parga}, \citenamefont {Reyes},\ and\
  \citenamefont {Harris}}]{Renart10_587}%
  \BibitemOpen
  \bibfield  {author} {\bibinfo {author} {\bibfnamefont {A.}~\bibnamefont
  {Renart}}, \bibinfo {author} {\bibfnamefont {J.}~\bibnamefont {{De La
  Rocha}}}, \bibinfo {author} {\bibfnamefont {P.}~\bibnamefont {Bartho}},
  \bibinfo {author} {\bibfnamefont {L.}~\bibnamefont {Hollender}}, \bibinfo
  {author} {\bibfnamefont {N.}~\bibnamefont {Parga}}, \bibinfo {author}
  {\bibfnamefont {A.}~\bibnamefont {Reyes}},\ and\ \bibinfo {author}
  {\bibfnamefont {K.~D.}\ \bibnamefont {Harris}},\ }\bibfield  {title}
  {\bibinfo {title} {The asynchronous state in cortical circuits},\ }\href
  {https://doi.org/10.1126/science.1179850} {\bibfield  {journal} {\bibinfo
  {journal} {Science}\ }\textbf {\bibinfo {volume} {327}},\ \bibinfo {pages}
  {587} (\bibinfo {year} {2010})}\BibitemShut {NoStop}%
\bibitem [{\citenamefont {Gleeson}(2011)}]{Gleeson11_068701}%
  \BibitemOpen
  \bibfield  {author} {\bibinfo {author} {\bibfnamefont {J.~P.}\ \bibnamefont
  {Gleeson}},\ }\bibfield  {title} {\bibinfo {title} {High-accuracy
  approximation of binary-state dynamics on networks},\ }\href@noop {}
  {\bibfield  {journal} {\bibinfo  {journal} {Phys. Rev. Lett.}\ }\textbf
  {\bibinfo {volume} {107}},\ \bibinfo {pages} {068701} (\bibinfo {year}
  {2011})}\BibitemShut {NoStop}%
\bibitem [{\citenamefont {Gleeson}(2013)}]{Gleeson13_021004}%
  \BibitemOpen
  \bibfield  {author} {\bibinfo {author} {\bibfnamefont {J.~P.}\ \bibnamefont
  {Gleeson}},\ }\bibfield  {title} {\bibinfo {title} {Binary-state dynamics on
  complex networks: pair approximation and beyond},\ }\href@noop {} {\bibfield
  {journal} {\bibinfo  {journal} {Phys. Rev. X}\ }\textbf {\bibinfo {volume}
  {3}},\ \bibinfo {pages} {021004} (\bibinfo {year} {2013})}\BibitemShut
  {NoStop}%
\bibitem [{\citenamefont {Coolen}\ \emph {et~al.}(1996)\citenamefont {Coolen},
  \citenamefont {Laughton},\ and\ \citenamefont {Sherrington}}]{Coolen96_8184}%
  \BibitemOpen
  \bibfield  {author} {\bibinfo {author} {\bibfnamefont {A.~C.~C.}\
  \bibnamefont {Coolen}}, \bibinfo {author} {\bibfnamefont {S.~N.}\
  \bibnamefont {Laughton}},\ and\ \bibinfo {author} {\bibfnamefont
  {D.}~\bibnamefont {Sherrington}},\ }\bibfield  {title} {\bibinfo {title}
  {Dynamical replica theory for disordered spin systems},\ }\href
  {https://doi.org/10.1103/PhysRevB.53.8184} {\bibfield  {journal} {\bibinfo
  {journal} {Phys. Rev. B}\ }\textbf {\bibinfo {volume} {53}},\ \bibinfo
  {pages} {8184} (\bibinfo {year} {1996})}\BibitemShut {NoStop}%
\bibitem [{\citenamefont {Laughton}\ \emph {et~al.}(1996)\citenamefont
  {Laughton}, \citenamefont {Coolen},\ and\ \citenamefont
  {Sherrington}}]{Laughton96_763}%
  \BibitemOpen
  \bibfield  {author} {\bibinfo {author} {\bibfnamefont {S.}~\bibnamefont
  {Laughton}}, \bibinfo {author} {\bibfnamefont {A.}~\bibnamefont {Coolen}},\
  and\ \bibinfo {author} {\bibfnamefont {D.}~\bibnamefont {Sherrington}},\
  }\bibfield  {title} {\bibinfo {title} {Order-parameter flow in the sk
  spin-glass: Ii. inclusion of microscopic memory effects},\ }\href@noop {}
  {\bibfield  {journal} {\bibinfo  {journal} {Journal of Physics A:
  Mathematical and General}\ }\textbf {\bibinfo {volume} {29}},\ \bibinfo
  {pages} {763} (\bibinfo {year} {1996})}\BibitemShut {NoStop}%
\bibitem [{\citenamefont {Mozeika}\ and\ \citenamefont
  {Coolen}(2008)}]{Mozeika08_115003}%
  \BibitemOpen
  \bibfield  {author} {\bibinfo {author} {\bibfnamefont {A.}~\bibnamefont
  {Mozeika}}\ and\ \bibinfo {author} {\bibfnamefont {A.}~\bibnamefont
  {Coolen}},\ }\bibfield  {title} {\bibinfo {title} {Dynamical replica analysis
  of processes on finitely connected random graphs: I. vertex covering},\
  }\href@noop {} {\bibfield  {journal} {\bibinfo  {journal} {Journal of Physics
  A: Mathematical and Theoretical}\ }\textbf {\bibinfo {volume} {41}},\
  \bibinfo {pages} {115003} (\bibinfo {year} {2008})}\BibitemShut {NoStop}%
\bibitem [{\citenamefont {Mozeika}\ and\ \citenamefont
  {Coolen}(2009)}]{Mozeika09_195006}%
  \BibitemOpen
  \bibfield  {author} {\bibinfo {author} {\bibfnamefont {A.}~\bibnamefont
  {Mozeika}}\ and\ \bibinfo {author} {\bibfnamefont {A.}~\bibnamefont
  {Coolen}},\ }\bibfield  {title} {\bibinfo {title} {Dynamical replica analysis
  of processes on finitely connected random graphs: Ii. dynamics in the
  griffiths phase of the diluted ising ferromagnet},\ }\href@noop {} {\bibfield
   {journal} {\bibinfo  {journal} {Journal of Physics A: Mathematical and
  Theoretical}\ }\textbf {\bibinfo {volume} {42}},\ \bibinfo {pages} {195006}
  (\bibinfo {year} {2009})}\BibitemShut {NoStop}%
\bibitem [{\citenamefont {M{\'e}zard}\ and\ \citenamefont
  {Sakellariou}(2011)}]{Mezard11_L07001}%
  \BibitemOpen
  \bibfield  {author} {\bibinfo {author} {\bibfnamefont {M.}~\bibnamefont
  {M{\'e}zard}}\ and\ \bibinfo {author} {\bibfnamefont {J.}~\bibnamefont
  {Sakellariou}},\ }\bibfield  {title} {\bibinfo {title} {Exact mean-field
  inference in asymmetric kinetic ising systems},\ }\href@noop {} {\bibfield
  {journal} {\bibinfo  {journal} {J. Stat. Mech. Theory Exp.}\ }\textbf
  {\bibinfo {volume} {2011}},\ \bibinfo {pages} {L07001} (\bibinfo {year}
  {2011})}\BibitemShut {NoStop}%
\bibitem [{\citenamefont {Plefka}(1982)}]{Plefka82_1971}%
  \BibitemOpen
  \bibfield  {author} {\bibinfo {author} {\bibfnamefont {T.}~\bibnamefont
  {Plefka}},\ }\bibfield  {title} {\bibinfo {title} {Convergence condition of
  the tap equation for the infinite-ranged ising spin glass model},\
  }\href@noop {} {\bibfield  {journal} {\bibinfo  {journal} {J. Phys. A Math.
  Gen.}\ }\textbf {\bibinfo {volume} {15}},\ \bibinfo {pages} {1971} (\bibinfo
  {year} {1982})}\BibitemShut {NoStop}%
\bibitem [{\citenamefont {Vasiliev}\ and\ \citenamefont
  {Radzhabov}(1974)}]{Vasiliev74}%
  \BibitemOpen
  \bibfield  {author} {\bibinfo {author} {\bibfnamefont {A.~N.}\ \bibnamefont
  {Vasiliev}}\ and\ \bibinfo {author} {\bibfnamefont {R.~A.}\ \bibnamefont
  {Radzhabov}},\ }\bibfield  {title} {\bibinfo {title} {Legendre transforms in
  the ising model},\ }\href {https://doi.org/10.1007/BF01035593} {\bibfield
  {journal} {\bibinfo  {journal} {Theor. Math. Phys.}\ }\textbf {\bibinfo
  {volume} {21}},\ \bibinfo {pages} {963} (\bibinfo {year} {1974})}\BibitemShut
  {NoStop}%
\bibitem [{\citenamefont {Vasiliev}\ and\ \citenamefont
  {Radzhabov}(1975)}]{Vasiliev75}%
  \BibitemOpen
  \bibfield  {author} {\bibinfo {author} {\bibfnamefont {A.~N.}\ \bibnamefont
  {Vasiliev}}\ and\ \bibinfo {author} {\bibfnamefont {R.~A.}\ \bibnamefont
  {Radzhabov}},\ }\bibfield  {title} {\bibinfo {title} {Analysis of the nonstar
  graphs of the legendre transform in the ising model},\ }\href
  {https://doi.org/10.1007/BF01041677} {\bibfield  {journal} {\bibinfo
  {journal} {Theor. Math. Phys.}\ }\textbf {\bibinfo {volume} {23}},\ \bibinfo
  {pages} {575} (\bibinfo {year} {1975})}\BibitemShut {NoStop}%
\bibitem [{\citenamefont {Thouless}\ \emph {et~al.}(1977)\citenamefont
  {Thouless}, \citenamefont {Anderson},\ and\ \citenamefont
  {Palmer}}]{Thouless77_593}%
  \BibitemOpen
  \bibfield  {author} {\bibinfo {author} {\bibfnamefont {D.~J.}\ \bibnamefont
  {Thouless}}, \bibinfo {author} {\bibfnamefont {P.~W.}\ \bibnamefont
  {Anderson}},\ and\ \bibinfo {author} {\bibfnamefont {R.~G.}\ \bibnamefont
  {Palmer}},\ }\bibfield  {title} {\bibinfo {title} {Solution of 'solvable
  model of a spin glass'},\ }\href@noop {} {\bibfield  {journal} {\bibinfo
  {journal} {Philos. Mag.}\ }\textbf {\bibinfo {volume} {35}},\ \bibinfo
  {pages} {593} (\bibinfo {year} {1977})}\BibitemShut {NoStop}%
\bibitem [{\citenamefont {Nakanishi}\ and\ \citenamefont
  {Takayama}(1997)}]{Nakanishi97_8085}%
  \BibitemOpen
  \bibfield  {author} {\bibinfo {author} {\bibfnamefont {K.}~\bibnamefont
  {Nakanishi}}\ and\ \bibinfo {author} {\bibfnamefont {H.}~\bibnamefont
  {Takayama}},\ }\bibfield  {title} {\bibinfo {title} {Mean-field theory for a
  spin-glass model of neural networks: Tap free energy and the paramagnetic to
  spin-glass transition},\ }\href@noop {} {\bibfield  {journal} {\bibinfo
  {journal} {J. Phys. A Math. Gen.}\ }\textbf {\bibinfo {volume} {30}},\
  \bibinfo {pages} {8085} (\bibinfo {year} {1997})}\BibitemShut {NoStop}%
\bibitem [{\citenamefont {Tanaka}(1998)}]{Tanaka98_2302}%
  \BibitemOpen
  \bibfield  {author} {\bibinfo {author} {\bibfnamefont {T.}~\bibnamefont
  {Tanaka}},\ }\bibfield  {title} {\bibinfo {title} {Mean-field theory of
  boltzmann machine learning},\ }\href@noop {} {\bibfield  {journal} {\bibinfo
  {journal} {Phys. Rev. E}\ }\textbf {\bibinfo {volume} {58}},\ \bibinfo
  {pages} {2302} (\bibinfo {year} {1998})}\BibitemShut {NoStop}%
\bibitem [{\citenamefont {Nishimori}(2001)}]{Nishimori01_01}%
  \BibitemOpen
  \bibfield  {author} {\bibinfo {author} {\bibfnamefont {H.}~\bibnamefont
  {Nishimori}},\ }\href@noop {} {\emph {\bibinfo {title} {Statistical Physics
  of Spin Glasses and Information Processing An Introduction}}}\ (\bibinfo
  {publisher} {Clarendron Press, Oxford},\ \bibinfo {year} {2001})\BibitemShut
  {NoStop}%
\bibitem [{\citenamefont {Roudi}\ and\ \citenamefont
  {Hertz}(2011)}]{Roudi11_P03031}%
  \BibitemOpen
  \bibfield  {author} {\bibinfo {author} {\bibfnamefont {Y.}~\bibnamefont
  {Roudi}}\ and\ \bibinfo {author} {\bibfnamefont {J.}~\bibnamefont {Hertz}},\
  }\bibfield  {title} {\bibinfo {title} {Dynamical tap equations for
  non-equilibrium ising spin glasses},\ }\href@noop {} {\bibfield  {journal}
  {\bibinfo  {journal} {Journal of Statistical Mechanics: Theory and
  Experiment}\ }\textbf {\bibinfo {volume} {2011}},\ \bibinfo {pages} {P03031}
  (\bibinfo {year} {2011})}\BibitemShut {NoStop}%
\bibitem [{\citenamefont {Pearl}(1982)}]{Pearl82_133}%
  \BibitemOpen
  \bibfield  {author} {\bibinfo {author} {\bibfnamefont {J.}~\bibnamefont
  {Pearl}},\ }\bibfield  {title} {\bibinfo {title} {Reverend bayes on inference
  engines: A distributed hierarchical approach},\ }in\ \href@noop {} {\emph
  {\bibinfo {booktitle} {AAAI-82 Proceedings}}}\ (\bibinfo {year}
  {1982})\BibitemShut {NoStop}%
\bibitem [{\citenamefont {Yedidia}\ \emph {et~al.}(2000)\citenamefont
  {Yedidia}, \citenamefont {Freeman},\ and\ \citenamefont {Weiss}}]{Yedidia00}%
  \BibitemOpen
  \bibfield  {author} {\bibinfo {author} {\bibfnamefont {J.~S.}\ \bibnamefont
  {Yedidia}}, \bibinfo {author} {\bibfnamefont {W.}~\bibnamefont {Freeman}},\
  and\ \bibinfo {author} {\bibfnamefont {Y.}~\bibnamefont {Weiss}},\ }\bibfield
   {title} {\bibinfo {title} {Generalized belief propagation},\ }\href@noop {}
  {\bibfield  {journal} {\bibinfo  {journal} {Advances in neural information
  processing systems}\ }\textbf {\bibinfo {volume} {13}} (\bibinfo {year}
  {2000})}\BibitemShut {NoStop}%
\bibitem [{\citenamefont {Altarelli}\ \emph {et~al.}(2013)\citenamefont
  {Altarelli}, \citenamefont {Braunstein}, \citenamefont {Dall’Asta},\ and\
  \citenamefont {Zecchina}}]{Altarelli13_062115}%
  \BibitemOpen
  \bibfield  {author} {\bibinfo {author} {\bibfnamefont {F.}~\bibnamefont
  {Altarelli}}, \bibinfo {author} {\bibfnamefont {A.}~\bibnamefont
  {Braunstein}}, \bibinfo {author} {\bibfnamefont {L.}~\bibnamefont
  {Dall’Asta}},\ and\ \bibinfo {author} {\bibfnamefont {R.}~\bibnamefont
  {Zecchina}},\ }\bibfield  {title} {\bibinfo {title} {Large deviations of
  cascade processes on graphs},\ }\href@noop {} {\bibfield  {journal} {\bibinfo
   {journal} {Phys. Rev. E}\ }\textbf {\bibinfo {volume} {87}},\ \bibinfo
  {pages} {062115} (\bibinfo {year} {2013})}\BibitemShut {NoStop}%
\bibitem [{\citenamefont {Braunstein}\ \emph {et~al.}(2005)\citenamefont
  {Braunstein}, \citenamefont {M{\'e}zard},\ and\ \citenamefont
  {Zecchina}}]{Braunstein05_201}%
  \BibitemOpen
  \bibfield  {author} {\bibinfo {author} {\bibfnamefont {A.}~\bibnamefont
  {Braunstein}}, \bibinfo {author} {\bibfnamefont {M.}~\bibnamefont
  {M{\'e}zard}},\ and\ \bibinfo {author} {\bibfnamefont {R.}~\bibnamefont
  {Zecchina}},\ }\bibfield  {title} {\bibinfo {title} {Survey propagation: An
  algorithm for satisfiability},\ }\href@noop {} {\bibfield  {journal}
  {\bibinfo  {journal} {Random Structures \& Algorithms}\ }\textbf {\bibinfo
  {volume} {27}},\ \bibinfo {pages} {201} (\bibinfo {year} {2005})}\BibitemShut
  {NoStop}%
\bibitem [{\citenamefont {Montanari}\ \emph {et~al.}(2007)\citenamefont
  {Montanari}, \citenamefont {Ricci-Tersenghi},\ and\ \citenamefont
  {Semerjian}}]{Montanari07_arxiv}%
  \BibitemOpen
  \bibfield  {author} {\bibinfo {author} {\bibfnamefont {A.}~\bibnamefont
  {Montanari}}, \bibinfo {author} {\bibfnamefont {F.}~\bibnamefont
  {Ricci-Tersenghi}},\ and\ \bibinfo {author} {\bibfnamefont {G.}~\bibnamefont
  {Semerjian}},\ }\bibfield  {title} {\bibinfo {title} {Solving constraint
  satisfaction problems through belief propagation-guided decimation},\
  }\href@noop {} {\bibfield  {journal} {\bibinfo  {journal} {arXiv preprint
  arXiv:0709.1667}\ } (\bibinfo {year} {2007})}\BibitemShut {NoStop}%
\bibitem [{\citenamefont {Ricci-Tersenghi}\ and\ \citenamefont
  {Semerjian}(2009)}]{Ricci09_P09001}%
  \BibitemOpen
  \bibfield  {author} {\bibinfo {author} {\bibfnamefont {F.}~\bibnamefont
  {Ricci-Tersenghi}}\ and\ \bibinfo {author} {\bibfnamefont {G.}~\bibnamefont
  {Semerjian}},\ }\bibfield  {title} {\bibinfo {title} {On the cavity method
  for decimated random constraint satisfaction problems and the analysis of
  belief propagation guided decimation algorithms},\ }\href@noop {} {\bibfield
  {journal} {\bibinfo  {journal} {Journal of Statistical Mechanics: Theory and
  Experiment}\ }\textbf {\bibinfo {volume} {2009}},\ \bibinfo {pages} {P09001}
  (\bibinfo {year} {2009})}\BibitemShut {NoStop}%
\bibitem [{\citenamefont {Del~Ferraro}\ and\ \citenamefont
  {Aurell}(2014)}]{DelFerraro14_arxiv}%
  \BibitemOpen
  \bibfield  {author} {\bibinfo {author} {\bibfnamefont {G.}~\bibnamefont
  {Del~Ferraro}}\ and\ \bibinfo {author} {\bibfnamefont {E.}~\bibnamefont
  {Aurell}},\ }\bibfield  {title} {\bibinfo {title} {Dynamic message-passing
  approach for kinetic spin models with reversible dynamics},\ }\href@noop {}
  {\bibfield  {journal} {\bibinfo  {journal} {arXiv preprint arXiv:1409.4684}\
  } (\bibinfo {year} {2014})}\BibitemShut {NoStop}%
\bibitem [{\citenamefont {Marino}\ \emph {et~al.}(2016)\citenamefont {Marino},
  \citenamefont {Parisi},\ and\ \citenamefont
  {Ricci-Tersenghi}}]{Marino16_12996}%
  \BibitemOpen
  \bibfield  {author} {\bibinfo {author} {\bibfnamefont {R.}~\bibnamefont
  {Marino}}, \bibinfo {author} {\bibfnamefont {G.}~\bibnamefont {Parisi}},\
  and\ \bibinfo {author} {\bibfnamefont {F.}~\bibnamefont {Ricci-Tersenghi}},\
  }\bibfield  {title} {\bibinfo {title} {The backtracking survey propagation
  algorithm for solving random k-sat problems},\ }\href@noop {} {\bibfield
  {journal} {\bibinfo  {journal} {Nature communications}\ }\textbf {\bibinfo
  {volume} {7}},\ \bibinfo {pages} {12996} (\bibinfo {year}
  {2016})}\BibitemShut {NoStop}%
\bibitem [{\citenamefont {Gabri{\'e}}(2020)}]{Gabrie20_223002}%
  \BibitemOpen
  \bibfield  {author} {\bibinfo {author} {\bibfnamefont {M.}~\bibnamefont
  {Gabri{\'e}}},\ }\bibfield  {title} {\bibinfo {title} {Mean-field inference
  methods for neural networks},\ }\href@noop {} {\bibfield  {journal} {\bibinfo
   {journal} {Journal of Physics A: Mathematical and Theoretical}\ }\textbf
  {\bibinfo {volume} {53}},\ \bibinfo {pages} {223002} (\bibinfo {year}
  {2020})}\BibitemShut {NoStop}%
\bibitem [{\citenamefont {Crotti}\ and\ \citenamefont
  {Braunstein}(2023)}]{Crotti23_e2307935120}%
  \BibitemOpen
  \bibfield  {author} {\bibinfo {author} {\bibfnamefont {S.}~\bibnamefont
  {Crotti}}\ and\ \bibinfo {author} {\bibfnamefont {A.}~\bibnamefont
  {Braunstein}},\ }\bibfield  {title} {\bibinfo {title} {Matrix product belief
  propagation for reweighted stochastic dynamics over graphs},\ }\href@noop {}
  {\bibfield  {journal} {\bibinfo  {journal} {Proc. Natl. Acad. Sci. USA}\
  }\textbf {\bibinfo {volume} {120}},\ \bibinfo {pages} {e2307935120} (\bibinfo
  {year} {2023})}\BibitemShut {NoStop}%
\bibitem [{\citenamefont {Pedretti}\ \emph {et~al.}(2025)\citenamefont
  {Pedretti}, \citenamefont {B{\"o}hm}, \citenamefont {Bhattacharya},
  \citenamefont {Heittmann}, \citenamefont {Zhang}, \citenamefont {Hizzani},
  \citenamefont {Hutchinson}, \citenamefont {Kwon}, \citenamefont {Moon},
  \citenamefont {Valiante} \emph {et~al.}}]{Pedretti25_7}%
  \BibitemOpen
  \bibfield  {author} {\bibinfo {author} {\bibfnamefont {G.}~\bibnamefont
  {Pedretti}}, \bibinfo {author} {\bibfnamefont {F.}~\bibnamefont {B{\"o}hm}},
  \bibinfo {author} {\bibfnamefont {T.}~\bibnamefont {Bhattacharya}}, \bibinfo
  {author} {\bibfnamefont {A.}~\bibnamefont {Heittmann}}, \bibinfo {author}
  {\bibfnamefont {X.}~\bibnamefont {Zhang}}, \bibinfo {author} {\bibfnamefont
  {M.}~\bibnamefont {Hizzani}}, \bibinfo {author} {\bibfnamefont
  {G.}~\bibnamefont {Hutchinson}}, \bibinfo {author} {\bibfnamefont
  {D.}~\bibnamefont {Kwon}}, \bibinfo {author} {\bibfnamefont {J.}~\bibnamefont
  {Moon}}, \bibinfo {author} {\bibfnamefont {E.}~\bibnamefont {Valiante}},
  \emph {et~al.},\ }\bibfield  {title} {\bibinfo {title} {Solving boolean
  satisfiability problems with resistive content addressable memories},\
  }\href@noop {} {\bibfield  {journal} {\bibinfo  {journal} {npj Unconventional
  Computing}\ }\textbf {\bibinfo {volume} {2}},\ \bibinfo {pages} {7} (\bibinfo
  {year} {2025})}\BibitemShut {NoStop}%
\bibitem [{\citenamefont {Machado}\ \emph {et~al.}(2023)\citenamefont
  {Machado}, \citenamefont {Mulet},\ and\ \citenamefont
  {Ricci-Tersenghi}}]{Machado23_123301}%
  \BibitemOpen
  \bibfield  {author} {\bibinfo {author} {\bibfnamefont {D.}~\bibnamefont
  {Machado}}, \bibinfo {author} {\bibfnamefont {R.}~\bibnamefont {Mulet}},\
  and\ \bibinfo {author} {\bibfnamefont {F.}~\bibnamefont {Ricci-Tersenghi}},\
  }\bibfield  {title} {\bibinfo {title} {Improved mean-field dynamical
  equations are able to detect the two-step relaxation in glassy dynamics at
  low temperatures},\ }\href@noop {} {\bibfield  {journal} {\bibinfo  {journal}
  {Journal of Statistical Mechanics: Theory and Experiment}\ }\textbf {\bibinfo
  {volume} {2023}},\ \bibinfo {pages} {123301} (\bibinfo {year}
  {2023})}\BibitemShut {NoStop}%
\bibitem [{\citenamefont {Machado}\ \emph {et~al.}(2025)\citenamefont
  {Machado}, \citenamefont {Gonz{\'a}lez-Garc{\'\i}a},\ and\ \citenamefont
  {Mulet}}]{Machado25_arxiv}%
  \BibitemOpen
  \bibfield  {author} {\bibinfo {author} {\bibfnamefont {D.}~\bibnamefont
  {Machado}}, \bibinfo {author} {\bibfnamefont {J.}~\bibnamefont
  {Gonz{\'a}lez-Garc{\'\i}a}},\ and\ \bibinfo {author} {\bibfnamefont
  {R.}~\bibnamefont {Mulet}},\ }\bibfield  {title} {\bibinfo {title} {Local
  equations describe unreasonably efficient stochastic algorithms in random
  k-sat},\ }\href@noop {} {\bibfield  {journal} {\bibinfo  {journal} {arXiv
  preprint arXiv:2504.06757}\ } (\bibinfo {year} {2025})}\BibitemShut {NoStop}%
\bibitem [{\citenamefont {Seitz}\ \emph {et~al.}(2005)\citenamefont {Seitz},
  \citenamefont {Alava},\ and\ \citenamefont {Orponen}}]{Seitz05_P06006}%
  \BibitemOpen
  \bibfield  {author} {\bibinfo {author} {\bibfnamefont {S.}~\bibnamefont
  {Seitz}}, \bibinfo {author} {\bibfnamefont {M.}~\bibnamefont {Alava}},\ and\
  \bibinfo {author} {\bibfnamefont {P.}~\bibnamefont {Orponen}},\ }\bibfield
  {title} {\bibinfo {title} {Focused local search for random
  3-satisfiability},\ }\href@noop {} {\bibfield  {journal} {\bibinfo  {journal}
  {Journal of Statistical Mechanics: Theory and Experiment}\ }\textbf {\bibinfo
  {volume} {2005}},\ \bibinfo {pages} {P06006} (\bibinfo {year}
  {2005})}\BibitemShut {NoStop}%
\bibitem [{\citenamefont {Toenshoff}\ \emph {et~al.}(2020)\citenamefont
  {Toenshoff}, \citenamefont {Ritzert}, \citenamefont {Wolf},\ and\
  \citenamefont {Grohe}}]{Toenshoff20_1909}%
  \BibitemOpen
  \bibfield  {author} {\bibinfo {author} {\bibfnamefont {J.}~\bibnamefont
  {Toenshoff}}, \bibinfo {author} {\bibfnamefont {M.}~\bibnamefont {Ritzert}},
  \bibinfo {author} {\bibfnamefont {H.}~\bibnamefont {Wolf}},\ and\ \bibinfo
  {author} {\bibfnamefont {M.}~\bibnamefont {Grohe}},\ }\href@noop {} {\bibinfo
  {title} {Graph neural networks for maximum constraint satisfaction}}
  (\bibinfo {year} {2020}),\ \Eprint {https://arxiv.org/abs/1909.08387}
  {arXiv:1909.08387 [cs.AI]} \BibitemShut {NoStop}%
\bibitem [{\citenamefont {M{\"u}ller}\ \emph {et~al.}(2025)\citenamefont
  {M{\"u}ller}, \citenamefont {Singh}, \citenamefont {Wilhelm},\ and\
  \citenamefont {Bode}}]{muller2025_023165}%
  \BibitemOpen
  \bibfield  {author} {\bibinfo {author} {\bibfnamefont {T.}~\bibnamefont
  {M{\"u}ller}}, \bibinfo {author} {\bibfnamefont {A.}~\bibnamefont {Singh}},
  \bibinfo {author} {\bibfnamefont {F.~K.}\ \bibnamefont {Wilhelm}},\ and\
  \bibinfo {author} {\bibfnamefont {T.}~\bibnamefont {Bode}},\ }\bibfield
  {title} {\bibinfo {title} {Limitations of quantum approximate optimization in
  solving generic higher-order constraint-satisfaction problems},\ }\href@noop
  {} {\bibfield  {journal} {\bibinfo  {journal} {Physical Review Research}\
  }\textbf {\bibinfo {volume} {7}},\ \bibinfo {pages} {023165} (\bibinfo {year}
  {2025})}\BibitemShut {NoStop}%
\bibitem [{\citenamefont {Angelini}\ \emph {et~al.}(2025)\citenamefont
  {Angelini}, \citenamefont {Avila-Gonz{\'a}lez}, \citenamefont {D'Amico},
  \citenamefont {Machado}, \citenamefont {Mulet},\ and\ \citenamefont
  {Ricci-Tersenghi}}]{angelini2025algorithmic}%
  \BibitemOpen
  \bibfield  {author} {\bibinfo {author} {\bibfnamefont {M.}~\bibnamefont
  {Angelini}}, \bibinfo {author} {\bibfnamefont {M.}~\bibnamefont
  {Avila-Gonz{\'a}lez}}, \bibinfo {author} {\bibfnamefont {F.}~\bibnamefont
  {D'Amico}}, \bibinfo {author} {\bibfnamefont {D.}~\bibnamefont {Machado}},
  \bibinfo {author} {\bibfnamefont {R.}~\bibnamefont {Mulet}},\ and\ \bibinfo
  {author} {\bibfnamefont {F.}~\bibnamefont {Ricci-Tersenghi}},\ }\bibfield
  {title} {\bibinfo {title} {Algorithmic thresholds in combinatorial
  optimization depend on the time scaling},\ }\href@noop {} {\bibfield
  {journal} {\bibinfo  {journal} {arXiv preprint arXiv:2504.11174}\ } (\bibinfo
  {year} {2025})}\BibitemShut {NoStop}%
\bibitem [{\citenamefont {Barthel}\ \emph {et~al.}(2003)\citenamefont
  {Barthel}, \citenamefont {Hartmann},\ and\ \citenamefont
  {Weigt}}]{Barthel03_066104}%
  \BibitemOpen
  \bibfield  {author} {\bibinfo {author} {\bibfnamefont {W.}~\bibnamefont
  {Barthel}}, \bibinfo {author} {\bibfnamefont {A.~K.}\ \bibnamefont
  {Hartmann}},\ and\ \bibinfo {author} {\bibfnamefont {M.}~\bibnamefont
  {Weigt}},\ }\bibfield  {title} {\bibinfo {title} {Solving satisfiability
  problems by fluctuations: The dynamics of stochastic local search
  algorithms},\ }\href@noop {} {\bibfield  {journal} {\bibinfo  {journal}
  {Phys. Rev. E}\ }\textbf {\bibinfo {volume} {67}},\ \bibinfo {pages} {066104}
  (\bibinfo {year} {2003})}\BibitemShut {NoStop}%
\bibitem [{\citenamefont {Swendsen}\ and\ \citenamefont
  {Wang}(1986)}]{swendsen1986_2607}%
  \BibitemOpen
  \bibfield  {author} {\bibinfo {author} {\bibfnamefont {R.~H.}\ \bibnamefont
  {Swendsen}}\ and\ \bibinfo {author} {\bibfnamefont {J.-S.}\ \bibnamefont
  {Wang}},\ }\bibfield  {title} {\bibinfo {title} {Replica monte carlo
  simulation of spin glasses},\ }\href@noop {} {\bibfield  {journal} {\bibinfo
  {journal} {Physical review letters}\ }\textbf {\bibinfo {volume} {57}},\
  \bibinfo {pages} {2607} (\bibinfo {year} {1986})}\BibitemShut {NoStop}%
\bibitem [{\citenamefont {Geyer}\ \emph {et~al.}(1991)\citenamefont {Geyer}
  \emph {et~al.}}]{geyer1991computing}%
  \BibitemOpen
  \bibfield  {author} {\bibinfo {author} {\bibfnamefont {C.~J.}\ \bibnamefont
  {Geyer}} \emph {et~al.},\ }\bibfield  {title} {\bibinfo {title} {Computing
  science and statistics: Proceedings of the 23rd symposium on the interface},\
  }\href@noop {} {\bibfield  {journal} {\bibinfo  {journal} {American
  Statistical Association, New York}\ }\textbf {\bibinfo {volume} {156}}
  (\bibinfo {year} {1991})}\BibitemShut {NoStop}%
\bibitem [{\citenamefont {Chowdhury}\ \emph {et~al.}(2025)\citenamefont
  {Chowdhury}, \citenamefont {Aadit}, \citenamefont {Grimaldi}, \citenamefont
  {Raimondo}, \citenamefont {Raut}, \citenamefont {Lott}, \citenamefont
  {Mentink}, \citenamefont {Rams}, \citenamefont {Ricci-Tersenghi},
  \citenamefont {Chiappini} \emph {et~al.}}]{Chowdhury25_9193}%
  \BibitemOpen
  \bibfield  {author} {\bibinfo {author} {\bibfnamefont {S.}~\bibnamefont
  {Chowdhury}}, \bibinfo {author} {\bibfnamefont {N.~A.}\ \bibnamefont
  {Aadit}}, \bibinfo {author} {\bibfnamefont {A.}~\bibnamefont {Grimaldi}},
  \bibinfo {author} {\bibfnamefont {E.}~\bibnamefont {Raimondo}}, \bibinfo
  {author} {\bibfnamefont {A.}~\bibnamefont {Raut}}, \bibinfo {author}
  {\bibfnamefont {P.~A.}\ \bibnamefont {Lott}}, \bibinfo {author}
  {\bibfnamefont {J.~H.}\ \bibnamefont {Mentink}}, \bibinfo {author}
  {\bibfnamefont {M.~M.}\ \bibnamefont {Rams}}, \bibinfo {author}
  {\bibfnamefont {F.}~\bibnamefont {Ricci-Tersenghi}}, \bibinfo {author}
  {\bibfnamefont {M.}~\bibnamefont {Chiappini}}, \emph {et~al.},\ }\bibfield
  {title} {\bibinfo {title} {Pushing the boundary of quantum advantage in hard
  combinatorial optimization with probabilistic computers},\ }\href@noop {}
  {\bibfield  {journal} {\bibinfo  {journal} {Nature Communications}\ }\textbf
  {\bibinfo {volume} {16}},\ \bibinfo {pages} {9193} (\bibinfo {year}
  {2025})}\BibitemShut {NoStop}%
\bibitem [{\citenamefont {Bertsimas}\ and\ \citenamefont
  {Tsitsiklis}(1993)}]{bertsimas1993_10}%
  \BibitemOpen
  \bibfield  {author} {\bibinfo {author} {\bibfnamefont {D.}~\bibnamefont
  {Bertsimas}}\ and\ \bibinfo {author} {\bibfnamefont {J.}~\bibnamefont
  {Tsitsiklis}},\ }\bibfield  {title} {\bibinfo {title} {Simulated annealing},\
  }\href@noop {} {\bibfield  {journal} {\bibinfo  {journal} {Statistical
  science}\ }\textbf {\bibinfo {volume} {8}},\ \bibinfo {pages} {10} (\bibinfo
  {year} {1993})}\BibitemShut {NoStop}%
\bibitem [{\citenamefont {Coolen}(2000)}]{Coolen00_arxiv_I}%
  \BibitemOpen
  \bibfield  {author} {\bibinfo {author} {\bibfnamefont {A.~C.~C.}\
  \bibnamefont {Coolen}},\ }\bibfield  {title} {\bibinfo {title} {Statistical
  mechanics of recurrent neural networks i. statics},\ }\href@noop {}
  {\bibfield  {journal} {\bibinfo  {journal} {ArXiv}\ } (\bibinfo {year}
  {2000})}\BibitemShut {NoStop}%
\end{thebibliography}
\end{document}